\documentclass[
    twocolumn,
]{aastex7}

\usepackage{amsmath}
\usepackage{microtype}

\providecommand{\sorthelp}[1]{}

\begin{document}

\title{Tracing Galactic Dust Reddening with Polycyclic Aromatic Hydrocarbon Emission: \\  SPHEREx Observations of the North Ecliptic Pole} 

% To use this file: simply \input{author_list_nep_pah_based_extinction.tex} in your main TeX file where you want the author list to go!
% generated for paper: NEP PAH Based Extinction
% 2026-11-12 16:45:26
\author[0000-0002-3455-1826]{Dennis~Lee}%
\affiliation{Jet Propulsion Laboratory, California Institute of Technology, 4800 Oak Grove Drive, Pasadena, CA 91109, USA}%
\affiliation{IPAC, California Institute of Technology, MC 100-22, 1200 E California Blvd, Pasadena, CA 91125, USA}%
\altaffiliation{\textcopyright~2026 California Institute of Technology. Government sponsorship acknowledged.}%
\email{dennisl@caltech.edu}%
\author[0000-0001-7449-4638]{Brandon~S.~Hensley}%
\affiliation{Jet Propulsion Laboratory, California Institute of Technology, 4800 Oak Grove Drive, Pasadena, CA 91109, USA}%
\email{brandon.s.hensley@jpl.nasa.gov}%
\author[0000-0001-5929-4187]{Tzu-Ching~Chang}%
\affiliation{Jet Propulsion Laboratory, California Institute of Technology, 4800 Oak Grove Drive, Pasadena, CA 91109, USA}%
\affiliation{Department of Physics, California Institute of Technology, 1200 E. California Boulevard, Pasadena, CA 91125, USA}%
\email{tzu@caltech.edu}%
\author[0000-0001-7432-2932]{Olivier~Dor\'{e}}%
\affiliation{Jet Propulsion Laboratory, California Institute of Technology, 4800 Oak Grove Drive, Pasadena, CA 91109, USA}%
\affiliation{Department of Physics, California Institute of Technology, 1200 E. California Boulevard, Pasadena, CA 91125, USA}%
\email{olivier.dore@caltech.edu }%
\author[0000-0002-3993-0745]{Matthew~L.~N.~Ashby}%
\affiliation{Center for Astrophysics $|$ Harvard \& Smithsonian, Optical and Infrared Astronomy Division, Cambridge, MA 02138, USA}%
\email{mashby@cfa.harvard.edu}%
\author[0000-0002-5710-5212]{James~J.~Bock}%
\affiliation{Department of Physics, California Institute of Technology, 1200 E. California Boulevard, Pasadena, CA 91125, USA}%
\affiliation{Jet Propulsion Laboratory, California Institute of Technology, 4800 Oak Grove Drive, Pasadena, CA 91109, USA}%
\email{jjb@astro.caltech.edu}%
\author[0009-0000-3415-2203]{Shuang-Shuang~Chen}%
\affiliation{Department of Physics, California Institute of Technology, 1200 E. California Boulevard, Pasadena, CA 91125, USA}%
\email{schen6@caltech.edu}%
\author[0000-0002-5437-0504]{Yun-Ting~Cheng}%
\affiliation{Department of Physics, California Institute of Technology, 1200 E. California Boulevard, Pasadena, CA 91125, USA}%
\affiliation{Jet Propulsion Laboratory, California Institute of Technology, 4800 Oak Grove Drive, Pasadena, CA 91109, USA}%
\email{ycheng3@caltech.edu}%
\author[0000-0001-6320-261X]{Yi-Kuan~Chiang}%
\affiliation{Academia Sinica Institute of Astronomy and Astrophysics (ASIAA), No. 1, Section 4, Roosevelt Road, Taipei 10617, Taiwan}%
\email{ykchiang@asiaa.sinica.edu.tw}%
\author[0000-0002-3892-0190]{Asantha~Cooray}%
\affiliation{Department of Physics \& Astronomy, University of California Irvine, Irvine, CA 92697, USA}%
\email{acooray@uci.edu}%
\author[0000-0002-4650-8518]{Brendan~P.~Crill}%
\affiliation{Jet Propulsion Laboratory, California Institute of Technology, 4800 Oak Grove Drive, Pasadena, CA 91109, USA}%
\affiliation{Department of Physics, California Institute of Technology, 1200 E. California Boulevard, Pasadena, CA 91125, USA}%
\email{bcrill@jpl.nasa.gov}%
\author[0000-0002-7471-719X]{Ari~J.~Cukierman}%
\affiliation{Department of Physics, California Institute of Technology, 1200 E. California Boulevard, Pasadena, CA 91125, USA}%
\email{ajcukier@caltech.edu}%
\author[0000-0002-9382-9832]{Andreas~L.~Faisst}%
\affiliation{IPAC, California Institute of Technology, MC 100-22, 1200 E California Blvd, Pasadena, CA 91125, USA}%
\email{afaisst@caltech.edu}%
\author[0000-0002-5599-4650]{Joseph~L.~Hora}%
\affiliation{Center for Astrophysics $|$ Harvard \& Smithsonian, Optical and Infrared Astronomy Division, Cambridge, MA 02138, USA}%
\email{jhora@cfa.harvard.edu}%
\author[0009-0009-1219-5128]{Zhaoyu~Huai}%
\affiliation{Department of Physics, California Institute of Technology, 1200 E. California Boulevard, Pasadena, CA 91125, USA}%
\email{zhuai@caltech.edu}%
\author[0000-0001-5812-1903]{Howard~Hui}%
\affiliation{Department of Physics, California Institute of Technology, 1200 E. California Boulevard, Pasadena, CA 91125, USA}%
\affiliation{Jet Propulsion Laboratory, California Institute of Technology, 4800 Oak Grove Drive, Pasadena, CA 91109, USA}%
\email{hhui@caltech.edu}%
\author[0000-0002-2770-808X]{Woong-Seob~Jeong}%
\affiliation{Korea Astronomy and Space Science Institute (KASI), 776 Daedeok-daero, Yuseong-gu, Daejeon 34055, Republic of Korea}%
\email{jeongws@kasi.re.kr}%
\author[0000-0002-5016-050X]{Miju~Kang}%
\affiliation{Korea Astronomy and Space Science Institute (KASI), 776 Daedeok-daero, Yuseong-gu, Daejeon 34055, Republic of Korea}%
\email{mjkang@kasi.re.kr}%
\author[0000-0002-9548-1526]{Carey~M.~Lisse}%
\affiliation{Johns Hopkins University, 3400 N Charles St, Baltimore, MD 21218, USA}%
\affiliation{Johns Hopkins University Applied Physics Laboratory, Laurel, MD 20723, USA}%
\email{carey.lisse@jhuapl.edu}%
\author[0000-0001-5382-6138]{Daniel~C.~Masters}%
\affiliation{IPAC, California Institute of Technology, MC 100-22, 1200 E California Blvd, Pasadena, CA 91125, USA}%
\email{dmasters@ipac.caltech.edu}%
\author[0000-0002-6025-0680]{Gary~J.~Melnick}%
\affiliation{Center for Astrophysics $|$ Harvard \& Smithsonian, Optical and Infrared Astronomy Division, Cambridge, MA 02138, USA}%
\email{gmelnick@cfa.harvard.edu}%
\author[0000-0001-9368-3186]{Chi~H.~Nguyen}%
\affiliation{Department of Physics, California Institute of Technology, 1200 E. California Boulevard, Pasadena, CA 91125, USA}%
\email{chnguyen@caltech.edu}%
\author[0000-0002-5158-243X]{Roberta~Paladini}%
\affiliation{IPAC, California Institute of Technology, MC 100-22, 1200 E California Blvd, Pasadena, CA 91125, USA}%
\email{paladini@ipac.caltech.edu}%
\author[0000-0002-0070-3246]{Ji~Yeon~Seok}%
\affiliation{Korea Astronomy and Space Science Institute (KASI), 776 Daedeok-daero, Yuseong-gu, Daejeon 34055, Republic of Korea}%
\email{jyseok@kasi.re.kr}%
\author[0000-0003-1841-2241]{Volker~Tolls}%
\affiliation{Center for Astrophysics $|$ Harvard \& Smithsonian, Optical and Infrared Astronomy Division, Cambridge, MA 02138, USA}%
\email{vtolls@cfa.harvard.edu}%
\author[0000-0003-4990-189X]{Michael~W.~Werner}%
\affiliation{Jet Propulsion Laboratory, California Institute of Technology, 4800 Oak Grove Drive, Pasadena, CA 91109, USA}%
\email{michael.w.werner@jpl.nasa.gov}%
\author[0000-0001-8253-1451]{Michael~Zemcov}%
\affiliation{School of Physics and Astronomy, Rochester Institute of Technology, 1 Lomb Memorial Dr., Rochester, NY 14623, USA}%
\affiliation{Jet Propulsion Laboratory, California Institute of Technology, 4800 Oak Grove Drive, Pasadena, CA 91109, USA}%
\email{mbzsps@rit.edu}
    

\begin{abstract}
We investigate the ability of SPHEREx observations of the $3.3$ $\mu$m polycyclic aromatic hydrocarbon (PAH) emission feature to trace Galactic dust reddening.
Applying a spectral template-based method to SPHEREx data, we measure the strength of the $3.3$ $\mu$m PAH feature in each $6''.2$ pixel of a $22^{\circ}$ by $22^{\circ}$ region centered on the north ecliptic pole (NEP).
We convert the PAH feature strength to reddening by applying a radiation field correction and a linear scale factor.
The resulting $E(B-V)$ map is consistent with far-infrared, stellar, and \ion{H}{1}-based reddening maps at arcminute scales, exhibiting a typical $E(B-V)$ difference of $\lesssim 7$ mmag.
A galaxy count test finds that all reddening maps have a dispersion of $\approx11$ mmag in $E(B-V)$ relative to what would be needed to produce spatially uniform galaxy counts.
While there is some evidence that reddening is overestimated by our map at high PAH abundances, this effect is smaller than the uncertainties on the applied radiation field corrections and limited to $\sim$10 mmag in the most extreme regions of the NEP.
For a diffuse, high Galactic latitude region like the NEP and at the sensitivities of our validation metrics, the SPHEREx PAH-emission based reddening map matches the accuracy of existing reddening maps while being capable of accessing a higher angular resolution of $\sim6''$.
\end{abstract}

\section{Introduction}\label{sec:intro}

Observations of extragalactic and Galactic objects are subjected to both absorption and scattering by interstellar dust in our Galaxy~\citep[][]{1930PASP...42..214T}. % 1930LicOB..14..154T ????
This Galactic dust extinction not only makes the object appear dimmer but also redder.
In order for the intrinsic brightness and color to be inferred, correction for this effect is required~\citep{2011piim.book.....D}.

Extinction correction is particularly important for galaxy surveys such as those by the Euclid~\citep[][]{2022A&A...662A.112E, 2025A&A...697A...1E}, the Nancy Grace Roman Space Telescope~\citep[][]{2015arXiv150303757S, 2019arXiv190205569A}, and the Vera C. Rubin Observatory~\citep[][]{2012arXiv1211.0310L, 2019ApJ...873..111I}.
Aiming to precisely constrain cosmological parameters, these experiments require accurate tracing of Galactic dust and its effects~\citep{2012MNRAS.424..564R, 2013MNRAS.432.2945H}.
Inaccuracies and uncertainties present in available Galactic extinction correction templates have been shown to introduce systematics into cosmological analyses~\citep[e.g.,][]{2025JCAP...02..045K}.

To address this problem, a variety of different techniques have been developed and employed to characterize the impact of interstellar dust.
These methods range from directly measuring dust reddening using background stars or galaxies \citep[e.g.,][]{2006A&A...453..635M, 2010ApJ...719..415P, 2011ApJ...737..103S, 2014ApJ...783..114G, 2015ApJ...810...25G, 2019ApJ...887...93G, 2014MNRAS.443.1192C, 2014MNRAS.445..256S, 2019A&A...631A..32L, 2020A&A...639A.138L, 2024A&A...685A..82E, 2025OJAp....8E..83Z, 2025Sci...387.1209Z, 2025ApJ...992...39Z} to inferring dust extinction from dust emission using observations at infrared and millimeter wavelengths~\cite[e.g.,][]{1998ApJ...500..525S, 2014ApJ...781....5M, 2015ApJ...798...88M} to predicting dust reddening using measurements of 21 cm~\ion{H}{1} emission line~\cite[e.g.,][]{1978ApJ...224..132B, 2017ApJ...846...38L}.

Historically, the most commonly used reddening map is the Schlegel-Finkbeiner-Davis (hereafter SFD) map~\citep[][]{1998ApJ...500..525S}.
Based on far-infrared observations from the Infrared Astronomical Satellite~\citep[IRAS;][]{1984ApJ...278L...1N} and the Diffuse Infrared Background Experiment (DIRBE) on the Cosmic Background Explorer \citep[COBE;][]{1992ApJ...397..420B}, SFD inferred the dust reddening by correlating the temperature-corrected thermal emission from Galactic dust and the extinction from Galactic dust.
This tight relation as well as the all-sky coverage and $6'.1$ angular resolution of the resulting reddening map made it a convenient and effective template for Galactic dust correction.

Nonetheless, reddening maps based on far-infrared dust emission, such as SFD, have been shown to have their own set of limitations.
As most of these maps were derived from relatively broad photometric filters, they have been known to suffer from contamination from the cosmic infrared background \citep[e.g.,][]{2007PASJ...59..205Y, 2019ApJ...870..120C}.
Furthermore, far-infrared emission has been found to be insensitive to a population of small dust grains known as polycyclic aromatic hydrocarbons (PAHs), a family of carbonaceous nanoparticles.
Recent work has shown that changes in PAH abundance can change reddening without affecting far-infrared emission~\citep{2025ApJ...994...61L}.

To address this limitation, we consider the capabilities of the Spectro-Photometer for the History of the universe, Epoch of Reionization, and Ices Explorer~\citep[SPHEREx;][]{2026ApJ...999..139B}.
SPHEREx is a NASA Medium-class Explorer (MIDEX) mission that observes the entire sky at near-infrared wavelengths. 
Beyond its primary scientific goals, SPHEREx's all-sky continuous spectral coverage between $0.75$ and $5.0\,\mu$m enables it to be a powerful probe of the Galactic interstellar medium~\citep[e.g.,][]{2026ApJ..1001..165H, 2026ApJ..1004..176M}.
In particular, as the prominent $3.3\,\mu$m feature---largely believed to originate from PAHs\footnote{In this work, we refer to the 3.3~$\mu$m emission feature as a PAH feature as it is commonly done so in the literature. However, we note that the exact carrier(s) of this feature remains under some debate \citep[see e.g.,][]{2025MNRAS.544.3280T}.}---lies within its observed wavelength range, SPHEREx is capable of  mapping and producing an all-sky map of PAH emission.
In addition, we also consider observations from the Wide-field Infrared Survey Explorer~\cite[WISE,][]{2010AJ....140.1868W}---specifically the full-sky map of diffuse Galactic dust emission from \citet{2014ApJ...781....5M}.
Based on Band 3 observations centered at 12\,$\mu$m (a wavelength near or in many prominent PAH features), such observations are also sensitive to PAH emission.

We expect a tight relation between dust reddening and PAH emission strength by virtue of both being strongly correlated with the dust column.
However, should PAH abundance change relative to the greater dust grain population, then the amount of reddening per unit PAH emission would change, introducing biases into our reddening map.
On the other hand, at higher Galactic latitudes dominated by diffuse emission (regions relevant to cosmological surveys), dust composition is likely more uniform, possibly rendering these concerns inconsequential.

In this work, we explore the ability of PAH observations from SPHEREx to form the basis of an improved dust reddening map toward diffuse regions of our Galaxy.
We focus our investigation on SPHEREx's northern deep field centered on the north ecliptic pole (NEP).
As a consequence of SPHEREx's orbit and survey strategy, two locations near the ecliptic poles receive significantly more visits.
The first of these two ``deep fields'' is centered on the NEP.
This region is relatively diffuse (typical $E(B-V)\sim60$~mmag); especially compared to regions such as those studied in \citet{2026ApJ..1004..176M} or \citet{2026ApJ..1001..165H}. 
Thus, it is more characteristic of the higher Galactic latitude regions observed by cosmological galaxy surveys, making it ideal for understanding SPHEREx's capability to construct a PAH-based reddening map suitable for cosmological applications.

Assessing our SPHEREx reddening map, we find that it traces dust reddening in the NEP as accurately as any existing map we test.
While we do find evidence for biases in the map that are correlated with tracers of PAH abundance, they are highly dependent on the method used to trace PAH abundance.
Regardless of the method, we find that these effects are limited in magnitude and scope.
At the level we are capable of assessing the data, a SPHEREx reddening map of the NEP matches the accuracy of existing maps while being available at higher angular resolution. 

This paper is organized as follows:
In Section~\ref{sec:data}, we describe our spectral mapmaking technique designed to extract spectral features from SPHEREx data and its application to a $22^{\circ}$ by $22^{\circ}$ region toward the NEP. In addition, we describe the method by which we construct the reddening map using this procedure.
Section~\ref{sec:comparison} describes the consistency of our SPHEREx reddening map in the context of a number of existing reddening maps available in the literature.
In Section~\ref{sec:gal_backlight_test}, we apply our galactic backlight test to our SPHEREx reddening map and compare its performance to literature reddening maps. 
In Section~\ref{sec:discussion}, we compare and discuss the performance of our SPHEREx reddening map. We also investigate and characterize sources of biases that may be relevant for a reddening based on PAH emission.
Finally, we conclude and summarize our results in Section~\ref{sec:conclusion}.

\section{Data}\label{sec:data}

\subsection{Spectral Mosaicking Method}
\label{sec:spec_feat_extract}
To construct our map of 3.3\,$\mu$m PAH emission toward the NEP, we begin with the Level-2 (L2) images produced by the SPHEREx Science Data Center (SSDC).
The details of this data product are described in~\citet{2026ApJS..285...67A}.
In summary, the L2 images have undergone astrometric and photometric calibrations and are accompanied with pixel-level flags to indicate issues such as transients, persistence or nonfunctional pixels.
The L2 processing also completes the absolute gain calibration resulting in units of surface brightness (MJy\,sr$^{-1}$).

We collect the L2 images that intersect a $22^{\circ}$ by $22^{\circ}$ region centered at the NEP.
As we are focused on the 3.3\,$\mu$m feature, we only use images from array 4 spanning the $2.42$--$3.82$\,$\mu$m wavelength range. 
We exclude any images that do not meet the highest astrometry standard~\citep[alignment RMS $<1''$, ][]{2026ApJS..285...67A} or were flagged due to contamination from shuttle glow or auroral lines.
After these exclusions, our dataset contains 20,641 individual L2 images in total from the beginning of SPHEREx's science observations in 2025 April 24 to 2026 May 14.

At the wavelengths observed by SPHEREx, the zodiacal light (sunlight scattered by interplanetary dust) is a time-variable foreground. 
Each SPHEREx L2 image is accompanied with a prediction of the diffuse zodiacal light for that image. 
These predictions are based on an updated variant of the model from \citet{1998ApJ...508...44K} \citep[see][for details]{2025ApJS..281...10C}.
In our procedure, we account for contamination from zodiacal light by taking this provided prediction for each L2 image and subtracting it.

We then reproject each L2 image onto a $22^{\circ}$ by $22^{\circ}$ grid with a pixel scale of $6''.2$ centered on the NEP.
After reprojection, each spatial mosaic pixel has multiple measurements of surface brightness at various wavelengths.
A spectral template fitting method that fits the amplitude of multiple linear components is then used to extract the PAH emission feature.
Each component corresponds to a spectral line, a broad spectral feature, or a component of the continuum.
In this work, we model the dust features in the vicinity of 3.3\,$\mu$m with Drude profiles with a linear continuum.
Specifically, following \citet{2020ApJ...905...55L}, we fit for an aromatic feature at 3.3\,$\mu$m, an aliphatic feature 3.4\,$\mu$m, and a plateau feature at 3.47\,$\mu$m.
Our model is:
\begin{equation}\label{eq:pah_model}
I_{\nu}(\lambda) = \theta_\alpha + \theta_\beta \lambda + \sum_{i} \theta_i \left(\frac{\gamma_i^2}{\left(\lambda/\lambda_i - \lambda_i/\lambda\right)^2 + \gamma_i^2}\right)\,,
\end{equation}
where $i \in \{3.3,\,3.4,\,3.47\}$ and $\theta_\alpha$ and $\theta_\beta$ are the parameters for the linear continuum.
Each Drude profile is parameterized by a central wavelength $\lambda_i$ and fractional FWHM $\gamma_i$ where $\gamma_i$ is defined as the ratio of FWHM to central wavelength (i.e., $\gamma_i\equiv {\rm FWHM}_i /\lambda_i$).
Using the values provided in \citet{2020ApJ...905...55L}, we fix the 3.3\,$\mu$m feature's central wavelength and fractional FWHM as $\lambda_{3.3}=3.29$\,$\mu$m and $\gamma_{3.3}=0.013$ respectively. 
For the aliphatic feature at 3.4\,$\mu$m, we use $\lambda_{3.4}=3.40$\,$\mu$m and $\gamma_{3.4}=0.009$. For the plateau feature, we fix $\lambda_{3.47}=3.47$\,$\mu$m and $\gamma_{3.47}=0.10$.
For our spectral template, by fixing each feature's central wavelength and fractional FWHM, we do not consider any possible variations to the profiles of our features. 
Overall, studies have shown that PAH profiles vary by about $\sim0.2\%$ and are overall stable across a variety of Galactic sources~\citep{1991ApJ...380..452T, 2002A&A...390.1089P, 2025ApJ...993...30C, 2025MNRAS.544.3280T}.

\begin{figure*}
    \centering
    \includegraphics[width=0.99\textwidth]{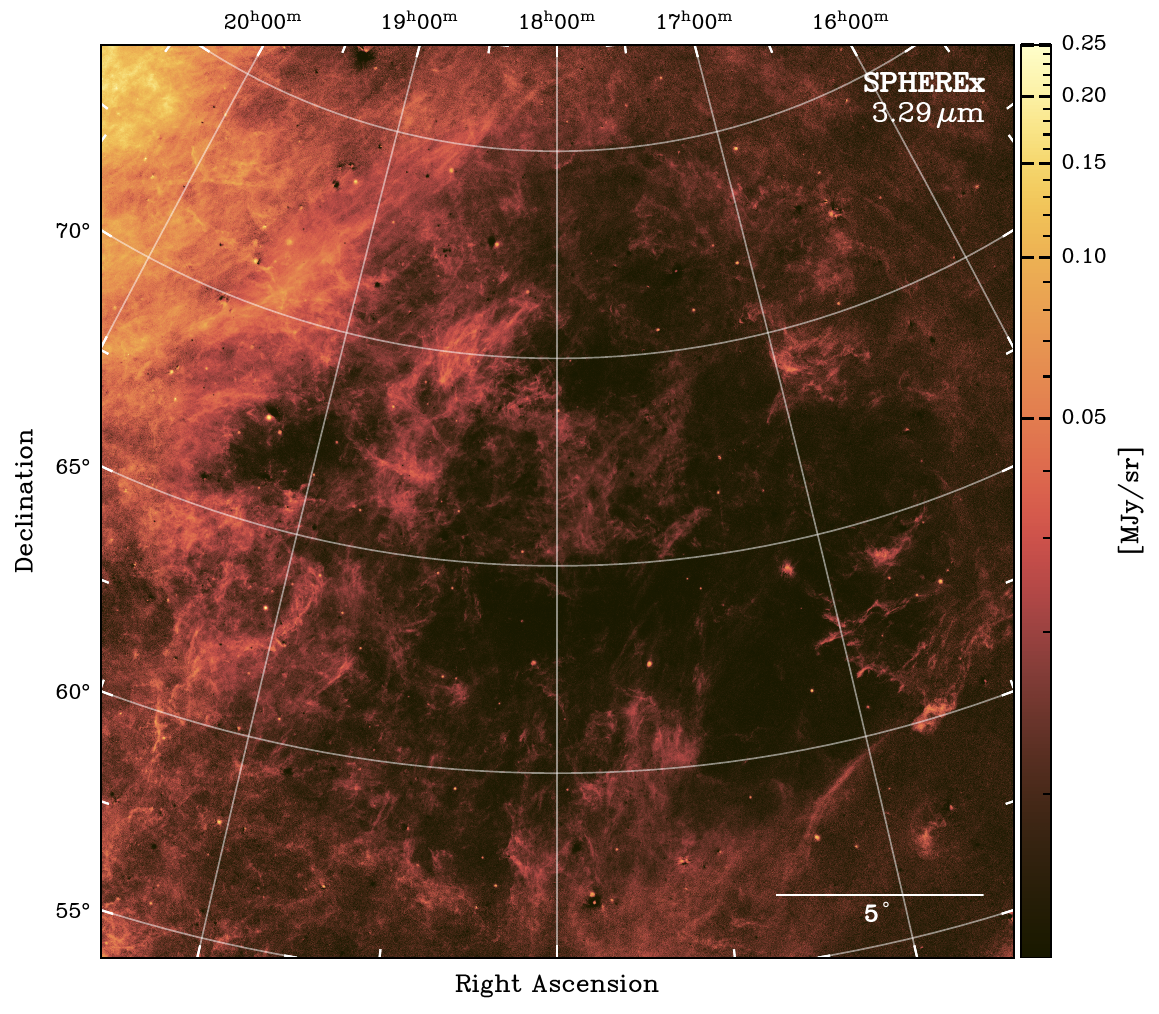}
    \caption{SPHEREx 3.3\,$\mu$m PAH emission in a $22^{\circ}$ by $22^{\circ}$ centered on the north ecliptic pole. The pixel scale is $6''.2$. Values are computed based on the method described in Section~\ref{sec:spec_feat_extract} and many, though not all, of the point sources are removed in that procedure. For the purpose of visualization in this figure, we have used~\texttt{CloudClean.jl} to inpaint masked pixels such as those vacated by point source masking~\citep{2022ApJ...933..155S}.}
    \label{fig:spherex_nep}
\end{figure*}

We then perform a fit for every sight line for the parameters to obtain the amplitude for each component.
When performing this fit, we do not aggregate or bin any observations spectrally. 
Instead, we convolve the model with the passband measured for each detector pixel of each observation~\citep{2026ApJS..284...10H}.
This is necessary as the detailed shape of the passband is required to accurately model the emission of the spectral features.
This approach differs from the procedure described in \citet{2026ApJS..286...22C} and used in \citet{2026ApJ..1004..176M} where observations are binned into specified narrow-band spectral channels.

To account for point sources, we use a point source mask based on WISE~\citep{2014ApJ...781....5M}.
In addition, we mask any pixels with reduced chi-square $\chi^2_{\nu} > 20$. These are typically stars that are not well represented by our PAH model.
Together, these two masks account for many, though not all, of the point sources.
In total $\approx5\%$ of pixels are masked, with the median reduced chi-square for the remaining map being $\chi^2_{\nu} = 1.29$.
We choose not to use the standard source flag provided in the L2 images as, due to the number of point sources present, this flag removes significant portions of the images.

Although we fit simultaneously for all three dust features in this wavelength range, this work is primarily concerned with the aromatic feature centered at 3.3\,$\mu$m. 
As such, for the remainder of the work, we will focus only on the 3.3\,$\mu$m feature (corresponding to $\theta_{3.3}$ in Equation~\ref{eq:pah_model}).
The resulting SPHEREx 3.3\,$\mu$m PAH emission map is shown in Figure~\ref{fig:spherex_nep}.

\begin{figure*}
    \centering
    \includegraphics[width=0.99\textwidth]{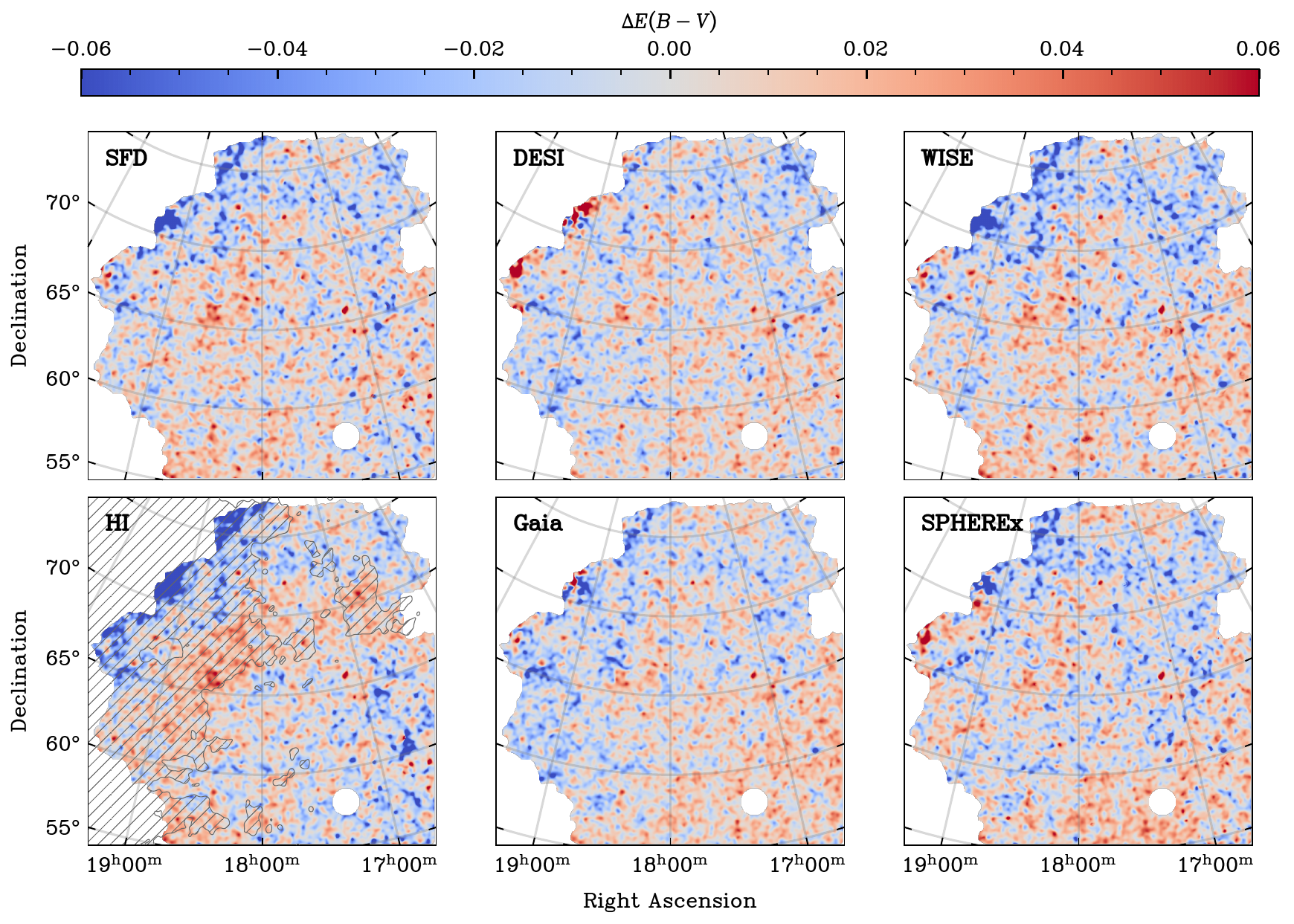}
    \caption{
    Results of the galaxy density test for a suite of reddening maps at a common resolution of $16'1$. 
    Only the region common to the coverage of all reddening maps is shown. 
    A small circular region ($0.86^\circ$ radius) in the lower right is also masked due to a high density of sources. This is likely spurious and may potentially be attributed to stars in a nearby galaxy that pass the ELG color cut.
    The only clear large scale spatial feature is the underestimation of reddening seen toward the northeastern region of the maps.
    This coincides with increased interstellar medium density in a region that is closer to the Galactic plane.
    In the result for the HI reddening map, the hatched region indicates where $N_{\text{\ion{H}{1}}}> 4\times10^{20}$~cm$^{-2}$.
    }
    \label{fig:galdense_eff_ebv}
\end{figure*}

\subsection{Reddening Map Construction}\label{sec:data:red_cnstr}
Dust reddening can be derived from dust emission as both quantities are proportional to the amount of dust along a given sight line. In principle, the only remaining factor that needs to be accounted for is the strength of the radiation field as this changes the dust emission but not the reddening. To account for the different populations of dust grains, we can describe the dust reddening as a combination of the smaller PAH dust grains traced by near-/mid-infrared emission, and larger non-PAH dust grains traced by far-infrared emission:
\begin{equation}
E(B-V) = \alpha_{\text{near}} I_{\nu, \text{near}}^{T} + \alpha_{\text{far}} I_{\nu, \text{far}}^{T}\, ,
\end{equation}
where $\alpha_{\text{near}}$ and $\alpha_{\text{far}}$ are calibration factors that can be derived empirically.
Here, the superscript $T$ indicates that this emission has been corrected for the variations in local radiation field. 
SFD, for example, used temperature-corrected $100$~$\mu$m emission (i.e., $I_{\nu, \text{far}}^{T} = I_{\nu, \text{100}\,\mu\text{m}}^{T}$) to construct their dust reddening map, but, by only tracing the far-infrared emitting larger dust grains, such a map has been shown to be insensitive to changes in reddening due to PAH abundance~\citep{2025ApJ...994...61L}.
In this work, we will use $I_{\nu, \text{near}}^{T} = I_{\nu}^{\rm 3.3, T}$.
Like SFD, by only using observations that trace emission from a portion of the dust population (i.e., only PAHs), it is possible that biases will arise due to varying abundances of either grain population. 
However, if the NEP is a region where such variations are limited in magnitude, the impact may be insignificant.

To obtain a reddening map based on the 3.3\,$\mu$m PAH emission, 
%we follow a similar procedure to \citet{2025ApJ...994...61L} as applied to the WISE 12\,$\mu$m emission data. 
we first perform a correction to the SPHEREx emission data to account for the aforementioned varying dust emissivity as the strength of the radiation field changes. Specifically, following~\citet{2025ApJ...994...61L}, we use:
\begin{equation}\label{eq:pah_temp_corr}
     I_{\nu}^{\rm 3.3, T} =  I_{\nu}^{\rm 3.3} \times \left(\frac{T}{T_0}\right)^{-(4 + \beta_{\rm FIR})}\,,
\end{equation}
where $\beta_{\rm FIR}=1.5$ \citep{planck2016-l11A}. 
For $T$ we use the SFD dust temperature map, which is derived from the 100\,$\mu$m~and~240\,$\mu$m~maps from the DIRBE instrument on COBE \citep{1992ApJ...397..420B}.
We set $T_0=17.9$\,K, the median of $T_{\rm SFD}$ for the whole sky, though the resulting reddening map is independent of choice of $T_0$ since it is degenerate with the scaling factor between $I_\nu^{3.3,{\rm T}}$ and reddening $E(B-V)$.
We do note that the temperature correction map is available at coarser resolution ($\approx 1^\circ.1$) than the resolution of our analysis.

We then fit a simple linear relation between the SPHEREx 3.3\,$\mu$m emission and SFD:
\begin{equation}\label{eq:linear_em_to_red}
    E(B-V)_{\rm SFD} = \alpha+\left(\beta\times I_\nu^{3.3,{\rm T}}\right)\,.
\end{equation}
The resulting coefficients are $\alpha = 0.016$~mag and $\beta = 2.658$ mag (MJy~sr$^{-1}$)$^{-1}$.

\subsection{Literature Reddening Maps}
We compare our SPHEREx PAH emission-derived reddening map with several reddening maps available in the literature, briefly described here:

\begin{enumerate}

    \item \citet[][SFD]{1998ApJ...500..525S} produced the most widely used reddening map based on the all-sky 100~$\mu$m observations from IRAS. The temperature-dependent emissivity of the IRAS data is corrected using the 100~$\mu$m and 240 $\mu$m maps from the DIRBE instrument on COBE. This temperature-corrected IRAS 100\,$\mu$m~emission is then converted to reddening by calibrating to the observed reddening of external elliptical galaxies. 
    However, subsequent work found this calibration to overestimate reddening~\citep{2010ApJ...725.1175S, 2011ApJ...737..103S}.
    To account for this, based on the conversion coefficients tabulated in \citet{2011ApJ...737..103S}, we apply a multiplicative correction factor of 0.884 to the SFD reddening map. 
    We use this updated calibration of SFD for all analyses in this work. 
    
    \item We use a reddening map constructed from WISE observations in Band~3.
    This band, centered at approximately $12$\,$\mu$m, spans numerous PAH features.
    Specifically, we base our reddening map on the processing from \citet[][]{2014ApJ...781....5M}. 
    We use the 15$"$ tiles that cover the NEP and aggressively mask using the provided bit-mask flags to remove contaminated pixels.
    With the exception of the bit-mask values 17 and 22, which do not apply here\footnote{Bit 17 corresponds to masking M31, the SMC, and the LMC while bit 22 corresponds to flagging of the ecliptic plane. Neither are relevant here.}, we use all of the bit-mask pixels.
    These correspond to pixels contaminated by the emission from the Moon or solar system objects, latent ghosts, as well as other instrument effects \citep[see][for details]{2014ApJ...781....5M}.
    To produce a reddening map from the W3 emission, we follow the procedure described in Section~\ref{sec:data:red_cnstr}. We again use the SFD dust temperature map to correct for varying dust emissivity. We then fit the WISE emission with a linear relation analogous to Equation~\ref{eq:linear_em_to_red}. The resulting coefficients for WISE are $\alpha = 0.003$~mag and $\beta = 0.481$ mag (MJy~sr$^{-1}$)$^{-1}$. Note that these coefficients are derived based only on the $22^{\circ}$ by $22^{\circ}$ observations toward the NEP rather than from fitting WISE observations of the full sky to the full-sky SFD map.
    
    \item \citet[][DESI]{2025OJAp....8E..83Z} constructed Galactic reddening maps using extinction measurements toward 2.6 million stars based on stellar spectroscopy from the first two years of the DESI spectroscopic survey. 
    Comparing these observations with DESI imaging, \citet{2025OJAp....8E..83Z} computed reddening in $g-r$ and $r-z$ bands. We use the $g-r$ reddening measurements binned on a  $N_{\rm side}=512$\footnote{\texttt{desi\_dust\_gr\_512.fits}} HEALPix\footnote{\href{http://healpix.sourceforge.net}{http://healpix.sourceforge.net}}~\citep{2005ApJ...622..759G}~map (corresponding to a resolution of $\approx6'.87$). 
    
    \item Using Gaia XP spectra and near-infrared photometry, \citet[][Gaia]{2025Sci...387.1209Z} obtained the dust extinction toward $\sim130$\,million stars. This work modeled both reddening $E(B-V)$ as well as $R_V$, a parameter describing wavelength dependence of  extinction at optical wavelengths.
    To produce 2D reddening maps from these measurements, we select all measurements toward the NEP. 
    We follow the identical set of selection criteria and quality cuts that are described in detail in \citet{2025Sci...387.1209Z} and \citet{2025ApJ...994...61L}.
    In summary, we only use stars beyond 1.0~kpc and exclude unreliable measurements (e.g., poor stellar fitting).
    From the selected stars in the NEP region, we convolve the measurements with a $16'.1$ Gaussian kernel and compute the inverse-variance weighted $E(B-V)$ and the $E(B-V)$ weighted $R_V$ values to create the final 2D Gaia $E(B-V)$ and $R_V$ maps at $16'.1$ resolution.
    
    \item We use a variant of the reddening map created by \citet[][HI]{2017ApJ...846...38L} with the \ion{H}{1} 21-cm emission from the HI4PI Survey~\citep{2016A&A...594A.116H}. In their work, \citet{2017ApJ...846...38L} constructed a reddening map by deriving a single parameter linear correlation between \ion{H}{1} emission\footnote{\texttt{mom0\_-90\_90\_1024.hpx.fits}} with $|v_{\rm LSR}|< 90$\,km\,s$^{-1}$ and observed reddening (from SFD specifically). However, the reddening map was limited to $N_{\text{\ion{H}{1}}}< 4\times10^{20}$~cm$^{-2}$ as above this threshold the relationship between \ion{H}{1} column density and reddening became strongly non-linear. In this work, using the scaling factor from \citet{2017ApJ...846...38L}, we construct a reddening map that includes regions of higher column density as well. Although we expect our extended map to now include regions that underestimate reddening, creating such an extended map will provide us with an additional independent tracer of dust reddening across a larger footprint. 
    
\end{enumerate}

\section{Comparison of Reddening Maps}
\label{sec:comparison}

In this section, we briefly compare the suite of reddening maps toward the NEP---including our SPHEREx map---directly against each other.
With this comparison, we assess the consistency of the various maps.
We degrade our maps to a common resolution of $16.'1$, the resolution of the HI reddening map, for comparison.
For each pair of reddening maps, we compute the difference in $E(B-V)$, subtract the median, and then compute the median absolute deviation (MAD). 
These values are listed in Table~\ref{tab:mad_diffs}. 

While we have degraded our SPHEREx resolution to the coarsest common angular resolution, we note that the MAD values change only slightly when conducting comparisons with pairs of reddening maps at the highest shared resolution.
The MAD between SPHEREx and WISE at $15"$ or between SPHEREx and SFD at $6'.1$ are larger, suggesting that there is increased variations as more structure is resolved, but such increase is approximately $\sim1$\,mmag.
We also note that these trends are robust to the specific choice of statistic. Using the root mean square in place of MAD, for example, changes the exact numerical values, but not the overall trend.

We find the MAD of the differences between the SPHEREx reddening map and the other reddening maps to range between $\approx 4.0$--$6.5$~mmag. 
The largest difference (6.5~mmag) is against the HI reddening map while the smallest difference is with SFD at 4.0~mmag.
The latter is likely somewhat by construction as SFD is the reddening used to calibrate our map of PAH emission.
The difference with WISE, the other PAH tracer, is at 4.6~mmag.

For comparison, the difference between SFD and other reddening maps is somewhat tighter, in the range of $\approx 2.6 - 5.1$\,mmag.
The largest differences among our reddening maps are those computed with the HI reddening map.
The comparison between HI and Gaia reddening map represents the largest value seen in our comparison at $7.5$\,mmag.
When restricting our HI map to regions where $N_{\text{\ion{H}{1}}}< 4\times10^{20}$~cm$^{-2}$, we find that the differences with HI are reduced by $\approx1-1.8$\,mmag.
Notably, the $7.5$\,mmag difference between HI and Gaia becomes $5.7$\,mmag with the restriction applied, closer to the variations between SFD and Gaia for example. 

Though there is certainly evidence that some pairs agree more---SFD and WISE at $2.6$~mmag---the magnitude of differences are small.
All of the reddening maps agree at $<8$\,mmag with typical levels falling closer to $5-6$~mmag.
The differences seen with SPHEREx are consistent and fall squarely within this range.

\begin{deluxetable}{lccccc}
\tabletypesize{\footnotesize}
\tablecaption{Median absolute deviation (MAD) of the differences in $E(B-V)$ between pairs of reddening maps at the common resolution of $16'.1$. Differences are in units of mmag. These values are computed after median subtraction.~\label{tab:mad_diffs}}
\tablehead{
\colhead{} & \colhead{SFD} & \colhead{DESI} & \colhead{WISE} &
\colhead{Gaia} & \colhead{SPHEREx}
}
\startdata
DESI       & 4.6 & $-$ & $-$ & $-$ & $-$ \\
WISE       & 2.6 & 4.6 & $-$ & $-$ & $-$ \\
Gaia       & 5.1 & 4.6 & 4.7 & $-$ & $-$ \\
SPHEREx    & 4.1 & 4.6 & 4.7 & 4.7 & $-$ \\
\hline
HI            & 3.3 & 6.6 & 5.6 & 7.5 & 6.6 \\
HI (diffuse)  & 2.3 & 5.2 & 4.5 & 5.7 & 5.5 \\
\enddata
\tablecomments{``HI (diffuse)'' are comparisons restricted to regions where $N_{\text{\ion{H}{1}}}< 4\times10^{20}$~cm$^{-2}$, the threshold above which \citet{2017ApJ...846...38L} finds that the relation between \ion{H}{1} column density and reddening begins to break down.}
\end{deluxetable}

\section{Galaxy Backlight Test}\label{sec:gal_backlight_test}
Although there are differences between our reddening maps, none of these reddening maps are \textit{a priori} known to be the `true' reddening map to assess against. As such, we employ the galaxy density test as an objective validation for our reddening maps. In principle, an accurate reddening map should yield a result that is independent of any Galactic structure.
A detailed description of the technique is available in \citet{2025OJAp....8E..83Z} and \citet{2025ApJ...994...61L}. 
In brief, this method uses the DESI emission line galaxy (ELG) sample~\citep{2023AJ....165..126R} selected from the Legacy Surveys (LS) Data Release 9 imaging catalog~\citep{2019AJ....157..168D}.
Using the ELG selection criteria which necessitates a reddening correction, we obtain a spatial map of the ELG galaxy density.
We then compute the median subtracted ELG galaxy density $\Delta \Sigma_{\rm gal}$.
Since the ELG distribution should be uniform over the sky, large-scale spatial over- or underdensities (indicated by positive or negative $\Delta \Sigma_{\rm gal}$ respectively) indicate possible underlying systematics in the reddening map.
In this analysis, we again degrade our reddening maps to a common resolution of $16.'1$ for comparison.

From our residual galaxy density map $\Delta \Sigma_{\rm gal}$, we then compute a reddening error field $\Delta E(B-V)$ using:
\begin{equation}
\Delta E(B-V) = \Delta \Sigma_{\rm gal} \cdot \left. \left( \frac{d E(B-V)}{d\Sigma_{\rm gal}} \right) \right\vert_{E(B-V)_{\rm ext}}\ .
\end{equation}
Here, ${d E(B-V)}/{d\Sigma_{\rm gal}}$ is computed empirically, by deriving the relation between ELG density and dust reddening for each pixel for each reddening map. We then derive $\Delta E(B-V)$ at each spatial pixel for each reddening map.
In essence, what this value $\Delta E(B-V)$ captures is the change in reddening required to cause a galaxy density deviation of $\Delta \Sigma_{\rm gal}$.
A positive value of $\Delta E(B-V)$ indicates overestimates of the reddening by the reddening map, while a negative value of $\Delta E(B-V)$ indicates an underestimate.
The result of our galaxy backlight test is shown in Figure~\ref{fig:galdense_eff_ebv}.
Only the spatial footprint shared by all six reddening maps are shown.

Qualitatively, the primary spatial feature observable is toward the northeastern portion of the region (upper left in the figures).
The SFD, HI, WISE, and SPHEREx reddening maps all show underestimates in this region. 
The stellar reddening maps---DESI and Gaia---show this to a lesser extent, with the DESI reddening map even exhibiting some evidence of overestimation.
Moving toward this direction from the NEP, the interstellar medium becomes denser as the Galactic plane is approached (Figure~\ref{fig:spherex_nep}).
Quantitatively, all six reddening maps show a comparable median absolute deviation (MAD) of $\Delta E(B-V)\approx11$\,mmag ranging from $10.7$\,mmag for the Gaia reddening map to $11.8$\,mmag for the WISE reddening map.

\subsection{Power Spectrum Analysis}

\begin{figure}
    \centering
    \includegraphics[width=0.99\columnwidth]{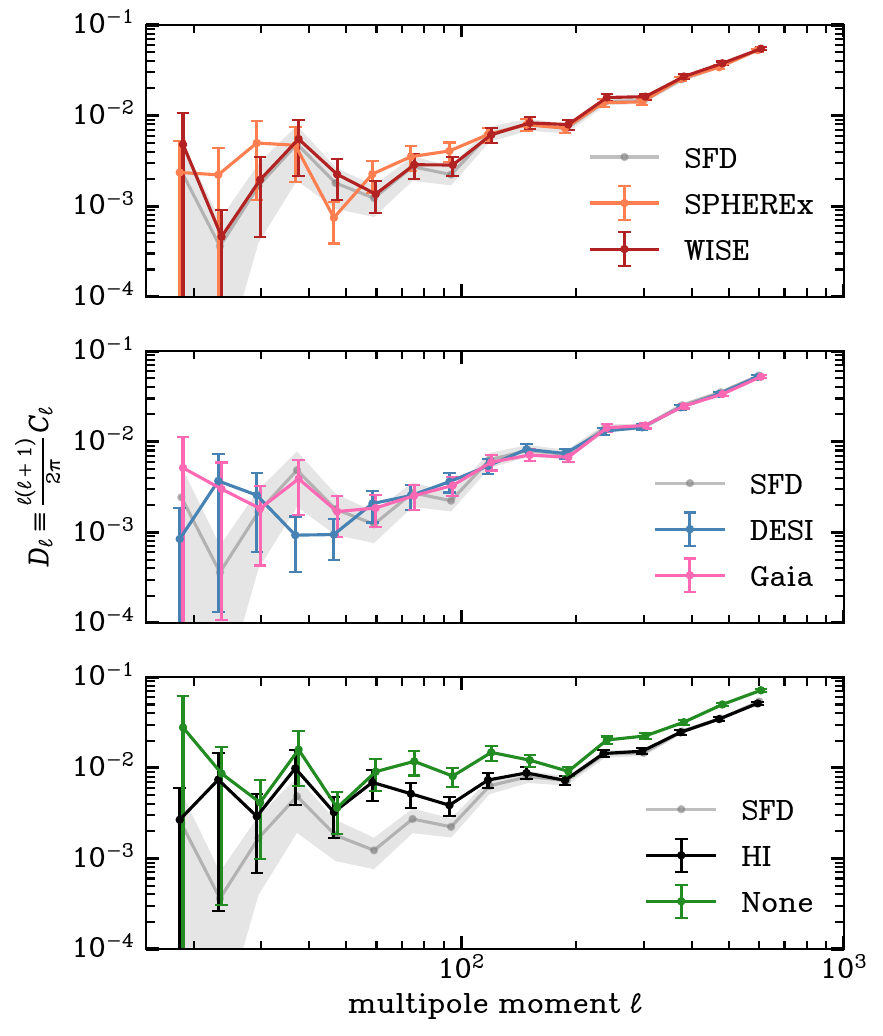}
    \caption{The power spectrum for the galaxy sample selected using the SPHEREx and WISE reddening maps are shown in the top panel. The middle and the bottom panel show for stellar reddening maps (DESI, Gaia) and HI, respectively. The gray curve and band corresponds to the results for SFD and is shown in all three panels for comparison. 
    On the bottom panel, we also show the galaxy sample selected without correction for dust contamination (``None''). For better visibility, each spectrum is plotted with a slight offset in $\ell$.
    No power spectrum corrected for dust contamination shows strong evidence of Galactic contamination. The spectrum of the uncorrected sample is nearly flat before aligning more closely with the corrected samples at higher $\ell$.}    
    \label{fig:galdense_power_spec}
\end{figure}

Overall, from Figure~\ref{fig:galdense_eff_ebv}, with the exception of the denser regions toward the northeast, each reddening map is largely comparable.
Furthermore, the variance of the galaxy counts in each reddening map is similar as well ($\approx 11$ mmag).
To further assess the quality of each reddening map in the NEP, we compute the power spectrum of the resulting galaxy density samples produced from each reddening map.
Specifically, we compute the overdensity, defined as $\delta\equiv n/\bar{n} - 1$ where $\bar{n}$ is the mean density of each sample from each reddening map.
The power spectrum of an overdensity that is uncontaminated is expected to follow a specific structure representing clustering and a Poisson-like component. 
Departure from this spectrum would suggest that our reddening map used to produce that sample is contaminated in some way.

We use the pseudo-$C_\ell$ estimator in {NaMaster} to compute the angular power spectra of each of our galaxy density maps~\citep{2019MNRAS.484.4127A}.
The measurements are binned into 16 equally spaced logarithmic bins from $\ell=16$ to $670$. %($\sim11^\circ$ to $16'$).
The uncertainty in our binned auto-spectrum $C_\ell$ is computed analytically using the Gaussian (sample-variance) approximation~\citep{1995PhRvD..52.4307K}.
The power spectrum of the overdensity map for each of our reddening maps is shown in Figure~\ref{fig:galdense_power_spec}.
We also include the power spectrum of an overdensity map constructed without applying a dust reddening correction.

Qualitatively, the power spectra for all of our galaxy density maps are similar.
All show the expected rise toward higher multipoles that is characteristic of the extragalactic background.
This rise is a combination of an extragalactic clustered component as well as a Poisson-like component.
The power spectrum for the overdensity map constructed without applying a dust reddening correction is higher at high-$\ell$ due to increased shot noise. 
This is expected as the uncorrected sample contains fewer galaxies.

We see no clear presence of a Galactic dust component in any of the spectra corrected for dust reddening.
Galactic dust contamination is expected to manifest as a power law of the form $C_\ell\propto \ell^{\alpha_{\rm d}}$, with 
$\alpha_{\rm d}\sim-3$~\citep[][]{planck2013-pip56, planck2016-l11A}, producing a falling spectrum in $\mathcal{D}_\ell \equiv \ell(\ell+1)C_\ell/2\pi$ toward higher multipoles.
Compared to the corrected samples, we find that the power spectrum for an overdensity computed without a dust reddening correction shows evidence of excess power at low $\ell$. 
While the corrected samples rise throughout the range, the spectrum of the uncorrected sample is nearly flat before aligning more closely with the corrected samples at higher $\ell$.

We quantitatively investigate this by modeling the power spectrum and assess the contributions from different components. 
Following work such as \citet{planck2016-l11A} and \citet{2024ApJ...960...96C}, we model the power spectrum as the sum of three components: a Galactic component, an extragalactic clustering component, and a Poisson-like extragalactic component:
\begin{equation}\label{eq:cell_model}
    C_\ell = A_{\rm d}\left(\frac{\ell}{\ell_0}\right)^{\alpha_{\rm d}} + A_{\rm C}\left(\frac{\ell}{\ell_0}\right)^{\alpha_{\rm C}} + A_{\rm P} ~, 
\end{equation}
where $A_{\rm d}$ is the amplitude of the Galactic component, $A_{\rm C}$ is the amplitude of the clustering component, and $A_{\rm P}$ is the amplitude of the Poisson-like component. $\alpha_{\rm d}$ and $\alpha_{\rm C}$ are the power-law index for the Galactic dust and clustering components, respectively. We set $\ell_0=100$ and fix $\alpha_{\rm d}=-3$. We fit for $A_{\rm C}^i$, $A_{\rm d}^i$, and $A_{\rm P}^i$ jointly where $i$ indicates each of our reddening maps. For all of our reddening maps, we fit for a fixed shared $\alpha_{\rm C}$ as the underlying galaxy sample is the same for all reddening maps.

\begin{deluxetable}{lccc}
\tabletypesize{\footnotesize}
\tablecaption{Best-fit parameters for the joint fit of $C_\ell$ to Equation~\ref{eq:cell_model}. The shared clustering slope is $\alpha_C = -1.01 \pm 0.18$. Amplitudes are in units of $10^{-7}$~sr.~\label{tab:pspec_fit}}
\tablehead{\colhead{} & \colhead{$A_d$} & \colhead{$A_C$} & \colhead{$A_P$}}
\startdata
SFD           & $0.0 \pm 0.6$ & $13.0 \pm 2.2$ & $7.2 \pm 0.9$ \\
DESI          & $0.1 \pm 0.9$ & $15.4 \pm 2.5$ & $6.3 \pm 1.0$ \\
WISE          & $0.0 \pm 0.7$ & $16.9 \pm 2.5$ & $6.8 \pm 1.1$ \\
Gaia          & $2.0 \pm 1.5$ & $13.9 \pm 2.7$ & $6.7 \pm 0.9$ \\
SPHEREx       & $0.0 \pm 1.2$ & $15.3 \pm 2.7$ & $6.5 \pm 1.0$ \\
\hline
HI            & $5.6 \pm 2.2$ & $16.4 \pm 2.9$ & $6.1 \pm 1.0$ \\
HI (diffuse)  & $0.0 \pm 0.6$ & \phn$9.4 \pm 3.1$ & $7.4 \pm 0.9$ \\
\hline
None          & $10.7 \pm 3.6$ & $20.4 \pm 4.2$ & $9.1 \pm 1.3$ \\
\enddata
\tablecomments{``None'' applies no correction. ``HI (diffuse)'' is measured over the smaller footprint where $N_{\text{\ion{H}{1}}}< 4\times10^{20}$~cm$^{-2}$.}
\end{deluxetable}

We perform a positive definite least-squares fit to our data including the overdensity map constructed without applying a dust reddening correction.
The results are shown in Table~\ref{tab:pspec_fit}.
We find a best-fit $\alpha_C = -1.01 \pm 0.18$. 
For our uncorrected sample, we find that its power spectrum requires a dust-like component (i.e., $A_{\rm d}^{\rm None}=(10.7 \pm 3.6)\times10^{-7}$).
On the other hand, we find that the best-fit $A_{\rm d}$ for most of our reddening maps (with the exception of Gaia and HI) are consistent with zero. 
In other words, there is no evidence of a dust component in the power spectrum for samples selected with the SFD, DESI, WISE, and SPHEREx reddening map. 
Gaia, with $A_{\rm d}^{\rm Gaia}=(2.0 \pm 1.5)\times10^{-7}$, shows only tenuous evidence ($<2\sigma$) for a dust component.
Similarly, HI, with an amplitude of {$A_{\rm d}^{\rm HI}=(5.6 \pm 2.2)\times10^{-7}$}, indicates a stronger presence of Milky Way dust in the power spectrum. 
However, when we restrict our analysis to lower column densities using the threshold from \citet{2017ApJ...846...38L}, the $A_{\rm d}^{\rm HI (diffuse)}$ becomes statistically equivalent to zero as well.

\begin{figure*}
    \centering
    \includegraphics[width=0.99\textwidth]{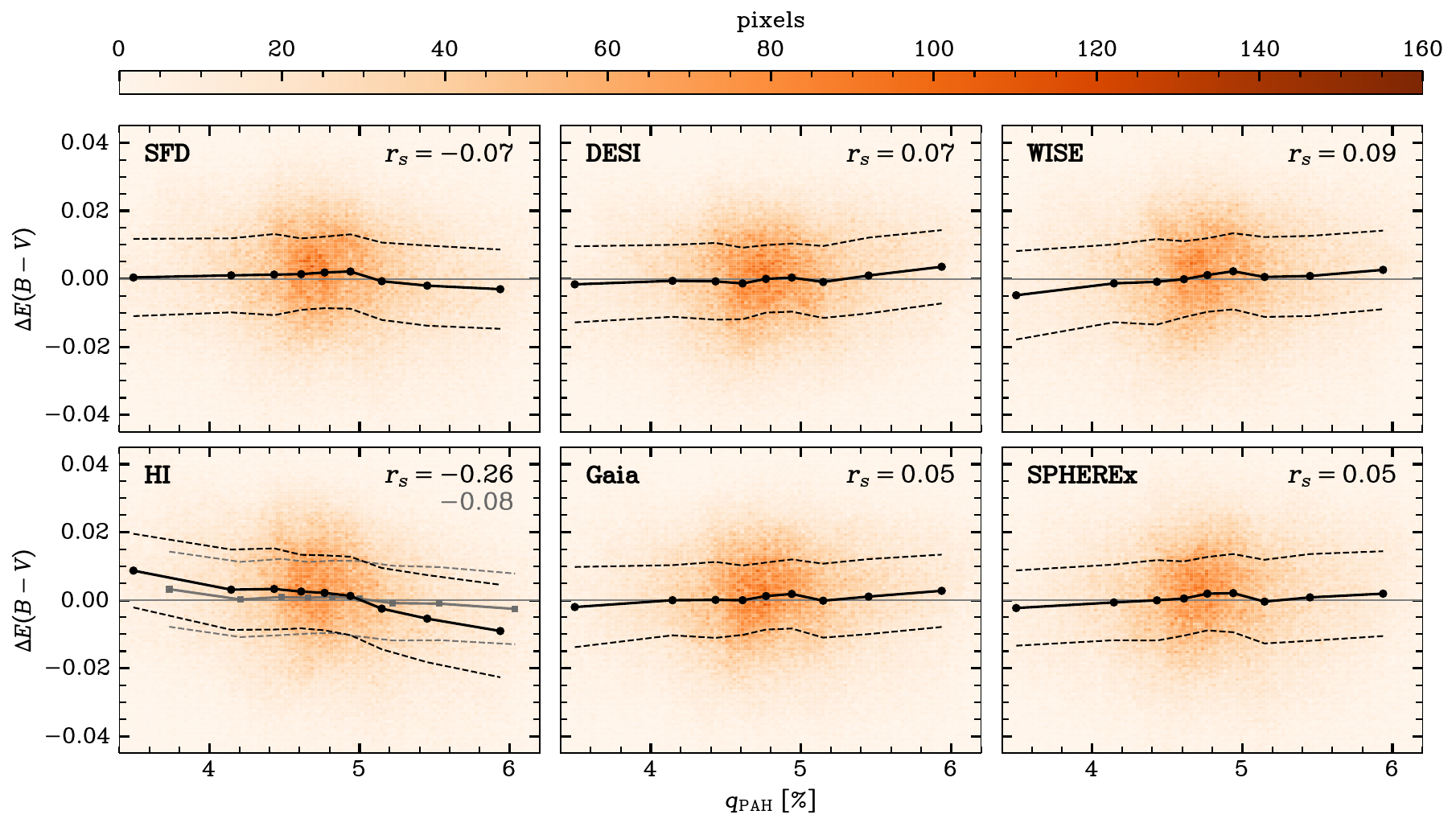}
    \caption{
    $\Delta E(B-V)$ vs. $q_{\rm PAH}$ for each reddening map. 
    The solid black line indicates the median values of nine equally sized bins. 
    The dashed black line indicates the corresponding median absolute deviations.
    The Spearman correlation coefficient $r_s$ is also shown. 
    Estimating the uncertainty using bootstrapping, we find $\sigma_{r_s} \approx 0.04$.
    For the HI reddening map, the gray number and trend indicates the relation when limited to only regions with $N_{\text{\ion{H}{1}}}< 4\times10^{20}$~cm$^{-2}$ \citep[i.e., the threshold from][]{2017ApJ...846...38L}. All reddening maps show weak correlations with PAH mass fraction ($|r_s| < 0.1$) with the exception of the HI reddening map. However, this negative correlation largely disappears when applying the column density threshold ($r_s=-0.26 \rightarrow -0.08$).
    }
    \label{fig:galdense_compare_qPAH}
\end{figure*}

\subsection{Comparison with Polycyclic Aromatic Hydrocarbon Abundance}
Given that our SPHEREx reddening map is based on PAH emission, we investigate whether biases arise when the abundance of PAHs relative to larger dust grains varies.
To assess our SPHEREx reddening map---as well as the ones from the literature---in this regard, we plot the accuracy of our reddening maps against two measures of PAH abundance.

The first is the mass fraction of dust in the form of PAHs $q_{\rm PAH}$\footnote{\texttt{COM\_CompMap\_Dust-DL07-Parameters\_2048\_R2.00.fits}}.
The $q_{\rm PAH}$ values are obtained by fitting observations from Planck, IRAS, and WISE to the \citet{2007ApJ...657..810D} dust model \citep{planck2014-XXIX}.
In addition to fitting for the PAH abundance in the form of $q_{\rm PAH}$, the dust modeling in \citet{planck2014-XXIX} also constrains properties such as starlight radiation intensity and dust column density.
Notably, the work in \citet{planck2014-XXIX} accounts for the radiative heating of dust grains using both an overall starlight intensity parameter as well as an additional intense radiation field parameter. 
The comparison with $q_{\rm PAH}$ is shown in Figure~\ref{fig:galdense_compare_qPAH}.

\begin{figure*}
    \centering
    \includegraphics[width=0.99\textwidth]{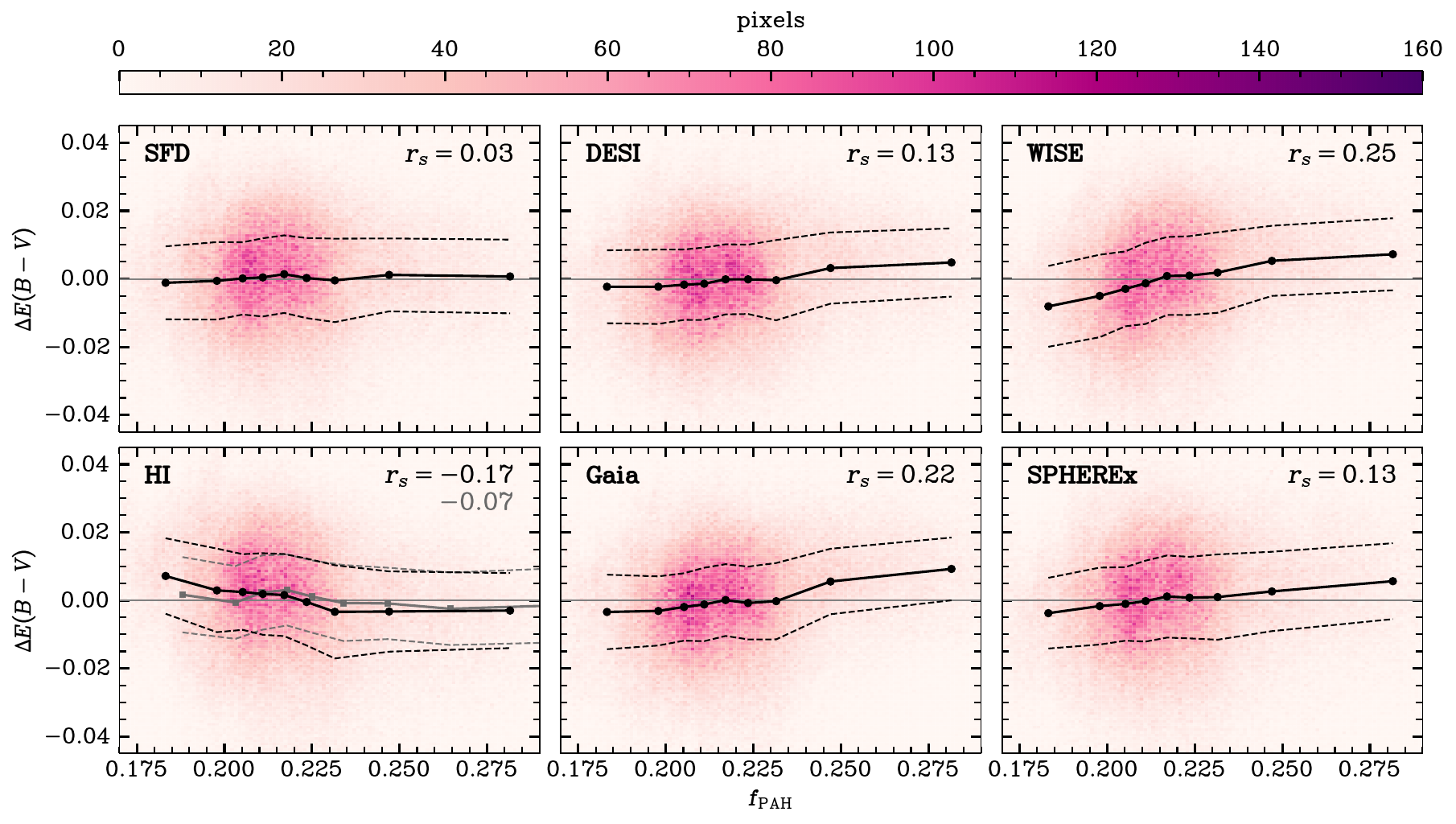}
    \caption{$\Delta E(B-V)$ vs. $f_{\rm PAH}$ for each reddening map. 
    Like Figure~\ref{fig:galdense_compare_qPAH}, the solid black line indicates the median values of nine equally sized bins, while the dashed black line indicates the corresponding median absolute deviations.   
    Estimating the uncertainty using bootstrapping, we find $\sigma_{r_s} \approx 0.04$.
    For the HI reddening map, the gray number and trend indicates the relation when limited to only regions with $N_{\text{\ion{H}{1}}}< 4\times10^{20}$~cm$^{-2}$ \citep[i.e., the threshold from][]{2017ApJ...846...38L}. 
    Compared to the minimal trends observed with $q_{\rm PAH}$ (Figure~\ref{fig:galdense_compare_qPAH}), the correlations for  DESI, WISE, Gaia, and SPHEREx when compared against $f_{\rm PAH}$ are stronger. Again, we find that the negative correlation for the HI reddening map is substantially reduced when applying the column density threshold.
    } 
    \label{fig:galdense_compare_wPAH}
\end{figure*}

Overall, we find only weak correlations between $q_{\rm PAH}$ with the accuracy of our reddening maps.
The SFD, DESI, WISE, Gaia, and SPHEREx reddening maps are only slightly correlated with $q_{\rm PAH}$.
SFD shows a slight negative correlation with a Spearman's rank correlation coefficient computed of $r_s=-0.07$ suggesting a slight underestimate at higher PAH abundance.
On the other hand, DESI and Gaia demonstrate weak evidence of underestimation at low PAH abundance and overestimation at high PAH abundance ($r_s=0.07$ and $r_s=0.05$, respectively).
This is also seen for WISE and SPHEREx with Spearman rank coefficients of $r_s=0.09$ and $r_s=0.05$, respectively.
By both visual inspection and by the Spearman rank coefficient $r_s$, only the HI map shows a stronger correlation with PAH mass fraction displaying a negative correlation and $r_s=-0.26$.
However, when excluding regions where \ion{H}{1} column density no longer traces reddening, this value falls to $r_s=-0.08$.
Overall, estimating the uncertainty of $r_s$ using bootstrapping, we find $\sigma_{r_s} \approx 0.04$ for all of our correlations.

The second method by which we trace PAH abundance is the ratio of PAH emission to dust radiance~\citep{2016ApJ...827...45H}.
Dust radiance represents the integrated dust emission and, in essence, traces both the dust column density and the strength of the radiation field responsible for heating the dust. 
For a given sight line with dust column density of $\Sigma_{\rm d}$ heated by radiation field with strength $U$, the dust radiance is proportional as $\mathcal{R}\propto U\Sigma_{\rm d}$.
For PAH emission given as $I_{\nu, \rm PAH}$ in a sightline with the mass fraction of dust that is PAHs as $q_{\rm PAH}$, the PAH emission is proportional as $I_{\nu, \rm PAH}\propto Uq_{\rm PAH}\Sigma_{\rm d}$.
Thus the ratio of $I_{\nu, \rm PAH}$ to $\mathcal{R}$ represents a value proportional to PAH mass fraction.
Specifically, we use the ratio of $\nu I_{\nu, \rm PAH}$ to $\mathcal{R}$ and, following \citet{2016ApJ...827...45H}, refer to the dimensionless quantity as $f_{\rm PAH}$, i.e., $f_{\rm PAH} \equiv \nu I_{\nu, \rm PAH} / \mathcal{R}$ with $\nu$ evaluated at 3.3\,$\mu$m for SPHEREx and 12\,$\mu$m for WISE.
In principle, this ratio should be similar to $q_{\rm PAH}$ described above without the benefits or drawbacks of incorporating a specific dust model.
For the PAH emission, we specifically use the WISE W3 emission as described in Section~\ref{sec:data}, but without the temperature correction (i.e., $\nu I_{\nu, \rm PAH} = \nu I_{\nu, \rm W3}$). 
This lends itself to a more direct comparison with $q_{\rm PAH}$.
For the dust radiance, we use the Planck dust radiance\footnote{\texttt{COM\_CompMap\_Dust-GNILC-Radiance\_2048\_R2.00.fits}} $\mathcal{R}$ from~\citet{planck2016-XLVIII} obtained by fitting a modified blackbody to observations at 353, 545, and 857\,GHz from Planck along with 100\,$\mu$m emission measurements from IRAS.
The comparison of reddening map inaccuracy with $f_{\rm PAH}$ is shown in Figure~\ref{fig:galdense_compare_wPAH}.

There are distinct differences in the behavior of the correlations between $q_{\rm PAH}$ and $f_{\rm PAH}$. 
When using $f_{\rm PAH}$ as our tracer for PAH abundance, we generally find stronger correlations. 
DESI, Gaia, WISE, and SPHEREx show positive correlations with larger values computed for $r_s$ ($r_s \geq 0.1)$.
For SFD, a change in correlation direction occurs changing to a very slight positive correlation with $f_{\rm PAH}$ ($r_s = 0.03$) from a negative correlation with $q_{\rm PAH}$ ($r_s = -0.07$).
Finally, HI continues to show a negative correlation, albeit weaker with $r_s=-0.17$.
Again, however, when excluding regions where \ion{H}{1} column density no longer traces reddening, this Spearman rank coefficient reduces to $r_s=-0.07$.

\section{Discussion}
\label{sec:discussion}
Based on our analysis in Section~\ref{sec:gal_backlight_test}, we find that our SPHEREx 3.3\,$\mu$m PAH feature-based reddening map compares favorably to existing reddening maps available in the literature.

Overall, we find good consistency, as shown in Table~\ref{tab:mad_diffs}, with variations at the MAD\,$<10$ mmag level. 
We do note two particularly consistent comparisons beyond our SPHEREx-based reddening map: between SFD and HI when limited to the diffuse regions (MAD$=2.3$ mmag), and between SFD and WISE (MAD$=2.6$ mmag),.
The former is expected as \citet{2017ApJ...846...38L} has demonstrated extremely tight correlation between the two.
The latter could potentially be attributed to the use of Planck 350\,$\mu$m (857 GHz) observations to set the large-scale zero-point of the WISE emission map.
In this analysis, we note that we have also opted to calibrate our map of temperature-corrected PAH emission from SPHEREx to SFD for the construction of our reddening map. 
Given the tight range in variations as well as the simple relation used to calibrate our emission to reddening (i.e., Equation~\ref{eq:linear_em_to_red}), the choice of calibration target does not appear to alter these results.

Using the galaxy backlight test, we find the median scatter of our SPHEREx reddening map to be 11.6\,mmag, consistent within 10\% of values obtained from the literature reddening maps.
Furthermore, the results of the power spectrum analysis show no evidence of contamination in the ELG galaxy sample selected using the SPHEREx reddening map. 
Its power spectrum is statistically indistinguishable from those obtained from the other reddening maps. 

\begin{figure*}
    \centering
    \includegraphics[width=0.99\textwidth]{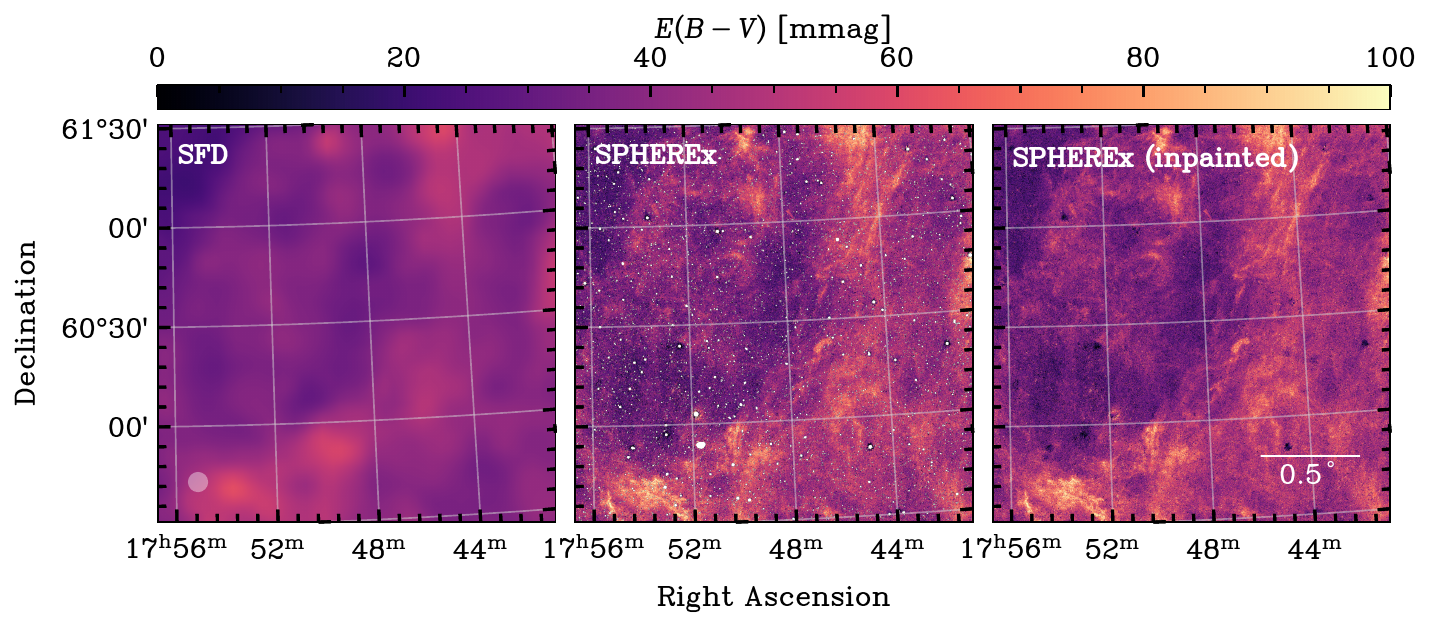}
    \caption{
    Comparison of dust reddening from SFD and SPHEREx (this work) in a $2^\circ$ by $2^\circ$ region within the NEP. 
    SFD, with a nominal angular resolution of $6'.1$, is shown on the left. This SFD nominal beam is shown as a white circle in the lower left.
    SPHEREx is shown with a pixel scale of $6".2$ in the middle. 
    On the right, for the purpose of visualization, we also show the SPHEREx $E(B-V)$ map where we have infilled masked pixels (typically point sources) using~\texttt{CloudClean.jl}~\citep{2022ApJ...933..155S}.}
    \label{fig:spherex_sfd_detail}
\end{figure*}

Although we have degraded our SPHEREx reddening map to a coarser shared resolution for the analysis, the SPHEREx observations are available at near the native pixel scale of $6''.2$. 
A comparison of this increased resolution is shown in Figure~\ref{fig:spherex_sfd_detail}.
While SPHEREx clearly traces the same larger scale Galactic dust structure, the improved resolution shows substructure that is below the nominal $6'.1$ resolution of SFD. 
Since no other reddening map is available at a resolution of $\approx 6''$ for such a large area of the NEP, we cannot directly compare our reddening map against such products. 
As a result, we cannot definitely state that a full resolution SPHEREx reddening map at $\approx 6''$ will be without any systematics.
Nonetheless, at the resolutions we have studied, we find that a SPHEREx reddening map offers accuracy comparable to that of existing reddening maps, while being capable of accessing angular scales beyond what is currently available with other maps.

\subsection{Expected Systematics for a PAH emission-based Reddening Map}
In this work, we constructed our PAH-based reddening map with a calibration that---ultimately---reduces to one key factor: $\beta$ in Equation~\ref{eq:linear_em_to_red}.
By using this single factor throughout our map, we have assumed that the ratio of 3.3\,$\mu$m PAH emission to $E(B-V)$ is fixed.
As PAHs are not the only constituent of interstellar dust, this implicitly assumes the PAH fraction of the dust is also fixed (i.e., $q_{\rm PAH}$ and $f_{\rm PAH}$ do not vary spatially).
Should $q_{\rm PAH}$ vary across the NEP, assuming a fixed $E(B-V) / I_{\nu,3.3}$ results in underestimates when true PAH abundance drop below the implied PAH abundance in our calibration factor and overestimate when PAH abundance increases above.

We can estimate the magnitude of such an effect in the NEP using the astrodust+PAH dust model from \citet{2023ApJ...948...55H}.
For a $q_{\rm PAH}$ change of $\approx3.5\rightarrow 6\%$, we expect the change in 3.3\,$\mu$m to $E(B-V)$ ratio to increase by $\approx40\%$ (depending slightly on how the size distribution changes as PAH mass is increased).
Similarly, the WISE W3 to $E(B-V)$ ratio increases by $\approx45\%$.
Given that the median reddening in the NEP is approximately $\approx40$\,mmag, this expected $\sim40\%$ change corresponds to an $E(B-V)$ increase of approximately 18\,mmag.

Despite, in principle, tracing the same physical quantity, our analysis for SPHEREx shows no strong correlation with $q_{\rm PAH}$ whereas it shows a more robust positive correlation with $f_{\rm PAH}$.
While there is no strong evidence of a change or correlation of this magnitude in $q_{\rm PAH}$ ($r_s = 0.05 \pm 0.04$), observing the correlations for our SPHEREx map with $f_{\rm PAH}$ ($r_s = 0.13 \pm 0.04 $), we find that the positive correlation is what is expected.
Given these seemingly conflicting results, we investigate whether the correlation does indeed exist with $q_{\rm PAH}$---but is obscured---or that the correlation with $f_{\rm PAH}$ arises from another source.

One possibility for this seeming lack of correlation in $q_{\rm PAH}$ could be the uncertainty in the estimated $q_{\rm PAH}$. 
In the NEP, the provided uncertainty values from \citet{planck2014-XXIX} indicate that fractional error for $q_{\rm PAH}$ is approximately $18\%$.
Using this value, we generate 1000 realizations of $q_{\rm PAH}$ measurements based on the reported uncertainties, and compute the correlations against our SPHEREx reddening map errors.
Despite this seemingly large value $q_{\rm PAH}$ fractional uncertainty, we find that there is no evidence that these uncertainties are affecting the correlation errors. % cannot account for the absence of a correlation.
Indeed, the magnitude of change in correlation induced by the $q_{\rm PAH}$ uncertainty is approximately $\pm0.005$, insignificant compared to, for example, correlations of SPHEREx reddening errors against $f_{\rm PAH}$.

Another possible origin for this discrepancy between our two measures of PAH abundance could lie in assumptions underlying each quantity's derivation. 
Though both values rely primarily on WISE W3 emission centered at 12\,$\mu$m, they differ greatly otherwise.
$q_{\rm PAH}$ is derived by fitting the WISE data along with Planck and IRAS observations to the \citet{2007ApJ...657..810D} dust model.
In addition to accounting for the dust column density, radiation field, and, of course, $q_{\rm PAH}$, the model also includes a free parameter $f_{\rm PDR}$ representing the fraction of emission due to intense radiation fields ($U>10^2$). 
This parameter is in addition to a general radiation field modeled as $U_{\rm min}$.
On the other hand, the $f_{\rm PAH}$ value, as described in Section~\ref{sec:gal_backlight_test}, is a much less complex quantity and is simply the ratio of WISE W3 emission to Planck radiance $\mathcal{R}$. 
Notably, this ratio does not explicitly model the radiation field and implicitly assumes that all dust grains (PAHs or otherwise) are subjected to the same radiation field.

\begin{figure}
    \centering
    \includegraphics[width=0.99\columnwidth]{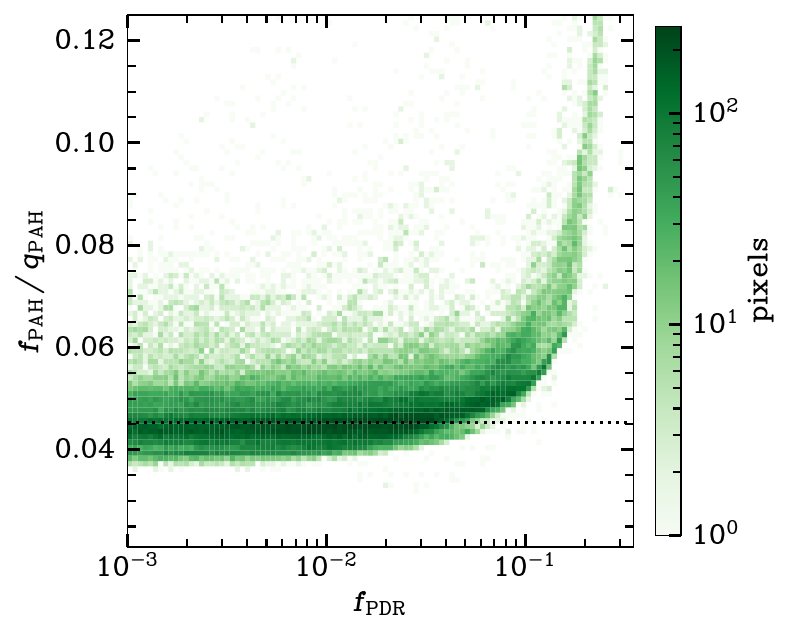}
    \caption{
    Ratio of $f_{\rm PAH}$/$q_{\rm PAH}$ against $f_{\rm PDR}$. 
    $f_{\rm PAH}$ and $q_{\rm PAH}$ remain proportional until enough WISE W3 emission is assigned to dust heated by intense radiation fields (i.e., larger values of $f_{\rm PDR}$) resulting in diverging predictions of PAH abundances. The dashed horizontal line at $\sim0.045$ represents the typical $f_{\rm PAH}$/$q_{\rm PAH}$ ratio.}
    \label{fig:wPAHqPAH_fPDR}
\end{figure}

As a consequence of this, when deriving $q_{\rm PAH}$ from the WISE W3 emission, it is possible for the model to attribute W3 emission either to a certain abundance of PAHs $q_{\rm PAH}$ or to the result of strong radiation fields $f_{\rm PDR}$.
On the other hand for $f_{\rm PAH}$, since radiation field is already accounted for implicitly, the WISE W3 emission can only be attributed to the abundance of PAHs $f_{\rm PAH}$.
Indeed, as shown in Figure~\ref{fig:wPAHqPAH_fPDR}, comparing the ratio of $f_{\rm PAH}$ to $q_{\rm PAH}$ against $f_{\rm PDR}$, we can clearly see that with increasing $f_{\rm PDR}$, the ratio of $f_{\rm PAH}$ to $q_{\rm PAH}$ increases.

\begin{figure}
    \centering
    \includegraphics[width=0.95\columnwidth]{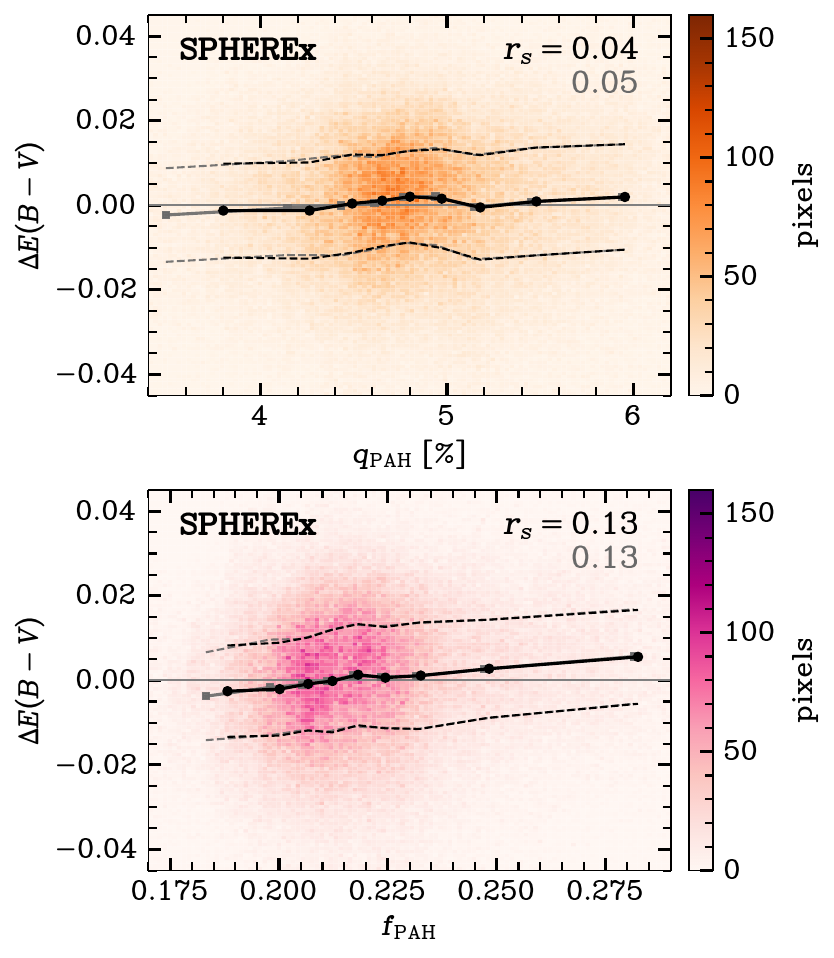}
    \caption{
    SPHEREx $\Delta E(B-V)$ vs. $q_{\rm PAH}$ and $\Delta E(B-V)$ vs. $f_{\rm PAH}$ correlations restricted to $f_{\rm PDR} < 0.075$.
    The result with the $f_{\rm PDR}$ cut is shown in black, while the correlation without the $f_{\rm PDR}$ cut is shown in gray.
    Regardless of whether the comparison is against $q_{\rm PAH}$ or against $f_{\rm PAH}$, no change in correlation is observed when restricting to regions with minimal predicted $f_{\rm PDR}$.
    }
    \label{fig:spherex_column_qPAH_wPAH_fpdr0.075}
\end{figure}

In light of this, we expect that the comparison of reddening errors with $q_{\rm PAH}$ and $f_{\rm PAH}$ to be consistent where $f_{\rm PDR}$ remains quite low. 
If we restrict our analysis to regions where the \citet{2007ApJ...657..810D} model fit did not weight $f_{\rm PDR}$ heavily, we expect $q_{\rm PAH}$ and $f_{\rm PAH}$ to be consistent.
Figure~\ref{fig:spherex_column_qPAH_wPAH_fpdr0.075} compares reddening errors against $q_{\rm PAH}$ and $f_{\rm PAH}$ only in regions where $f_{\rm PDR}$ is low. Specifically, we choose regions with $f_{\rm PDR} < 0.075$ where $q_{\rm PAH}$ and $f_{\rm PAH}$ should be strongly correlated. 
Doing so, however, we find that the values of the Spearman rank coefficient shift from $r_s=0.05 \rightarrow 0.04$ for comparison with $q_{\rm PAH}$ and no change when using $f_{\rm PAH}$.
In essence, we find minimal changes in the nature of the correlations.
This suggests that, while $q_{\rm PAH}$ and $f_{\rm PAH}$ do systematically differ at high $f_{\rm PDR}$, this is not the origin of our correlation discrepancy.

Finally, we consider whether this discrepancy is the result of the temperature-correction that is applied to our SPHEREx reddening map.
This procedure is done in order to account for dust grains being heated to a range of different temperatures depending on the local radiation field. 
As a result of these radiation field variations, dust grains exhibit varying observed emissivities.
This correction was done, as described in Section~\ref{sec:data}, using the DIRBE/COBE derived temperature correction map from \citet{1998ApJ...500..525S}.
In the event that this correction was incorrect or otherwise biased, it is possible that it has introduced spurious correlations or concealed existing ones. 

\begin{figure}
    \centering
    \includegraphics[width=0.95\columnwidth]{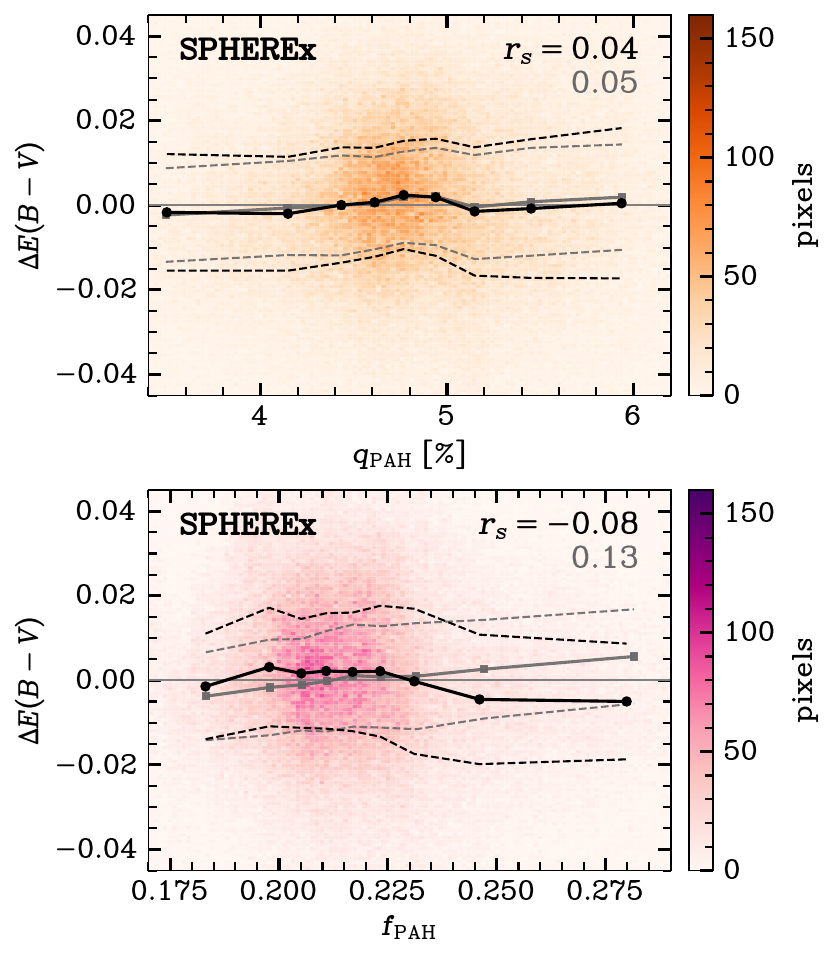}
    \caption{
    SPHEREx $\Delta E(B-V)$ vs. $q_{\rm PAH}$ and $\Delta E(B-V)$ vs. $f_{\rm PAH}$ correlations without temperature correction.
    The correlation without the temperature correction is shown in black while the correlation with the temperature correction (originally shown in Figure~\ref{fig:galdense_compare_qPAH} and~\ref{fig:galdense_compare_wPAH}) is shown in gray.
    With and without the temperature correction, correlations of the SPHEREx reddening map with $q_{\rm PAH}$ show minimal correlation change. 
    On the other hand, correlations of the SPHEREx reddening map with $f_{\rm PAH}$ show a reduction in correlation without a temperature correction ($r_s=0.13\rightarrow-0.08$).
    }
    \label{fig:spherex_column_qPAH_wPAH_tcorr}
\end{figure}

Figure~\ref{fig:spherex_column_qPAH_wPAH_tcorr} shows our comparison for SPHEREx reddening errors against $q_{\rm PAH}$ and $f_{\rm PAH}$ with and without the temperature correction applied.
Notably, the correlations against $q_{\rm PAH}$ remain largely unchanged. 
On the other hand, the correlations against $f_{\rm PAH}$ are clearly suppressed and flattened with a slight negative correlation with the Spearman rank coefficient decreasing from $r_s = 0.13$ with the temperature correction applied to $r_s = -0.08$ without.

This is potentially tied to two effects (temperature correction and PAH abundance) that act in opposite directions.
Without a temperature correction applied to the observed emission, cooler, denser regions will have their dust column, and thus reddening, underestimated.
These cooler, denser regions largely correlate with regions of higher PAH abundance and PAH emission relative to Planck radiance (i.e., $f_{\rm PAH}$). 
As a result, this effect alone will lead a PAH emission-based reddening map to overestimate $E(B-V)$ exactly where the map (without a temperature correction) would underestimate.
Thus, these two biases together may counteract and compensate for each other.
Application of the temperature correction removes one of these effects and thus reveals the trend initially seen in Figure~\ref{fig:galdense_compare_wPAH} for SPHEREx.
Although not applying the temperature correction appears to statistically improve our reddening map, it is not necessarily the physically correct choice.
The fact that these two effects may be comparable in magnitude, and thus cancel out, is not guaranteed across environments. 

Though we have explored several potential sources of disagreement, none fully resolves the discrepancy. 
If $q_{\rm PAH}$ accurately captures variations in PAH abundance, we are left with a somewhat paradoxical implication.
In essence, what this implies is that using PAH emission as a tracer for dust extinction is effective even as the PAH content of the dust population varies.
Since PAHs represent only a fraction of the dust in the Galaxy, a fixed $E(B-V) / I_{\nu, 3.3}$ ratio should fail as that fraction changes.
That it does not, across a $\sim40\%$ change in PAH abundance as measured by $q_{\rm PAH}$, is difficult to reconcile.
While we explored potential degeneracies between $q_{\rm PAH}$ and $f_{\rm PDR}$, $q_{\rm PAH}$ may be degenerate with other parameters in the \citet{2007ApJ...657..810D} dust model.
In that event, PAH emission changes due to PAH abundance variations could have been inadvertently and incorrectly attributed to these other parameters (e.g., $U_{\rm min}$).
In fact, \citet{planck2014-XXIX} note that $U_{\rm min}$, intended to model the general radiation field, likely absorbed details of the grain properties themselves and that, as a result, certain aspects of the dust model required revision.
Fitting and producing $q_{\rm PAH}$ using a more recent dust model \citep[e.g.,][]{2023ApJ...948...55H, 2024A&A...684A..34Y} would test whether the absence of correlation here is true, or a consequence of some modeling deficiency. 

On the other hand, if $f_{\rm PAH}$ is a more accurate tracer of PAH abundance variations than $q_{\rm PAH}$, then the PAH-driven bias as described and predicted is present.
The change observed in SPHEREx---an approximate 10\,mmag change from $\Delta E(B-V)= -4 \rightarrow 6$~mmag---is smaller than the expected 18\,mmag predicted.
Furthermore, the bias appears most significantly at $f_{\rm PAH} \gtrsim 0.235$ suggesting that in regions with a lower measured $f_{\rm PAH}$, the map is largely free of this bias.
For the NEP, this indicates the majority of the NEP ($\approx75\%$) is beneath this threshold.
Finally, we find that we can mitigate this correlation by not applying a temperature correction. 
Though we emphasize that this is not necessarily justified physically, we also note that the correction as applied here is not itself free of all systematics.
After all, the temperature correction template is itself a derived product from observations from COBE/DIRBE. 
As a result, investigations into improved or alternative methods of estimating the temperature/radiation field variations will be important in fully understanding the effects of this correction.

As we have noted, the PAH bias is predicted based on the astrodust+PAH dust model.
While the expected trend should be visible given the scatter from the galaxy density test ($\sim11$\,mmag),---and indeed a $\sim10$\,mmag trend is detectable in $f_{\rm PAH}$---it remains possible that an overestimate of the effect by the dust model inhibits detection.
Furthermore, the $q_{\rm PAH}$ range in the NEP spans the lower end of $q_{\rm PAH}$ values in the Galaxy.
Values in the Galaxy extend up to 10\% with Ursa Major, for example, having $q_{\rm PAH}$ values that range between $4\%$ to $8\%$.
Extension of this analysis to the full-sky will be important in testing this hypothesis.

Given that the $\sim11$\,mmag scatter is observed with the galaxy density test for all of our reddening maps, this could imply that we have reached the intrinsic sensitivity of the test using the ELG sample.
If reddening maps differ by less than this as they appear to (see Table~\ref{tab:mad_diffs}), this galaxy density test would not be able to distinguish.
While this did not preclude us from using it as an avenue to search for correlations, it does suggest that there is a clear limitation to the capabilities of the galaxy backlight test.

Finally, we emphasize that the correlation coefficients reported in this analysis are all fairly weak. 
The maximum magnitude of the correlation coefficient with PAH abundance ($q_{\rm PAH}$ or $f_{\rm PAH}$) observed for all our reddening maps is $|r_s| = 0.26$ with no SPHEREx correlation having $|r_s| > 0.13$ . 
As such, while we cannot conclude definitively about the origin of the possible reddening errors, from the analysis here, we find that our SPHEREx reddening map toward the NEP traces Galactic dust as accurately as any existing map we have tested.

\subsection{Other Reddening Maps}
Though the focus of this work is our SPHEREx reddening map, here we briefly discuss our analysis of the other reddening maps.

The WISE reddening map constructed from WISE W3 emission largely shows performance and correlations similar to SPHEREx.
Since both are derived from observations of PAH emission, this is expected.
However, given that both our measures of $q_{\rm PAH}$ and $f_{\rm PAH}$ rely heavily on the same observations (namely WISE W3 emission) to estimate the PAH abundance as well, comparisons and the resulting interpretations are somewhat circular.
As such, and given its comparable behavior to the SPHEREx reddening map, we will not analyze correlations with the WISE reddening map further. 

SFD shows a very slight underestimate of reddening at higher values of $q_{\rm PAH}$.
This is consistent with what is expected based on~\citet{2025ApJ...994...61L}.
SFD, a far-infrared based reddening map, underestimates reddening (negative $\Delta E(B-V)$) at higher PAH abundances (larger $q_{\rm PAH}$).
However, compared to the effect observed in Ursa Major by~\citet{2025ApJ...994...61L}, this effect is directionally the same, but of a much smaller magnitude.
Furthermore, when using a different measure of PAH abundance---namely $f_{\rm PAH}$---, we find that this correlation becomes an even weaker but \textit{positive} correlation.
In light of this change in correlation, it is likely that should SFD be susceptible to PAH-induced reddening variations, it does not exist at a presently detectable level at the NEP.

For the two reddening maps derived from stellar measurements, both DESI and Gaia show relatively weak correlations with $q_{\rm PAH}$ as expected from reddening maps not constructed from dust emission.
For our comparison with $f_{\rm PAH}$, Gaia exhibits a positive correlation comparable to WISE and SPHEREx, whereas DESI shows a correlation somewhat intermediate between WISE/SPHEREx and SFD. There is no physical reason why direct reddening measurements should have errors correlated with PAH abundance itself, but it may be the case that the maps are less reliable in regions associated with greater PAH abundance, e.g., having higher column densities or less standard extinction curves.

Finally, the HI reddening map shows a strong negative correlation with $q_{\rm PAH}$ and $f_{\rm PAH}$.
Given that the correlation can be considerably reduced to become the weakest correlation with the application of the column density cut from \citet{2017ApJ...846...38L}, it is likely that the deficiencies are primarily due to the breakdown of the relationship between \ion{H}{1} column density and reddening. 
In particular, this suggests that the threshold of $N_{\text{\ion{H}{1}}}< 4\times10^{20}$~cm$^{-2}$ set by \citet{2017ApJ...846...38L} appropriately marks this critical value and that PAH abundance is indeed correlated with column density.

\section{Conclusion}
\label{sec:conclusion}
In this work, we construct a reddening map based on SPHEREx measurements of the 3.3\,$\mu$m PAH feature toward the NEP.
We compare its performance against a sample of reddening maps from the literature constructed from infrared emission, stellar extinction measurements, as well as 21 cm~\ion{H}{1} emission line.

Our primary results are as follows:
\begin{enumerate}
    \item Using SPHEREx observations, we construct a reddening map in a $22^{\circ}$ by $22^{\circ}$ region centered at the NEP at the native pixel scale of $6''.2$. 
    We model the 3.3\,$\mu$m emission feature from PAHs by spectrally fitting each sightline.
    From this map of 3.3\,$\mu$m emission, we apply a temperature correction and derive a linear scaling with SFD to obtain reddening.

    \item For our SPHEREx reddening map of the NEP, we find a $\sim5$\,mmag agreement with the literature reddening maps when compared at the common resolution of $16''.1$. The galaxy density test also reports a median absolute deviation of reddening error $\Delta E(B-V)$ of $\approx11.6$\,mmag that is consistent with what is obtained for the literature reddening maps as well.
    
    \item The power spectrum of the galaxy sample selected using the SPHEREx reddening map is statistically indistinguishable from the galaxy samples obtained from the majority of literature reddening maps and shows no evidence of Galactic contamination. 

    \item We correlate our SPHEREx reddening errors $\Delta E(B-V)$ against two measures of PAH abundance---$q_{\rm PAH}$ and $f_{\rm PAH}$. 
    Modeling with the astrodust+PAH dust model shows the presence of an overestimation bias for PAH-emission based reddening maps in regions of elevated PAH abundance. 
    Correlating our SPHEREx reddening error against $q_{\rm PAH}$, we find no evidence of any significant correlations. 
    Correlating our SPHEREx reddening error against $f_{\rm PAH}$, we find a positive correlation consistent in direction with the theoretical expectations.
    
    \item This difference between the correlations seen between $q_{\rm PAH}$ and $f_{\rm PAH}$ cannot be attributed to a difference in modeling the presence or lack of intense radiation fields via the $f_{\rm PDR}$ parameter. On the other hand, the strong positive correlation of $\Delta E(B-V)$ with $f_{\rm PAH}$ can be greatly diminished without application of a temperature correction (i.e., Equation~\ref{eq:pah_temp_corr}) when constructing our SPHEREx reddening map. For $q_{\rm PAH}$, the lack of correlation persists regardless of whether a temperature correction is applied.

    \item Nonetheless, should this PAH-emission bias exist, we find that the magnitude of such an effect is $\sim 10$~mmag in $E(B-V)$ at the strongest, less than the expectation using the \citet{2023ApJ...948...55H} dust model. Furthermore, it is only detectable---at any level---, in $\lesssim 25\%$ of the NEP. 

\end{enumerate}

Looking ahead, this work suggests that SPHEREx 3.3\,$\mu$m emission is capable of forming the basis for an effective reddening map in regions of our Galaxy where PAH properties do not vary significantly.
By extending and applying our method, we can produce a large-scale map of SPHEREx 3.3\,$\mu$m emission, and, accordingly, of Galactic dust reddening.

Although the extraction of the 3.3\,$\mu$m PAH feature was effective, there remains avenues for improvement. 
Notably, both the zodiacal light subtraction and point source removal, as implemented in this work, are fairly simplistic. 
While we treat the zodiacal light as a foreground when constructing our map of SPHEREx 3.3\,$\mu$m emission, SPHEREx observations themselves can also be used to constrain and model this time-variable foreground.
Future work may benefit from such improved zodiacal light modeling.

As for point source removal, use of a point source catalog---external or built from SPHEREx observations---to model and subtract such objects would also improve our PAH emission maps.
The NEP region benefited in this regard by possessing a relatively low density of point sources.
Maps of regions closer to the Galactic plane, for example, may not only see improvement with careful point source subtraction, but in fact require it.

Furthermore, we have limited our initial study to only the first year of SPHEREx observations that were reprocessed using the same data reduction pipeline.
We expect analyses such as ours to improve with additional observations over the course of SPHEREx's mission.
This will be especially beneficial to regions, such as those further from the NEP, that have not received as many repeat observations.

Finally, in this work, we have focused on our SPHEREx map of 3.3\,$\mu$m PAH emission and its ability to trace Galactic dust reddening.
In addition to this, through fitting for a map of 3.3\,$\mu$m PAH emission, we have also simultaneously obtained maps of the aliphatic emission feature at 3.4\,$\mu$m and the plateau feature at 3.47\,$\mu$m.
Beyond this, the spectral feature extraction and mapmaking method can also be extended to map other diffuse emission such as various hydrogen recombination lines within the wavelength range observed by SPHEREx.
As such, in addition to dust reddening, extraction of spectral features and lines from SPHEREx observations represents a powerful tool to study the Galactic interstellar medium.

\begin{acknowledgments}
This research was carried out at the Jet Propulsion Laboratory, California Institute of Technology, under a contract with the National Aeronautics and Space Administration (80NM0018D0004). 
Some of the results in this paper have been derived using the healpy and HEALPix package. 
The High Performance Computing resources used in this investigation were provided by funding from the JPL Enterprise IT Division.
The authors acknowledge the Texas Advanced Computing Center (TACC)\footnote{\href{http://www.tacc.utexas.edu}{http://www.tacc.utexas.edu}} at The University of Texas at Austin for providing computational resources that have contributed to the research results reported within this paper.

The Legacy Surveys consist of three individual and complementary projects: the Dark Energy Camera Legacy Survey (DECaLS; Proposal ID \#2014B-0404; PIs: David Schlegel and Arjun Dey), the Beijing-Arizona Sky Survey (BASS; NOAO Prop. ID \#2015A-0801; PIs: Zhou Xu and Xiaohui Fan), and the Mayall z-band Legacy Survey (MzLS; Prop. ID \#2016A-0453; PI: Arjun Dey). DECaLS, BASS and MzLS together include data obtained, respectively, at the Blanco telescope, Cerro Tololo Inter-American Observatory, NSF’s NOIRLab; the Bok telescope, Steward Observatory, University of Arizona; and the Mayall telescope, Kitt Peak National Observatory, NOIRLab. Pipeline processing and analyses of the data were supported by NOIRLab and the Lawrence Berkeley National Laboratory (LBNL). The Legacy Surveys project is honored to be permitted to conduct astronomical research on Iolkam Du’ag (Kitt Peak), a mountain with particular significance to the Tohono O’odham Nation.

NOIRLab is operated by the Association of Universities for Research in Astronomy (AURA) under a cooperative agreement with the National Science Foundation. LBNL is managed by the Regents of the University of California under contract to the U.S. Department of Energy.

This project used data obtained with the Dark Energy Camera (DECam), which was constructed by the Dark Energy Survey (DES) collaboration. Funding for the DES Projects has been provided by the U.S. Department of Energy, the U.S. National Science Foundation, the Ministry of Science and Education of Spain, the Science and Technology Facilities Council of the United Kingdom, the Higher Education Funding Council for England, the National Center for Supercomputing Applications at the University of Illinois at Urbana-Champaign, the Kavli Institute of Cosmological Physics at the University of Chicago, Center for Cosmology and Astro-Particle Physics at the Ohio State University, the Mitchell Institute for Fundamental Physics and Astronomy at Texas A\&M University, Financiadora de Estudos e Projetos, Fundacao Carlos Chagas Filho de Amparo, Financiadora de Estudos e Projetos, Fundacao Carlos Chagas Filho de Amparo a Pesquisa do Estado do Rio de Janeiro, Conselho Nacional de Desenvolvimento Cientifico e Tecnologico and the Ministerio da Ciencia, Tecnologia e Inovacao, the Deutsche Forschungsgemeinschaft and the Collaborating Institutions in the Dark Energy Survey. The Collaborating Institutions are Argonne National Laboratory, the University of California at Santa Cruz, the University of Cambridge, Centro de Investigaciones Energeticas, Medioambientales y Tecnologicas-Madrid, the University of Chicago, University College London, the DES-Brazil Consortium, the University of Edinburgh, the Eidgenossische Technische Hochschule (ETH) Zurich, Fermi National Accelerator Laboratory, the University of Illinois at Urbana-Champaign, the Institut de Ciencies de l'Espai (IEEC/CSIC), the Institut de Fisica d’Altes Energies, Lawrence Berkeley National Laboratory, the Ludwig Maximilians Universitat Munchen and the associated Excellence Cluster Universe, the University of Michigan, NSF’s NOIRLab, the University of Nottingham, the Ohio State University, the University of Pennsylvania, the University of Portsmouth, SLAC National Accelerator Laboratory, Stanford University, the University of Sussex, and Texas A\&M University.

BASS is a key project of the Telescope Access Program (TAP), which has been funded by the National Astronomical Observatories of China, the Chinese Academy of Sciences (the Strategic Priority Research Program “The Emergence of Cosmological Structures” Grant \# XDB09000000), and the Special Fund for Astronomy from the Ministry of Finance. The BASS is also supported by the External Cooperation Program of Chinese Academy of Sciences (Grant \# 114A11KYSB20160057), and Chinese National Natural Science Foundation (Grant \# 12120101003, \# 11433005).

The Legacy Survey team makes use of data products from the Near-Earth Object Wide-field Infrared Survey Explorer (NEOWISE), which is a project of the Jet Propulsion Laboratory/California Institute of Technology. NEOWISE is funded by the National Aeronautics and Space Administration.

The Legacy Surveys imaging of the DESI footprint is supported by the Director, Office of Science, Office of High Energy Physics of the U.S. Department of Energy under Contract No. DE-AC02-05CH1123, by the National Energy Research Scientific Computing Center, a DOE Office of Science User Facility under the same contract; and by the U.S. National Science Foundation, Division of Astronomical Sciences under Contract No. AST-0950945 to NOAO.
\end{acknowledgments}
\facility{SPHEREx}

\software{
    \texttt{astropy} \citep{2013A&A...558A..33A, 2018AJ....156..123A, 2022ApJ...935..167A},
    \texttt{numpy} \citep{harris2020array}, 
    \texttt{scipy} \citep{2020NaMet..17..261V}, 
    \texttt{matplotlib} \citep{2007CSE.....9...90H},
    \texttt{NaMaster} \citep{2019MNRAS.484.4127A},
    \texttt{CloudClean.jl} \citep{2022ApJ...933..155S},
}

\bibliography{reference, Planck_bib, software}{}

@ARTICLE{planck2016-l11A,
author = {{\sorthelp{Planck Collaboration 2018K}}{Planck Collaboration XI}},
title = "{\textit{Planck} 2018 results. XI. Polarized dust foregrounds}",
journal = {\aap},
archivePrefix = "arXiv",
eprint = {1801.04945},
year = 2020,
volume = 641,
pages = {A11},
doi = {10.1051/0004-6361/201832618}
}

@ARTICLE{planck2013-pip56,
author = {{\sorthelp{Planck Collaboration 2014ZE}}{Planck Collaboration XXX}},
title = "{\textit{Planck} 2013 results. XXX. Cosmic infrared background
 measurements and implications for star formation}",
journal = {\aap},
archivePrefix = "arXiv",
eprint = {1309.0382},
year = 2014,
volume = 571,
pages = {A30},
doi = {10.1051/0004-6361/201322093}
}

@ARTICLE{planck2014-XXIX,
author = {{\sorthelp{Planck Collaboration IntZD}}{Planck Collaboration Int.
 XXIX}},
title = "{\textit{Planck} intermediate results. XXIX. All-sky dust modelling
 with \textit{Planck}, IRAS, and WISE observations}",
journal = {\aap},
archivePrefix = "arXiv",
eprint = {1409.2495},
year = 2016,
volume = 586,
pages = {A132},
doi = {10.1051/0004-6361/201424945}
}

@ARTICLE{planck2016-XLVIII,
author = {{\sorthelp{Planck Collaboration IntZW}}{Planck Collaboration Int.
 XLVIII}},
title = "{\textit{Planck} intermediate results. XLVIII. Disentangling Galactic
 dust emission and cosmic infrared background anisotropies}",
journal = {\aap},
archivePrefix = "arXiv",
eprint = {1605.09387},
year = 2016,
volume = 596,
pages = {A109},
doi = {10.1051/0004-6361/201629022}
}

@ARTICLE{2007CSE.....9...90H,
       author = {{Hunter}, John D.},
        title = "{Matplotlib: A 2D Graphics Environment}",
      journal = {Computing in Science and Engineering},
         year = 2007,
        month = jan,
       volume = {9},
       number = {3},
        pages = {90-95},
          doi = {10.1109/MCSE.2007.55},
       adsurl = {https://ui.adsabs.harvard.edu/abs/2007CSE.....9...90H}
}

@ARTICLE{2020NaMet..17..261V,
       author = {{Virtanen}, Pauli and {Gommers}, Ralf and {Oliphant}, Travis E. and {Haberland}, Matt and {Reddy}, Tyler and {Cournapeau}, David and {Burovski}, Evgeni and {Peterson}, Pearu and {Weckesser}, Warren and {Bright}, Jonathan and {van der Walt}, St{\'e}fan J. and {Brett}, Matthew and {Wilson}, Joshua and {Millman}, K. Jarrod and {Mayorov}, Nikolay and {Nelson}, Andrew R.~J. and {Jones}, Eric and {Kern}, Robert and {Larson}, Eric and {Carey}, C.~J. and {Polat}, {\.I}lhan and {Feng}, Yu and {Moore}, Eric W. and {VanderPlas}, Jake and {Laxalde}, Denis and {Perktold}, Josef and {Cimrman}, Robert and {Henriksen}, Ian and {Quintero}, E.~A. and {Harris}, Charles R. and {Archibald}, Anne M. and {Ribeiro}, Ant{\^o}nio H. and {Pedregosa}, Fabian and {van Mulbregt}, Paul and {SciPy 1.  0 Contributors}},
        title = "{SciPy 1.0: fundamental algorithms for scientific computing in Python}",
      journal = {Nature Methods},
         year = 2020,
        month = feb,
       volume = {17},
        pages = {261-272},
          doi = {10.1038/s41592-019-0686-2},
archivePrefix = {arXiv},
       eprint = {1907.10121},
 primaryClass = {cs.MS},
       adsurl = {https://ui.adsabs.harvard.edu/abs/2020NaMet..17..261V}
}

@ARTICLE{2013A&A...558A..33A,
       author = {{Astropy Collaboration} and {Robitaille}, Thomas P. and {Tollerud}, Erik J. and {Greenfield}, Perry and {Droettboom}, Michael and {Bray}, Erik and {Aldcroft}, Tom and {Davis}, Matt and {Ginsburg}, Adam and {Price-Whelan}, Adrian M. and {Kerzendorf}, Wolfgang E. and {Conley}, Alexander and {Crighton}, Neil and {Barbary}, Kyle and {Muna}, Demitri and {Ferguson}, Henry and {Grollier}, Fr{\'e}d{\'e}ric and {Parikh}, Madhura M. and {Nair}, Prasanth H. and {Unther}, Hans M. and {Deil}, Christoph and {Woillez}, Julien and {Conseil}, Simon and {Kramer}, Roban and {Turner}, James E.~H. and {Singer}, Leo and {Fox}, Ryan and {Weaver}, Benjamin A. and {Zabalza}, Victor and {Edwards}, Zachary I. and {Azalee Bostroem}, K. and {Burke}, D.~J. and {Casey}, Andrew R. and {Crawford}, Steven M. and {Dencheva}, Nadia and {Ely}, Justin and {Jenness}, Tim and {Labrie}, Kathleen and {Lim}, Pey Lian and {Pierfederici}, Francesco and {Pontzen}, Andrew and {Ptak}, Andy and {Refsdal}, Brian and {Servillat}, Mathieu and {Streicher}, Ole},
        title = "{Astropy: A community Python package for astronomy}",
      journal = {\aap},
         year = 2013,
        month = oct,
       volume = {558},
          eid = {A33},
        pages = {A33},
          doi = {10.1051/0004-6361/201322068},
archivePrefix = {arXiv},
       eprint = {1307.6212},
 primaryClass = {astro-ph.IM},
       adsurl = {https://ui.adsabs.harvard.edu/abs/2013A&A...558A..33A}
}

@ARTICLE{1998ApJ...500..525S,
       author = {{Schlegel}, David J. and {Finkbeiner}, Douglas P. and {Davis}, Marc},
        title = "{Maps of Dust Infrared Emission for Use in Estimation of Reddening and Cosmic Microwave Background Radiation Foregrounds}",
      journal = {\apj},
         year = 1998,
        month = jun,
       volume = {500},
       number = {2},
        pages = {525-553},
          doi = {10.1086/305772},
archivePrefix = {arXiv},
       eprint = {astro-ph/9710327},
 primaryClass = {astro-ph},
       adsurl = {https://ui.adsabs.harvard.edu/abs/1998ApJ...500..525S}
}

@ARTICLE{2011ApJ...737..103S,
       author = {{Schlafly}, Edward F. and {Finkbeiner}, Douglas P.},
        title = "{Measuring Reddening with Sloan Digital Sky Survey Stellar Spectra and Recalibrating SFD}",
      journal = {\apj},
         year = 2011,
        month = aug,
       volume = {737},
       number = {2},
          eid = {103},
        pages = {103},
          doi = {10.1088/0004-637X/737/2/103},
archivePrefix = {arXiv},
       eprint = {1012.4804},
 primaryClass = {astro-ph.GA},
       adsurl = {https://ui.adsabs.harvard.edu/abs/2011ApJ...737..103S}
}

@ARTICLE{2010AJ....140.1868W,
       author = {{Wright}, Edward L. and {Eisenhardt}, Peter R.~M. and {Mainzer}, Amy K. and {Ressler}, Michael E. and {Cutri}, Roc M. and {Jarrett}, Thomas and {Kirkpatrick}, J. Davy and {Padgett}, Deborah and {McMillan}, Robert S. and {Skrutskie}, Michael and {Stanford}, S.~A. and {Cohen}, Martin and {Walker}, Russell G. and {Mather}, John C. and {Leisawitz}, David and {Gautier}, III, Thomas N. and {McLean}, Ian and {Benford}, Dominic and {Lonsdale}, Carol J. and {Blain}, Andrew and {Mendez}, Bryan and {Irace}, William R. and {Duval}, Valerie and {Liu}, Fengchuan and {Royer}, Don and {Heinrichsen}, Ingolf and {Howard}, Joan and {Shannon}, Mark and {Kendall}, Martha and {Walsh}, Amy L. and {Larsen}, Mark and {Cardon}, Joel G. and {Schick}, Scott and {Schwalm}, Mark and {Abid}, Mohamed and {Fabinsky}, Beth and {Naes}, Larry and {Tsai}, Chao-Wei},
        title = "{The Wide-field Infrared Survey Explorer (WISE): Mission Description and Initial On-orbit Performance}",
      journal = {\aj},
         year = 2010,
        month = dec,
       volume = {140},
       number = {6},
        pages = {1868-1881},
          doi = {10.1088/0004-6256/140/6/1868},
archivePrefix = {arXiv},
       eprint = {1008.0031},
 primaryClass = {astro-ph.IM},
       adsurl = {https://ui.adsabs.harvard.edu/abs/2010AJ....140.1868W}
}

@ARTICLE{2019ApJ...873..111I,
       author = {{Ivezi{\'c}}, {\v{Z}}eljko and {Kahn}, Steven M. and {Tyson}, J. Anthony and {Abel}, Bob and {Acosta}, Emily and {Allsman}, Robyn and {Alonso}, David and {AlSayyad}, Yusra and {Anderson}, Scott F. and {Andrew}, John and {Angel}, James Roger P. and {Angeli}, George Z. and {Ansari}, Reza and {Antilogus}, Pierre and {Araujo}, Constanza and {Armstrong}, Robert and {Arndt}, Kirk T. and {Astier}, Pierre and {Aubourg}, {\'E}ric and {Auza}, Nicole and {Axelrod}, Tim S. and {Bard}, Deborah J. and {Barr}, Jeff D. and {Barrau}, Aurelian and {Bartlett}, James G. and {Bauer}, Amanda E. and {Bauman}, Brian J. and {Baumont}, Sylvain and {Bechtol}, Ellen and {Bechtol}, Keith and {Becker}, Andrew C. and {Becla}, Jacek and {Beldica}, Cristina and {Bellavia}, Steve and {Bianco}, Federica B. and {Biswas}, Rahul and {Blanc}, Guillaume and {Blazek}, Jonathan and {Blandford}, Roger D. and {Bloom}, Josh S. and {Bogart}, Joanne and {Bond}, Tim W. and {Booth}, Michael T. and {Borgland}, Anders W. and {Borne}, Kirk and {Bosch}, James F. and {Boutigny}, Dominique and {Brackett}, Craig A. and {Bradshaw}, Andrew and {Brandt}, William Nielsen and {Brown}, Michael E. and {Bullock}, James S. and {Burchat}, Patricia and {Burke}, David L. and {Cagnoli}, Gianpietro and {Calabrese}, Daniel and {Callahan}, Shawn and {Callen}, Alice L. and {Carlin}, Jeffrey L. and {Carlson}, Erin L. and {Chandrasekharan}, Srinivasan and {Charles-Emerson}, Glenaver and {Chesley}, Steve and {Cheu}, Elliott C. and {Chiang}, Hsin-Fang and {Chiang}, James and {Chirino}, Carol and {Chow}, Derek and {Ciardi}, David R. and {Claver}, Charles F. and {Cohen-Tanugi}, Johann and {Cockrum}, Joseph J. and {Coles}, Rebecca and {Connolly}, Andrew J. and {Cook}, Kem H. and {Cooray}, Asantha and {Covey}, Kevin R. and {Cribbs}, Chris and {Cui}, Wei and {Cutri}, Roc and {Daly}, Philip N. and {Daniel}, Scott F. and {Daruich}, Felipe and {Daubard}, Guillaume and {Daues}, Greg and {Dawson}, William and {Delgado}, Francisco and {Dellapenna}, Alfred and {de Peyster}, Robert and {de Val-Borro}, Miguel and {Digel}, Seth W. and {Doherty}, Peter and {Dubois}, Richard and {Dubois-Felsmann}, Gregory P. and {Durech}, Josef and {Economou}, Frossie and {Eifler}, Tim and {Eracleous}, Michael and {Emmons}, Benjamin L. and {Fausti Neto}, Angelo and {Ferguson}, Henry and {Figueroa}, Enrique and {Fisher-Levine}, Merlin and {Focke}, Warren and {Foss}, Michael D. and {Frank}, James and {Freemon}, Michael D. and {Gangler}, Emmanuel and {Gawiser}, Eric and {Geary}, John C. and {Gee}, Perry and {Geha}, Marla and {Gessner}, Charles J.~B. and {Gibson}, Robert R. and {Gilmore}, D. Kirk and {Glanzman}, Thomas and {Glick}, William and {Goldina}, Tatiana and {Goldstein}, Daniel A. and {Goodenow}, Iain and {Graham}, Melissa L. and {Gressler}, William J. and {Gris}, Philippe and {Guy}, Leanne P. and {Guyonnet}, Augustin and {Haller}, Gunther and {Harris}, Ron and {Hascall}, Patrick A. and {Haupt}, Justine and {Hernandez}, Fabio and {Herrmann}, Sven and {Hileman}, Edward and {Hoblitt}, Joshua and {Hodgson}, John A. and {Hogan}, Craig and {Howard}, James D. and {Huang}, Dajun and {Huffer}, Michael E. and {Ingraham}, Patrick and {Innes}, Walter R. and {Jacoby}, Suzanne H. and {Jain}, Bhuvnesh and {Jammes}, Fabrice and {Jee}, M. James and {Jenness}, Tim and {Jernigan}, Garrett and {Jevremovi{\'c}}, Darko and {Johns}, Kenneth and {Johnson}, Anthony S. and {Johnson}, Margaret W.~G. and {Jones}, R. Lynne and {Juramy-Gilles}, Claire and {Juri{\'c}}, Mario and {Kalirai}, Jason S. and {Kallivayalil}, Nitya J. and {Kalmbach}, Bryce and {Kantor}, Jeffrey P. and {Karst}, Pierre and {Kasliwal}, Mansi M. and {Kelly}, Heather and {Kessler}, Richard and {Kinnison}, Veronica and {Kirkby}, David and {Knox}, Lloyd and {Kotov}, Ivan V. and {Krabbendam}, Victor L. and {Krughoff}, K. Simon and {Kub{\'a}nek}, Petr and {Kuczewski}, John and {Kulkarni}, Shri and {Ku}, John and {Kurita}, Nadine R. and {Lage}, Craig S. and {Lambert}, Ron and {Lange}, Travis and {Langton}, J. Brian and {Le Guillou}, Laurent and {Levine}, Deborah and {Liang}, Ming and {Lim}, Kian-Tat and {Lintott}, Chris J. and {Long}, Kevin E. and {Lopez}, Margaux and {Lotz}, Paul J. and {Lupton}, Robert H. and {Lust}, Nate B. and {MacArthur}, Lauren A. and {Mahabal}, Ashish and {Mandelbaum}, Rachel and {Markiewicz}, Thomas W. and {Marsh}, Darren S. and {Marshall}, Philip J. and {Marshall}, Stuart and {May}, Morgan and {McKercher}, Robert and {McQueen}, Michelle and {Meyers}, Joshua and {Migliore}, Myriam and {Miller}, Michelle and {Mills}, David J.},
        title = "{LSST: From Science Drivers to Reference Design and Anticipated Data Products}",
      journal = {\apj},
         year = 2019,
        month = mar,
       volume = {873},
       number = {2},
          eid = {111},
        pages = {111},
          doi = {10.3847/1538-4357/ab042c},
archivePrefix = {arXiv},
       eprint = {0805.2366},
 primaryClass = {astro-ph},
       adsurl = {https://ui.adsabs.harvard.edu/abs/2019ApJ...873..111I}
}

@ARTICLE{2018AJ....156..123A,
       author = {{Astropy Collaboration} and {Price-Whelan}, A.~M. and {Sip{\H{o}}cz}, B.~M. and {G{\"u}nther}, H.~M. and {Lim}, P.~L. and {Crawford}, S.~M. and {Conseil}, S. and {Shupe}, D.~L. and {Craig}, M.~W. and {Dencheva}, N. and {Ginsburg}, A. and {VanderPlas}, J.~T. and {Bradley}, L.~D. and {P{\'e}rez-Su{\'a}rez}, D. and {de Val-Borro}, M. and {Aldcroft}, T.~L. and {Cruz}, K.~L. and {Robitaille}, T.~P. and {Tollerud}, E.~J. and {Ardelean}, C. and {Babej}, T. and {Bach}, Y.~P. and {Bachetti}, M. and {Bakanov}, A.~V. and {Bamford}, S.~P. and {Barentsen}, G. and {Barmby}, P. and {Baumbach}, A. and {Berry}, K.~L. and {Biscani}, F. and {Boquien}, M. and {Bostroem}, K.~A. and {Bouma}, L.~G. and {Brammer}, G.~B. and {Bray}, E.~M. and {Breytenbach}, H. and {Buddelmeijer}, H. and {Burke}, D.~J. and {Calderone}, G. and {Cano Rodr{\'\i}guez}, J.~L. and {Cara}, M. and {Cardoso}, J.~V.~M. and {Cheedella}, S. and {Copin}, Y. and {Corrales}, L. and {Crichton}, D. and {D'Avella}, D. and {Deil}, C. and {Depagne}, {\'E}. and {Dietrich}, J.~P. and {Donath}, A. and {Droettboom}, M. and {Earl}, N. and {Erben}, T. and {Fabbro}, S. and {Ferreira}, L.~A. and {Finethy}, T. and {Fox}, R.~T. and {Garrison}, L.~H. and {Gibbons}, S.~L.~J. and {Goldstein}, D.~A. and {Gommers}, R. and {Greco}, J.~P. and {Greenfield}, P. and {Groener}, A.~M. and {Grollier}, F. and {Hagen}, A. and {Hirst}, P. and {Homeier}, D. and {Horton}, A.~J. and {Hosseinzadeh}, G. and {Hu}, L. and {Hunkeler}, J.~S. and {Ivezi{\'c}}, {\v{Z}}. and {Jain}, A. and {Jenness}, T. and {Kanarek}, G. and {Kendrew}, S. and {Kern}, N.~S. and {Kerzendorf}, W.~E. and {Khvalko}, A. and {King}, J. and {Kirkby}, D. and {Kulkarni}, A.~M. and {Kumar}, A. and {Lee}, A. and {Lenz}, D. and {Littlefair}, S.~P. and {Ma}, Z. and {Macleod}, D.~M. and {Mastropietro}, M. and {McCully}, C. and {Montagnac}, S. and {Morris}, B.~M. and {Mueller}, M. and {Mumford}, S.~J. and {Muna}, D. and {Murphy}, N.~A. and {Nelson}, S. and {Nguyen}, G.~H. and {Ninan}, J.~P. and {N{\"o}the}, M. and {Ogaz}, S. and {Oh}, S. and {Parejko}, J.~K. and {Parley}, N. and {Pascual}, S. and {Patil}, R. and {Patil}, A.~A. and {Plunkett}, A.~L. and {Prochaska}, J.~X. and {Rastogi}, T. and {Reddy Janga}, V. and {Sabater}, J. and {Sakurikar}, P. and {Seifert}, M. and {Sherbert}, L.~E. and {Sherwood-Taylor}, H. and {Shih}, A.~Y. and {Sick}, J. and {Silbiger}, M.~T. and {Singanamalla}, S. and {Singer}, L.~P. and {Sladen}, P.~H. and {Sooley}, K.~A. and {Sornarajah}, S. and {Streicher}, O. and {Teuben}, P. and {Thomas}, S.~W. and {Tremblay}, G.~R. and {Turner}, J.~E.~H. and {Terr{\'o}n}, V. and {van Kerkwijk}, M.~H. and {de la Vega}, A. and {Watkins}, L.~L. and {Weaver}, B.~A. and {Whitmore}, J.~B. and {Woillez}, J. and {Zabalza}, V. and {Astropy Contributors}},
        title = "{The Astropy Project: Building an Open-science Project and Status of the v2.0 Core Package}",
      journal = {\aj},
         year = 2018,
        month = sep,
       volume = {156},
       number = {3},
          eid = {123},
        pages = {123},
          doi = {10.3847/1538-3881/aabc4f},
archivePrefix = {arXiv},
       eprint = {1801.02634},
 primaryClass = {astro-ph.IM},
       adsurl = {https://ui.adsabs.harvard.edu/abs/2018AJ....156..123A}
}

@ARTICLE{2022ApJ...935..167A,
       author = {{Astropy Collaboration} and {Price-Whelan}, Adrian M. and {Lim}, Pey Lian and {Earl}, Nicholas and {Starkman}, Nathaniel and {Bradley}, Larry and {Shupe}, David L. and {Patil}, Aarya A. and {Corrales}, Lia and {Brasseur}, C.~E. and {N{\"o}the}, Maximilian and {Donath}, Axel and {Tollerud}, Erik and {Morris}, Brett M. and {Ginsburg}, Adam and {Vaher}, Eero and {Weaver}, Benjamin A. and {Tocknell}, James and {Jamieson}, William and {van Kerkwijk}, Marten H. and {Robitaille}, Thomas P. and {Merry}, Bruce and {Bachetti}, Matteo and {G{\"u}nther}, H. Moritz and {Aldcroft}, Thomas L. and {Alvarado-Montes}, Jaime A. and {Archibald}, Anne M. and {B{\'o}di}, Attila and {Bapat}, Shreyas and {Barentsen}, Geert and {Baz{\'a}n}, Juanjo and {Biswas}, Manish and {Boquien}, M{\'e}d{\'e}ric and {Burke}, D.~J. and {Cara}, Daria and {Cara}, Mihai and {Conroy}, Kyle E. and {Conseil}, Simon and {Craig}, Matthew W. and {Cross}, Robert M. and {Cruz}, Kelle L. and {D'Eugenio}, Francesco and {Dencheva}, Nadia and {Devillepoix}, Hadrien A.~R. and {Dietrich}, J{\"o}rg P. and {Eigenbrot}, Arthur Davis and {Erben}, Thomas and {Ferreira}, Leonardo and {Foreman-Mackey}, Daniel and {Fox}, Ryan and {Freij}, Nabil and {Garg}, Suyog and {Geda}, Robel and {Glattly}, Lauren and {Gondhalekar}, Yash and {Gordon}, Karl D. and {Grant}, David and {Greenfield}, Perry and {Groener}, Austen M. and {Guest}, Steve and {Gurovich}, Sebastian and {Handberg}, Rasmus and {Hart}, Akeem and {Hatfield-Dodds}, Zac and {Homeier}, Derek and {Hosseinzadeh}, Griffin and {Jenness}, Tim and {Jones}, Craig K. and {Joseph}, Prajwel and {Kalmbach}, J. Bryce and {Karamehmetoglu}, Emir and {Ka{\l}uszy{\'n}ski}, Miko{\l}aj and {Kelley}, Michael S.~P. and {Kern}, Nicholas and {Kerzendorf}, Wolfgang E. and {Koch}, Eric W. and {Kulumani}, Shankar and {Lee}, Antony and {Ly}, Chun and {Ma}, Zhiyuan and {MacBride}, Conor and {Maljaars}, Jakob M. and {Muna}, Demitri and {Murphy}, N.~A. and {Norman}, Henrik and {O'Steen}, Richard and {Oman}, Kyle A. and {Pacifici}, Camilla and {Pascual}, Sergio and {Pascual-Granado}, J. and {Patil}, Rohit R. and {Perren}, Gabriel I. and {Pickering}, Timothy E. and {Rastogi}, Tanuj and {Roulston}, Benjamin R. and {Ryan}, Daniel F. and {Rykoff}, Eli S. and {Sabater}, Jose and {Sakurikar}, Parikshit and {Salgado}, Jes{\'u}s and {Sanghi}, Aniket and {Saunders}, Nicholas and {Savchenko}, Volodymyr and {Schwardt}, Ludwig and {Seifert-Eckert}, Michael and {Shih}, Albert Y. and {Jain}, Anany Shrey and {Shukla}, Gyanendra and {Sick}, Jonathan and {Simpson}, Chris and {Singanamalla}, Sudheesh and {Singer}, Leo P. and {Singhal}, Jaladh and {Sinha}, Manodeep and {Sip{\H{o}}cz}, Brigitta M. and {Spitler}, Lee R. and {Stansby}, David and {Streicher}, Ole and {{\v{S}}umak}, Jani and {Swinbank}, John D. and {Taranu}, Dan S. and {Tewary}, Nikita and {Tremblay}, Grant R. and {de Val-Borro}, Miguel and {Van Kooten}, Samuel J. and {Vasovi{\'c}}, Zlatan and {Verma}, Shresth and {de Miranda Cardoso}, Jos{\'e} Vin{\'\i}cius and {Williams}, Peter K.~G. and {Wilson}, Tom J. and {Winkel}, Benjamin and {Wood-Vasey}, W.~M. and {Xue}, Rui and {Yoachim}, Peter and {Zhang}, Chen and {Zonca}, Andrea and {Astropy Project Contributors}},
        title = "{The Astropy Project: Sustaining and Growing a Community-oriented Open-source Project and the Latest Major Release (v5.0) of the Core Package}",
      journal = {\apj},
         year = 2022,
        month = aug,
       volume = {935},
       number = {2},
          eid = {167},
        pages = {167},
          doi = {10.3847/1538-4357/ac7c74},
archivePrefix = {arXiv},
       eprint = {2206.14220},
 primaryClass = {astro-ph.IM},
       adsurl = {https://ui.adsabs.harvard.edu/abs/2022ApJ...935..167A}
}

@ARTICLE{2025A&A...697A...1E,
       author = {{Euclid Collaboration} and {Mellier}, Y. and {Abdurro'uf} and {Acevedo Barroso}, J.~A. and {Ach{\'u}carro}, A. and {Adamek}, J. and {Adam}, R. and {Addison}, G.~E. and {Aghanim}, N. and {Aguena}, M. and {Ajani}, V. and {Akrami}, Y. and {Al-Bahlawan}, A. and {Alavi}, A. and {Albuquerque}, I.~S. and {Alestas}, G. and {Alguero}, G. and {Allaoui}, A. and {Allen}, S.~W. and {Allevato}, V. and {Alonso-Tetilla}, A.~V. and {Altieri}, B. and {Alvarez-Candal}, A. and {Alvi}, S. and {Amara}, A. and {Amendola}, L. and {Amiaux}, J. and {Andika}, I.~T. and {Andreon}, S. and {Andrews}, A. and {Angora}, G. and {Angulo}, R.~E. and {Annibali}, F. and {Anselmi}, A. and {Anselmi}, S. and {Arcari}, S. and {Archidiacono}, M. and {Aric{\`o}}, G. and {Arnaud}, M. and {Arnouts}, S. and {Asgari}, M. and {Asorey}, J. and {Atayde}, L. and {Atek}, H. and {Atrio-Barandela}, F. and {Aubert}, M. and {Aubourg}, E. and {Auphan}, T. and {Auricchio}, N. and {Aussel}, B. and {Aussel}, H. and {Avelino}, P.~P. and {Avgoustidis}, A. and {Avila}, S. and {Awan}, S. and {Azzollini}, R. and {Baccigalupi}, C. and {Bachelet}, E. and {Bacon}, D. and {Baes}, M. and {Bagley}, M.~B. and {Bahr-Kalus}, B. and {Balaguera-Antolinez}, A. and {Balbinot}, E. and {Balcells}, M. and {Baldi}, M. and {Baldry}, I. and {Balestra}, A. and {Ballardini}, M. and {Ballester}, O. and {Balogh}, M. and {Ba{\~n}ados}, E. and {Barbier}, R. and {Bardelli}, S. and {Baron}, M. and {Barreiro}, T. and {Barrena}, R. and {Barriere}, J.-C. and {Barros}, B.~J. and {Barthelemy}, A. and {Bartolo}, N. and {Basset}, A. and {Battaglia}, P. and {Battisti}, A.~J. and {Baugh}, C.~M. and {Baumont}, L. and {Bazzanini}, L. and {Beaulieu}, J.-P. and {Beckmann}, V. and {Belikov}, A.~N. and {Bel}, J. and {Bellagamba}, F. and {Bella}, M. and {Bellini}, E. and {Benabed}, K. and {Bender}, R. and {Benevento}, G. and {Bennett}, C.~L. and {Benson}, K. and {Bergamini}, P. and {Bermejo-Climent}, J.~R. and {Bernardeau}, F. and {Bertacca}, D. and {Berthe}, M. and {Berthier}, J. and {Bethermin}, M. and {Beutler}, F. and {Bevillon}, C. and {Bhargava}, S. and {Bhatawdekar}, R. and {Bianchi}, D. and {Bisigello}, L. and {Biviano}, A. and {Blake}, R.~P. and {Blanchard}, A. and {Blazek}, J. and {Blot}, L. and {Bosco}, A. and {Bodendorf}, C. and {Boenke}, T. and {B{\"o}hringer}, H. and {Boldrini}, P. and {Bolzonella}, M. and {Bonchi}, A. and {Bonici}, M. and {Bonino}, D. and {Bonino}, L. and {Bonvin}, C. and {Bon}, W. and {Booth}, J.~T. and {Borgani}, S. and {Borlaff}, A.~S. and {Borsato}, E. and {Bose}, B. and {Botticella}, M.~T. and {Boucaud}, A. and {Bouche}, F. and {Boucher}, J.~S. and {Boutigny}, D. and {Bouvard}, T. and {Bouwens}, R. and {Bouy}, H. and {Bowler}, R.~A.~A. and {Bozza}, V. and {Bozzo}, E. and {Branchini}, E. and {Brando}, G. and {Brau-Nogue}, S. and {Brekke}, P. and {Bremer}, M.~N. and {Brescia}, M. and {Breton}, M.-A. and {Brinchmann}, J. and {Brinckmann}, T. and {Brockley-Blatt}, C. and {Brodwin}, M. and {Brouard}, L. and {Brown}, M.~L. and {Bruton}, S. and {Bucko}, J. and {Buddelmeijer}, H. and {Buenadicha}, G. and {Buitrago}, F. and {Burger}, P. and {Burigana}, C. and {Busillo}, V. and {Busonero}, D. and {Cabanac}, R. and {Cabayol-Garcia}, L. and {Cagliari}, M.~S. and {Caillat}, A. and {Caillat}, L. and {Calabrese}, M. and {Calabro}, A. and {Calderone}, G. and {Calura}, F. and {Camacho Quevedo}, B. and {Camera}, S. and {Campos}, L. and {Ca{\~n}as-Herrera}, G. and {Candini}, G.~P. and {Cantiello}, M. and {Capobianco}, V. and {Cappellaro}, E. and {Cappelluti}, N. and {Cappi}, A. and {Caputi}, K.~I. and {Cara}, C. and {Carbone}, C. and {Cardone}, V.~F. and {Carella}, E. and {Carlberg}, R.~G. and {Carle}, M. and {Carminati}, L. and {Caro}, F. and {Carrasco}, J.~M. and {Carretero}, J. and {Carrilho}, P. and {Carron Duque}, J. and {Carry}, B.},
        title = "{Euclid: I. Overview of the Euclid mission}",
      journal = {\aap},
         year = 2025,
        month = may,
       volume = {697},
          eid = {A1},
        pages = {A1},
          doi = {10.1051/0004-6361/202450810},
archivePrefix = {arXiv},
       eprint = {2405.13491},
 primaryClass = {astro-ph.CO},
       adsurl = {https://ui.adsabs.harvard.edu/abs/2025A&A...697A...1E}
}

@ARTICLE{2005ApJ...622..759G,
       author = {{G{\'o}rski}, K.~M. and {Hivon}, E. and {Banday}, A.~J. and {Wandelt}, B.~D. and {Hansen}, F.~K. and {Reinecke}, M. and {Bartelmann}, M.},
        title = "{HEALPix: A Framework for High-Resolution Discretization and Fast Analysis of Data Distributed on the Sphere}",
      journal = {\apj},
         year = 2005,
        month = apr,
       volume = {622},
       number = {2},
        pages = {759-771},
          doi = {10.1086/427976},
archivePrefix = {arXiv},
       eprint = {astro-ph/0409513},
 primaryClass = {astro-ph},
       adsurl = {https://ui.adsabs.harvard.edu/abs/2005ApJ...622..759G}
}

@BOOK{2011piim.book.....D,
       author = {{Draine}, Bruce T.},
        title = "{Physics of the Interstellar and Intergalactic Medium}",
         year = 2011,
       adsurl = {https://ui.adsabs.harvard.edu/abs/2011piim.book.....D}
}

@ARTICLE{2019AJ....157..168D,
       author = {{Dey}, Arjun and {Schlegel}, David J. and {Lang}, Dustin and {Blum}, Robert and {Burleigh}, Kaylan and {Fan}, Xiaohui and {Findlay}, Joseph R. and {Finkbeiner}, Doug and {Herrera}, David and {Juneau}, St{\'e}phanie and {Landriau}, Martin and {Levi}, Michael and {McGreer}, Ian and {Meisner}, Aaron and {Myers}, Adam D. and {Moustakas}, John and {Nugent}, Peter and {Patej}, Anna and {Schlafly}, Edward F. and {Walker}, Alistair R. and {Valdes}, Francisco and {Weaver}, Benjamin A. and {Y{\`e}che}, Christophe and {Zou}, Hu and {Zhou}, Xu and {Abareshi}, Behzad and {Abbott}, T.~M.~C. and {Abolfathi}, Bela and {Aguilera}, C. and {Alam}, Shadab and {Allen}, Lori and {Alvarez}, A. and {Annis}, James and {Ansarinejad}, Behzad and {Aubert}, Marie and {Beechert}, Jacqueline and {Bell}, Eric F. and {BenZvi}, Segev Y. and {Beutler}, Florian and {Bielby}, Richard M. and {Bolton}, Adam S. and {Brice{\~n}o}, C{\'e}sar and {Buckley-Geer}, Elizabeth J. and {Butler}, Karen and {Calamida}, Annalisa and {Carlberg}, Raymond G. and {Carter}, Paul and {Casas}, Ricard and {Castander}, Francisco J. and {Choi}, Yumi and {Comparat}, Johan and {Cukanovaite}, Elena and {Delubac}, Timoth{\'e}e and {DeVries}, Kaitlin and {Dey}, Sharmila and {Dhungana}, Govinda and {Dickinson}, Mark and {Ding}, Zhejie and {Donaldson}, John B. and {Duan}, Yutong and {Duckworth}, Christopher J. and {Eftekharzadeh}, Sarah and {Eisenstein}, Daniel J. and {Etourneau}, Thomas and {Fagrelius}, Parker A. and {Farihi}, Jay and {Fitzpatrick}, Mike and {Font-Ribera}, Andreu and {Fulmer}, Leah and {G{\"a}nsicke}, Boris T. and {Gaztanaga}, Enrique and {George}, Koshy and {Gerdes}, David W. and {Gontcho}, Satya Gontcho A. and {Gorgoni}, Claudio and {Green}, Gregory and {Guy}, Julien and {Harmer}, Diane and {Hernandez}, M. and {Honscheid}, Klaus and {Huang}, Lijuan Wendy and {James}, David J. and {Jannuzi}, Buell T. and {Jiang}, Linhua and {Joyce}, Richard and {Karcher}, Armin and {Karkar}, Sonia and {Kehoe}, Robert and {Kneib}, Jean-Paul and {Kueter-Young}, Andrea and {Lan}, Ting-Wen and {Lauer}, Tod R. and {Le Guillou}, Laurent and {Le Van Suu}, Auguste and {Lee}, Jae Hyeon and {Lesser}, Michael and {Perreault Levasseur}, Laurence and {Li}, Ting S. and {Mann}, Justin L. and {Marshall}, Robert and {Mart{\'\i}nez-V{\'a}zquez}, C.~E. and {Martini}, Paul and {du Mas des Bourboux}, H{\'e}lion and {McManus}, Sean and {Meier}, Tobias Gabriel and {M{\'e}nard}, Brice and {Metcalfe}, Nigel and {Mu{\~n}oz-Guti{\'e}rrez}, Andrea and {Najita}, Joan and {Napier}, Kevin and {Narayan}, Gautham and {Newman}, Jeffrey A. and {Nie}, Jundan and {Nord}, Brian and {Norman}, Dara J. and {Olsen}, Knut A.~G. and {Paat}, Anthony and {Palanque-Delabrouille}, Nathalie and {Peng}, Xiyan and {Poppett}, Claire L. and {Poremba}, Megan R. and {Prakash}, Abhishek and {Rabinowitz}, David and {Raichoor}, Anand and {Rezaie}, Mehdi and {Robertson}, A.~N. and {Roe}, Natalie A. and {Ross}, Ashley J. and {Ross}, Nicholas P. and {Rudnick}, Gregory and {Safonova}, Sasha and {Saha}, Abhijit and {S{\'a}nchez}, F. Javier and {Savary}, Elodie and {Schweiker}, Heidi and {Scott}, Adam and {Seo}, Hee-Jong and {Shan}, Huanyuan and {Silva}, David R. and {Slepian}, Zachary and {Soto}, Christian and {Sprayberry}, David and {Staten}, Ryan and {Stillman}, Coley M. and {Stupak}, Robert J. and {Summers}, David L. and {Sien Tie}, Suk and {Tirado}, H. and {Vargas-Maga{\~n}a}, Mariana and {Vivas}, A. Katherina and {Wechsler}, Risa H. and {Williams}, Doug and {Yang}, Jinyi and {Yang}, Qian and {Yapici}, Tolga and {Zaritsky}, Dennis and {Zenteno}, A. and {Zhang}, Kai and {Zhang}, Tianmeng and {Zhou}, Rongpu and {Zhou}, Zhimin},
        title = "{Overview of the DESI Legacy Imaging Surveys}",
      journal = {\aj},
         year = 2019,
        month = may,
       volume = {157},
       number = {5},
          eid = {168},
        pages = {168},
          doi = {10.3847/1538-3881/ab089d},
archivePrefix = {arXiv},
       eprint = {1804.08657},
 primaryClass = {astro-ph.IM},
       adsurl = {https://ui.adsabs.harvard.edu/abs/2019AJ....157..168D}
}

@ARTICLE{1978ApJ...224..132B,
       author = {{Bohlin}, R.~C. and {Savage}, B.~D. and {Drake}, J.~F.},
        title = "{A survey of interstellar H I from Lalpha absorption measurements. II.}",
      journal = {\apj},
         year = 1978,
        month = aug,
       volume = {224},
        pages = {132-142},
          doi = {10.1086/156357},
       adsurl = {https://ui.adsabs.harvard.edu/abs/1978ApJ...224..132B}
}

@ARTICLE{2019ApJ...887...93G,
       author = {{Green}, Gregory M. and {Schlafly}, Edward and {Zucker}, Catherine and {Speagle}, Joshua S. and {Finkbeiner}, Douglas},
        title = "{A 3D Dust Map Based on Gaia, Pan-STARRS 1, and 2MASS}",
      journal = {\apj},
         year = 2019,
        month = dec,
       volume = {887},
       number = {1},
          eid = {93},
        pages = {93},
          doi = {10.3847/1538-4357/ab5362},
archivePrefix = {arXiv},
       eprint = {1905.02734},
 primaryClass = {astro-ph.GA},
       adsurl = {https://ui.adsabs.harvard.edu/abs/2019ApJ...887...93G}
}

@ARTICLE{2007ApJ...657..810D,
       author = {{Draine}, B.~T. and {Li}, Aigen},
        title = "{Infrared Emission from Interstellar Dust. IV. The Silicate-Graphite-PAH Model in the Post-Spitzer Era}",
      journal = {\apj},
         year = 2007,
        month = mar,
       volume = {657},
       number = {2},
        pages = {810-837},
          doi = {10.1086/511055},
archivePrefix = {arXiv},
       eprint = {astro-ph/0608003},
 primaryClass = {astro-ph},
       adsurl = {https://ui.adsabs.harvard.edu/abs/2007ApJ...657..810D}
}

@ARTICLE{2016A&A...594A.116H,
       author = {{HI4PI Collaboration} and {Ben Bekhti}, N. and {Fl{\"o}er}, L. and {Keller}, R. and {Kerp}, J. and {Lenz}, D. and {Winkel}, B. and {Bailin}, J. and {Calabretta}, M.~R. and {Dedes}, L. and {Ford}, H.~A. and {Gibson}, B.~K. and {Haud}, U. and {Janowiecki}, S. and {Kalberla}, P.~M.~W. and {Lockman}, F.~J. and {McClure-Griffiths}, N.~M. and {Murphy}, T. and {Nakanishi}, H. and {Pisano}, D.~J. and {Staveley-Smith}, L.},
        title = "{HI4PI: A full-sky H I survey based on EBHIS and GASS}",
      journal = {\aap},
         year = 2016,
        month = oct,
       volume = {594},
          eid = {A116},
        pages = {A116},
          doi = {10.1051/0004-6361/201629178},
archivePrefix = {arXiv},
       eprint = {1610.06175},
 primaryClass = {astro-ph.GA},
       adsurl = {https://ui.adsabs.harvard.edu/abs/2016A&A...594A.116H}
}

@ARTICLE{2022A&A...662A.112E,
       author = {{Euclid Collaboration} and {Scaramella}, R. and {Amiaux}, J. and {Mellier}, Y. and {Burigana}, C. and {Carvalho}, C.~S. and {Cuillandre}, J.-C. and {Da Silva}, A. and {Derosa}, A. and {Dinis}, J. and {Maiorano}, E. and {Maris}, M. and {Tereno}, I. and {Laureijs}, R. and {Boenke}, T. and {Buenadicha}, G. and {Dupac}, X. and {Gaspar Venancio}, L.~M. and {G{\'o}mez-{\'A}lvarez}, P. and {Hoar}, J. and {Lorenzo Alvarez}, J. and {Racca}, G.~D. and {Saavedra-Criado}, G. and {Schwartz}, J. and {Vavrek}, R. and {Schirmer}, M. and {Aussel}, H. and {Azzollini}, R. and {Cardone}, V.~F. and {Cropper}, M. and {Ealet}, A. and {Garilli}, B. and {Gillard}, W. and {Granett}, B.~R. and {Guzzo}, L. and {Hoekstra}, H. and {Jahnke}, K. and {Kitching}, T. and {Maciaszek}, T. and {Meneghetti}, M. and {Miller}, L. and {Nakajima}, R. and {Niemi}, S.~M. and {Pasian}, F. and {Percival}, W.~J. and {Pottinger}, S. and {Sauvage}, M. and {Scodeggio}, M. and {Wachter}, S. and {Zacchei}, A. and {Aghanim}, N. and {Amara}, A. and {Auphan}, T. and {Auricchio}, N. and {Awan}, S. and {Balestra}, A. and {Bender}, R. and {Bodendorf}, C. and {Bonino}, D. and {Branchini}, E. and {Brau-Nogue}, S. and {Brescia}, M. and {Candini}, G.~P. and {Capobianco}, V. and {Carbone}, C. and {Carlberg}, R.~G. and {Carretero}, J. and {Casas}, R. and {Castander}, F.~J. and {Castellano}, M. and {Cavuoti}, S. and {Cimatti}, A. and {Cledassou}, R. and {Congedo}, G. and {Conselice}, C.~J. and {Conversi}, L. and {Copin}, Y. and {Corcione}, L. and {Costille}, A. and {Courbin}, F. and {Degaudenzi}, H. and {Douspis}, M. and {Dubath}, F. and {Duncan}, C.~A.~J. and {Dusini}, S. and {Farrens}, S. and {Ferriol}, S. and {Fosalba}, P. and {Fourmanoit}, N. and {Frailis}, M. and {Franceschi}, E. and {Franzetti}, P. and {Fumana}, M. and {Gillis}, B. and {Giocoli}, C. and {Grazian}, A. and {Grupp}, F. and {Haugan}, S.~V.~H. and {Holmes}, W. and {Hormuth}, F. and {Hudelot}, P. and {Kermiche}, S. and {Kiessling}, A. and {Kilbinger}, M. and {Kohley}, R. and {Kubik}, B. and {K{\"u}mmel}, M. and {Kunz}, M. and {Kurki-Suonio}, H. and {Lahav}, O. and {Ligori}, S. and {Lilje}, P.~B. and {Lloro}, I. and {Mansutti}, O. and {Marggraf}, O. and {Markovic}, K. and {Marulli}, F. and {Massey}, R. and {Maurogordato}, S. and {Melchior}, M. and {Merlin}, E. and {Meylan}, G. and {Mohr}, J.~J. and {Moresco}, M. and {Morin}, B. and {Moscardini}, L. and {Munari}, E. and {Nichol}, R.~C. and {Padilla}, C. and {Paltani}, S. and {Peacock}, J. and {Pedersen}, K. and {Pettorino}, V. and {Pires}, S. and {Poncet}, M. and {Popa}, L. and {Pozzetti}, L. and {Raison}, F. and {Rebolo}, R. and {Rhodes}, J. and {Rix}, H.-W. and {Roncarelli}, M. and {Rossetti}, E. and {Saglia}, R. and {Schneider}, P. and {Schrabback}, T. and {Secroun}, A. and {Seidel}, G. and {Serrano}, S. and {Sirignano}, C. and {Sirri}, G. and {Skottfelt}, J. and {Stanco}, L. and {Starck}, J.~L. and {Tallada-Cresp{\'\i}}, P. and {Tavagnacco}, D. and {Taylor}, A.~N. and {Teplitz}, H.~I. and {Toledo-Moreo}, R. and {Torradeflot}, F. and {Trifoglio}, M. and {Valentijn}, E.~A. and {Valenziano}, L. and {Verdoes Kleijn}, G.~A. and {Wang}, Y. and {Welikala}, N. and {Weller}, J. and {Wetzstein}, M. and {Zamorani}, G. and {Zoubian}, J. and {Andreon}, S. and {Baldi}, M. and {Bardelli}, S. and {Boucaud}, A. and {Camera}, S. and {Di Ferdinando}, D. and {Fabbian}, G. and {Farinelli}, R. and {Galeotta}, S. and {Graci{\'a}-Carpio}, J. and {Maino}, D. and {Medinaceli}, E. and {Mei}, S. and {Neissner}, C. and {Polenta}, G. and {Renzi}, A. and {Romelli}, E. and {Rosset}, C. and {Sureau}, F. and {Tenti}, M. and {Vassallo}, T. and {Zucca}, E. and {Baccigalupi}, C. and {Balaguera-Antol{\'\i}nez}, A. and {Battaglia}, P. and {Biviano}, A. and {Borgani}, S. and {Bozzo}, E. and {Cabanac}, R. and {Cappi}, A.},
        title = "{Euclid preparation. I. The Euclid Wide Survey}",
      journal = {\aap},
         year = 2022,
        month = jun,
       volume = {662},
          eid = {A112},
        pages = {A112},
          doi = {10.1051/0004-6361/202141938},
archivePrefix = {arXiv},
       eprint = {2108.01201},
 primaryClass = {astro-ph.CO},
       adsurl = {https://ui.adsabs.harvard.edu/abs/2022A&A...662A.112E}
}

@ARTICLE{2015arXiv150303757S,
       author = {{Spergel}, D. and {Gehrels}, N. and {Baltay}, C. and {Bennett}, D. and {Breckinridge}, J. and {Donahue}, M. and {Dressler}, A. and {Gaudi}, B.~S. and {Greene}, T. and {Guyon}, O. and {Hirata}, C. and {Kalirai}, J. and {Kasdin}, N.~J. and {Macintosh}, B. and {Moos}, W. and {Perlmutter}, S. and {Postman}, M. and {Rauscher}, B. and {Rhodes}, J. and {Wang}, Y. and {Weinberg}, D. and {Benford}, D. and {Hudson}, M. and {Jeong}, W.-S. and {Mellier}, Y. and {Traub}, W. and {Yamada}, T. and {Capak}, P. and {Colbert}, J. and {Masters}, D. and {Penny}, M. and {Savransky}, D. and {Stern}, D. and {Zimmerman}, N. and {Barry}, R. and {Bartusek}, L. and {Carpenter}, K. and {Cheng}, E. and {Content}, D. and {Dekens}, F. and {Demers}, R. and {Grady}, K. and {Jackson}, C. and {Kuan}, G. and {Kruk}, J. and {Melton}, M. and {Nemati}, B. and {Parvin}, B. and {Poberezhskiy}, I. and {Peddie}, C. and {Ruffa}, J. and {Wallace}, J.~K. and {Whipple}, A. and {Wollack}, E. and {Zhao}, F.},
        title = "{Wide-Field InfrarRed Survey Telescope-Astrophysics Focused Telescope Assets WFIRST-AFTA 2015 Report}",
      journal = {arXiv e-prints},
         year = 2015,
        month = mar,
          eid = {arXiv:1503.03757},
        pages = {arXiv:1503.03757},
          doi = {10.48550/arXiv.1503.03757},
archivePrefix = {arXiv},
       eprint = {1503.03757},
 primaryClass = {astro-ph.IM},
       adsurl = {https://ui.adsabs.harvard.edu/abs/2015arXiv150303757S}
}

@ARTICLE{2026ApJ...999..139B,
       author = {{Bock}, James J. and {Aboobaker}, Asad M. and {Adamo}, Joseph and {Akeson}, Rachel and {Alred}, John M. and {Alibay}, Farah and {Ashby}, Matthew L.~N. and {Bach}, Yoonsoo P. and {Bleem}, Lindsey E. and {Bolton}, Douglas and {Braun}, David F. and {Bruton}, Sean and {Bryan}, Sean A. and {Chang}, Tzu-Ching and {Chen}, Shuang-Shuang and {Cheng}, Yun-Ting and {Cheshire}, IV, James R. and {Chiang}, Yi-Kuan and {Choppin de Janvry}, Jean and {Condon}, Samuel and {Cook}, Walter R. and {Cooray}, Asantha and {Crill}, Brendan P. and {Cukierman}, Ari J. and {Dor{\'e}}, Olivier and {Dowell}, C. Darren and {Dubois-Felsmann}, Gregory P. and {Eifler}, Tim and {Everett}, Spencer and {Fabinsky}, Beth E. and {Faisst}, Andreas L. and {Fanson}, James L. and {Farrington}, Allen H. and {Fatahi}, Tamim and {Fazar}, Candice M. and {Feder}, Richard M. and {Frater}, Eric H. and {Grasshorn Gebhardt}, Henry S. and {Giri}, Utkarsh and {Goldina}, Tatiana and {Gorjian}, Varoujan and {Habib}, Salman and {Hart}, William G. and {Heinrich}, Chen and {Hora}, Joseph L. and {Huai}, Zhaoyu and {Hui}, Howard and {Jo}, Young-Soo and {Jeong}, Woong-Seob and {Kang}, Jae Hwan and {Kang}, Miju and {Kecman}, Branislav and {Kim}, Chul-Hwan and {Kim}, Jaeyeong and {Kim}, Minjin and {Kim}, Young-Jun and {Kim}, Yongjung and {Kirkpatrick}, J. Davy and {Kobayashi}, Yosuke and {Korngut}, Phil M. and {Krause}, Elisabeth and {Lee}, Bomee and {Lee}, Ho-Gyu and {Lee}, Jae-Joon and {Lee}, Jeong-Eun and {Lisse}, Carey M. and {Mariani}, Giacomo and {Masters}, Daniel C. and {Mauskopf}, Philip D. and {Melnick}, Gary J. and {Minasyan}, Mary H. and {Mirocha}, Jordan and {Miyasaka}, Hiromasa and {Moore}, Anne and {Moore}, Bradley D. and {Murgia}, Giulia and {Naylor}, Bret J. and {Nelson}, Christina and {Nguyen}, Chi H. and {Nguyen}, Hien T. and {Noh}, Jinyoung K. and {Padin}, Stephen and {Paladini}, Roberta and {Park}, Sung-Joon and {Penanen}, Konstantin I. and {Putnam}, Dustin S. and {Pyo}, Jeonghyun and {Ramachandra}, Nesar and {Ramanathan}, Keshav and {Rustamkulov}, Zafar and {Reiley}, Daniel J. and {Rice}, Eric B. and {Rocca}, Jennifer M. and {Seok}, Ji Yeon and {Smith}, Roger and {Stober}, Jeremy and {Susca}, Sara and {Teplitz}, Harry I. and {Thelen}, Michael P. and {Tolls}, Volker and {Torrini}, Gabriela and {Trangsrud}, Amy R. and {Unwin}, Stephen and {Velicheti}, Phani and {Wang}, Pao-Yu and {Wen}, Robin Y. and {Werner}, Michael W. and {Williams}, Abby E. and {Williamson}, Ross and {Wincentsen}, James and {Windhorst}, Rogier A. and {Yang}, Soung-Chul and {Yang}, Yujin and {Zemcov}, Michael},
        title = "{The SPHEREx Satellite Mission}",
      journal = {\apj},
         year = 2026,
        month = mar,
       volume = {999},
       number = {1},
          eid = {139},
        pages = {139},
          doi = {10.3847/1538-4357/ae2be2},
archivePrefix = {arXiv},
       eprint = {2511.02985},
 primaryClass = {astro-ph.IM},
       adsurl = {https://ui.adsabs.harvard.edu/abs/2026ApJ...999..139B}
}

@ARTICLE{2024A&A...685A..82E,
       author = {{Edenhofer}, Gordian and {Zucker}, Catherine and {Frank}, Philipp and {Saydjari}, Andrew K. and {Speagle}, Joshua S. and {Finkbeiner}, Douglas and {En{\ss}lin}, Torsten A.},
        title = "{A parsec-scale Galactic 3D dust map out to 1.25 kpc from the Sun}",
      journal = {\aap},
         year = 2024,
        month = may,
       volume = {685},
          eid = {A82},
        pages = {A82},
          doi = {10.1051/0004-6361/202347628},
archivePrefix = {arXiv},
       eprint = {2308.01295},
 primaryClass = {astro-ph.GA},
       adsurl = {https://ui.adsabs.harvard.edu/abs/2024A&A...685A..82E}
}

@ARTICLE{1984ApJ...278L...1N,
       author = {{Neugebauer}, G. and {Habing}, H.~J. and {van Duinen}, R. and {Aumann}, H.~H. and {Baud}, B. and {Beichman}, C.~A. and {Beintema}, D.~A. and {Boggess}, N. and {Clegg}, P.~E. and {de Jong}, T. and {Emerson}, J.~P. and {Gautier}, T.~N. and {Gillett}, F.~C. and {Harris}, S. and {Hauser}, M.~G. and {Houck}, J.~R. and {Jennings}, R.~E. and {Low}, F.~J. and {Marsden}, P.~L. and {Miley}, G. and {Olnon}, F.~M. and {Pottasch}, S.~R. and {Raimond}, E. and {Rowan-Robinson}, M. and {Soifer}, B.~T. and {Walker}, R.~G. and {Wesselius}, P.~R. and {Young}, E.},
        title = "{The Infrared Astronomical Satellite (IRAS) Mission.}",
      journal = {\apjl},
         year = 1984,
        month = mar,
       volume = {278},
        pages = {L1-L6},
          doi = {10.1086/184209},
       adsurl = {https://ui.adsabs.harvard.edu/abs/1984ApJ...278L...1N}
}

@ARTICLE{2019arXiv190205569A,
       author = {{Akeson}, Rachel and {Armus}, Lee and {Bachelet}, Etienne and {Bailey}, Vanessa and {Bartusek}, Lisa and {Bellini}, Andrea and {Benford}, Dominic and {Bennett}, David and {Bhattacharya}, Aparna and {Bohlin}, Ralph and {Boyer}, Martha and {Bozza}, Valerio and {Bryden}, Geoffrey and {Calchi Novati}, Sebastiano and {Carpenter}, Kenneth and {Casertano}, Stefano and {Choi}, Ami and {Content}, David and {Dayal}, Pratika and {Dressler}, Alan and {Dor{\'e}}, Olivier and {Fall}, S. Michael and {Fan}, Xiaohui and {Fang}, Xiao and {Filippenko}, Alexei and {Finkelstein}, Steven and {Foley}, Ryan and {Furlanetto}, Steven and {Kalirai}, Jason and {Gaudi}, B. Scott and {Gilbert}, Karoline and {Girard}, Julien and {Grady}, Kevin and {Greene}, Jenny and {Guhathakurta}, Puragra and {Heinrich}, Chen and {Hemmati}, Shoubaneh and {Hendel}, David and {Henderson}, Calen and {Henning}, Thomas and {Hirata}, Christopher and {Ho}, Shirley and {Huff}, Eric and {Hutter}, Anne and {Jansen}, Rolf and {Jha}, Saurabh and {Johnson}, Samson and {Jones}, David and {Kasdin}, Jeremy and {Kelly}, Patrick and {Kirshner}, Robert and {Koekemoer}, Anton and {Kruk}, Jeffrey and {Lewis}, Nikole and {Macintosh}, Bruce and {Madau}, Piero and {Malhotra}, Sangeeta and {Mandel}, Kaisey and {Massara}, Elena and {Masters}, Daniel and {McEnery}, Julie and {McQuinn}, Kristen and {Melchior}, Peter and {Melton}, Mark and {Mennesson}, Bertrand and {Peeples}, Molly and {Penny}, Matthew and {Perlmutter}, Saul and {Pisani}, Alice and {Plazas}, Andr{\'e}s and {Poleski}, Radek and {Postman}, Marc and {Ranc}, Cl{\'e}ment and {Rauscher}, Bernard and {Rest}, Armin and {Roberge}, Aki and {Robertson}, Brant and {Rodney}, Steven and {Rhoads}, James and {Rhodes}, Jason and {Ryan}, Jr., Russell and {Sahu}, Kailash and {Sand}, David and {Scolnic}, Dan and {Seth}, Anil and {Shvartzvald}, Yossi and {Siellez}, Karelle and {Smith}, Arfon and {Spergel}, David and {Stassun}, Keivan and {Street}, Rachel and {Strolger}, Louis-Gregory and {Szalay}, Alexander and {Trauger}, John and {Troxel}, M.~A. and {Turnbull}, Margaret and {van der Marel}, Roeland and {von der Linden}, Anja and {Wang}, Yun and {Weinberg}, David and {Williams}, Benjamin and {Windhorst}, Rogier and {Wollack}, Edward and {Wu}, Hao-Yi and {Yee}, Jennifer and {Zimmerman}, Neil},
        title = "{The Wide Field Infrared Survey Telescope: 100 Hubbles for the 2020s}",
      journal = {arXiv e-prints},
         year = 2019,
        month = feb,
          eid = {arXiv:1902.05569},
        pages = {arXiv:1902.05569},
          doi = {10.48550/arXiv.1902.05569},
archivePrefix = {arXiv},
       eprint = {1902.05569},
 primaryClass = {astro-ph.IM},
       adsurl = {https://ui.adsabs.harvard.edu/abs/2019arXiv190205569A}
}

@ARTICLE{2025ApJ...992...39Z,
       author = {{Zucker}, Catherine and {Saydjari}, Andrew K. and {Speagle}, Joshua S. and {Schlafly}, Edward F. and {Green}, Gregory M. and {Benjamin}, Robert and {Peek}, Joshua and {Edenhofer}, Gordian and {Goodman}, Alyssa and {Kuhn}, Michael A. and {Finkbeiner}, Douglas P.},
        title = "{A Deep, High-angular-resolution 3D Dust Map of the Southern Galactic Plane}",
      journal = {\apj},
         year = 2025,
        month = oct,
       volume = {992},
       number = {1},
          eid = {39},
        pages = {39},
          doi = {10.3847/1538-4357/adfbe6},
archivePrefix = {arXiv},
       eprint = {2503.02657},
 primaryClass = {astro-ph.GA},
       adsurl = {https://ui.adsabs.harvard.edu/abs/2025ApJ...992...39Z}
}

@ARTICLE{2023ApJ...948...55H,
       author = {{Hensley}, Brandon S. and {Draine}, B.~T.},
        title = "{The Astrodust+PAH Model: A Unified Description of the Extinction, Emission, and Polarization from Dust in the Diffuse Interstellar Medium}",
      journal = {\apj},
         year = 2023,
        month = may,
       volume = {948},
       number = {1},
          eid = {55},
        pages = {55},
          doi = {10.3847/1538-4357/acc4c2},
archivePrefix = {arXiv},
       eprint = {2208.12365},
 primaryClass = {astro-ph.GA},
       adsurl = {https://ui.adsabs.harvard.edu/abs/2023ApJ...948...55H}
}

@ARTICLE{2025Sci...387.1209Z,
       author = {{Zhang}, Xiangyu and {Green}, Gregory M.},
        title = "{Three-dimensional maps of the interstellar dust extinction curve within the Milky Way galaxy}",
      journal = {Science},
         year = 2025,
        month = mar,
       volume = {387},
       number = {6739},
        pages = {1209-1214},
          doi = {10.1126/science.ado9787},
archivePrefix = {arXiv},
       eprint = {2407.14594},
 primaryClass = {astro-ph.GA},
       adsurl = {https://ui.adsabs.harvard.edu/abs/2025Sci...387.1209Z}
}

@ARTICLE{2006A&A...453..635M,
       author = {{Marshall}, D.~J. and {Robin}, A.~C. and {Reyl{\'e}}, C. and {Schultheis}, M. and {Picaud}, S.},
        title = "{Modelling the Galactic interstellar extinction distribution in three dimensions}",
      journal = {\aap},
         year = 2006,
        month = jul,
       volume = {453},
       number = {2},
        pages = {635-651},
          doi = {10.1051/0004-6361:20053842},
archivePrefix = {arXiv},
       eprint = {astro-ph/0604427},
 primaryClass = {astro-ph},
       adsurl = {https://ui.adsabs.harvard.edu/abs/2006A&A...453..635M}
}

@ARTICLE{2002A&A...390.1089P,
       author = {{Peeters}, E. and {Hony}, S. and {Van Kerckhoven}, C. and {Tielens}, A.~G.~G.~M. and {Allamandola}, L.~J. and {Hudgins}, D.~M. and {Bauschlicher}, C.~W.},
        title = "{The rich 6 to 9 vec mu m spectrum of interstellar PAHs}",
      journal = {\aap},
         year = 2002,
        month = aug,
       volume = {390},
        pages = {1089-1113},
          doi = {10.1051/0004-6361:20020773},
archivePrefix = {arXiv},
       eprint = {astro-ph/0205400},
 primaryClass = {astro-ph},
       adsurl = {https://ui.adsabs.harvard.edu/abs/2002A&A...390.1089P}
}

@ARTICLE{1998ApJ...508...44K,
       author = {{Kelsall}, T. and {Weiland}, J.~L. and {Franz}, B.~A. and {Reach}, W.~T. and {Arendt}, R.~G. and {Dwek}, E. and {Freudenreich}, H.~T. and {Hauser}, M.~G. and {Moseley}, S.~H. and {Odegard}, N.~P. and {Silverberg}, R.~F. and {Wright}, E.~L.},
        title = "{The COBE Diffuse Infrared Background Experiment Search for the Cosmic Infrared Background. II. Model of the Interplanetary Dust Cloud}",
      journal = {\apj},
         year = 1998,
        month = nov,
       volume = {508},
       number = {1},
        pages = {44-73},
          doi = {10.1086/306380},
archivePrefix = {arXiv},
       eprint = {astro-ph/9806250},
 primaryClass = {astro-ph},
       adsurl = {https://ui.adsabs.harvard.edu/abs/1998ApJ...508...44K}
}

@ARTICLE{2015ApJ...810...25G,
       author = {{Green}, Gregory M. and {Schlafly}, Edward F. and {Finkbeiner}, Douglas P. and {Rix}, Hans-Walter and {Martin}, Nicolas and {Burgett}, William and {Draper}, Peter W. and {Flewelling}, Heather and {Hodapp}, Klaus and {Kaiser}, Nicholas and {Kudritzki}, Rolf Peter and {Magnier}, Eugene and {Metcalfe}, Nigel and {Price}, Paul and {Tonry}, John and {Wainscoat}, Richard},
        title = "{A Three-dimensional Map of Milky Way Dust}",
      journal = {\apj},
         year = 2015,
        month = sep,
       volume = {810},
       number = {1},
          eid = {25},
        pages = {25},
          doi = {10.1088/0004-637X/810/1/25},
archivePrefix = {arXiv},
       eprint = {1507.01005},
 primaryClass = {astro-ph.GA},
       adsurl = {https://ui.adsabs.harvard.edu/abs/2015ApJ...810...25G}
}

@ARTICLE{2026ApJ..1001..165H,
       author = {{Hora}, Joseph L. and {Noh}, Jinyoung K. and {Melnick}, Gary J. and {Hensley}, Brandon S. and {Paladini}, Roberta and {Lee}, Jeong-Eun and {Ashby}, Matthew L.~N. and {Tolls}, Volker and {Kim}, Jaeyeong and {Werner}, Michael W. and {Bock}, James J. and {Bruton}, Sean and {Chen}, Shuang-Shuang and {Chang}, Tzu-Ching and {Chiang}, Yi-Kuan and {Cooray}, Asantha and {Crill}, Brendan P. and {Cukierman}, Ari J. and {Dor{\'e}}, Olivier and {Faisst}, Andreas L. and {Huai}, Zhaoyu and {Hui}, Howard and {Jeong}, Woong-Seob and {Kang}, Miju and {Korngut}, Phil M. and {Lee}, Ho-Gyu and {Lisse}, Carey M. and {Masters}, Daniel C. and {Murgia}, Giulia and {Nguyen}, Chi H. and {Rustamkulov}, Zafar and {Seok}, Ji Yeon and {Wen}, Robin Y. and {Yang}, Yujin and {Zemcov}, Michael},
        title = "{SPHEREx Widefield Infrared Spectral Mapping of Interstellar Ices and Polycyclic Aromatic Hydrocarbons}",
      journal = {\apj},
         year = 2026,
        month = apr,
       volume = {1001},
       number = {2},
          eid = {165},
        pages = {165},
          doi = {10.3847/1538-4357/ae5180},
archivePrefix = {arXiv},
       eprint = {2603.12390},
 primaryClass = {astro-ph.GA},
       adsurl = {https://ui.adsabs.harvard.edu/abs/2026ApJ..1001..165H}
}

@ARTICLE{2023AJ....165..126R,
       author = {{Raichoor}, A. and {Moustakas}, J. and {Newman}, Jeffrey A. and {Karim}, T. and {Ahlen}, S. and {Alam}, Shadab and {Bailey}, S. and {Brooks}, D. and {Dawson}, K. and {de la Macorra}, A. and {de Mattia}, A. and {Dey}, A. and {Dey}, Biprateep and {Dhungana}, G. and {Eftekharzadeh}, S. and {Eisenstein}, D.~J. and {Fanning}, K. and {Font-Ribera}, A. and {Garc{\'\i}a-Bellido}, J. and {Gazta{\~n}aga}, E. and {A Gontcho}, S. Gontcho and {Guy}, J. and {Honscheid}, K. and {Ishak}, M. and {Kehoe}, R. and {Kisner}, T. and {Kremin}, Anthony and {Lan}, Ting-Wen and {Landriau}, M. and {Le Guillou}, L. and {Levi}, Michael E. and {Magneville}, C. and {Manera}, M. and {Martini}, P. and {Meisner}, Aaron M. and {Myers}, Adam D. and {Nie}, Jundan and {Palanque-Delabrouille}, N. and {Percival}, W.~J. and {Poppett}, C. and {Prada}, F. and {Ross}, A.~J. and {Ruhlmann-Kleider}, V. and {Sabiu}, C.~G. and {Schlafly}, E.~F. and {Schlegel}, D. and {Tarl{\'e}}, Gregory and {Weaver}, B.~A. and {Y{\`e}che}, Christophe and {Zhou}, Rongpu and {Zhou}, Zhimin and {Zou}, H.},
        title = "{Target Selection and Validation of DESI Emission Line Galaxies}",
      journal = {\aj},
         year = 2023,
        month = mar,
       volume = {165},
       number = {3},
          eid = {126},
        pages = {126},
          doi = {10.3847/1538-3881/acb213},
archivePrefix = {arXiv},
       eprint = {2208.08513},
 primaryClass = {astro-ph.CO},
       adsurl = {https://ui.adsabs.harvard.edu/abs/2023AJ....165..126R}
}

@ARTICLE{2019MNRAS.484.4127A,
       author = {{Alonso}, David and {Sanchez}, Javier and {Slosar}, An{\v{z}}e and {LSST Dark Energy Science Collaboration}},
        title = "{A unified pseudo-C$_{{\ensuremath{\ell}}}$ framework}",
      journal = {\mnras},
         year = 2019,
        month = apr,
       volume = {484},
       number = {3},
        pages = {4127-4151},
          doi = {10.1093/mnras/stz093},
archivePrefix = {arXiv},
       eprint = {1809.09603},
 primaryClass = {astro-ph.CO},
       adsurl = {https://ui.adsabs.harvard.edu/abs/2019MNRAS.484.4127A}
}

@ARTICLE{1992ApJ...397..420B,
       author = {{Boggess}, N.~W. and {Mather}, J.~C. and {Weiss}, R. and {Bennett}, C.~L. and {Cheng}, E.~S. and {Dwek}, E. and {Gulkis}, S. and {Hauser}, M.~G. and {Janssen}, M.~A. and {Kelsall}, T. and {Meyer}, S.~S. and {Moseley}, S.~H. and {Murdock}, T.~L. and {Shafer}, R.~A. and {Silverberg}, R.~F. and {Smoot}, G.~F. and {Wilkinson}, D.~T. and {Wright}, E.~L.},
        title = "{The COBE Mission: Its Design and Performance Two Years after Launch}",
      journal = {\apj},
         year = 1992,
        month = oct,
       volume = {397},
        pages = {420},
          doi = {10.1086/171797},
       adsurl = {https://ui.adsabs.harvard.edu/abs/1992ApJ...397..420B}
}

@ARTICLE{2024A&A...684A..34Y,
       author = {{Ysard}, N. and {Jones}, A.~P. and {Guillet}, V. and {Demyk}, K. and {Decleir}, M. and {Verstraete}, L. and {Choubani}, I. and {Miville-Desch{\^e}nes}, M.-A. and {Fanciullo}, L.},
        title = "{THEMIS 2.0: A self-consistent model for dust extinction, emission, and polarisation}",
      journal = {\aap},
         year = 2024,
        month = apr,
       volume = {684},
          eid = {A34},
        pages = {A34},
          doi = {10.1051/0004-6361/202348391},
archivePrefix = {arXiv},
       eprint = {2401.07739},
 primaryClass = {astro-ph.GA},
       adsurl = {https://ui.adsabs.harvard.edu/abs/2024A&A...684A..34Y}
}

@ARTICLE{2012arXiv1211.0310L,
       author = {{LSST Dark Energy Science Collaboration}},
        title = "{Large Synoptic Survey Telescope: Dark Energy Science Collaboration}",
      journal = {arXiv e-prints},
         year = 2012,
        month = nov,
          eid = {arXiv:1211.0310},
        pages = {arXiv:1211.0310},
          doi = {10.48550/arXiv.1211.0310},
archivePrefix = {arXiv},
       eprint = {1211.0310},
 primaryClass = {astro-ph.CO},
       adsurl = {https://ui.adsabs.harvard.edu/abs/2012arXiv1211.0310L}
}

@ARTICLE{1930PASP...42..214T,
       author = {{Trumpler}, Robert J.},
        title = "{Absorption of Light in the Galactic System}",
      journal = {\pasp},
         year = 1930,
        month = aug,
       volume = {42},
       number = {248},
        pages = {214},
          doi = {10.1086/124039},
       adsurl = {https://ui.adsabs.harvard.edu/abs/1930PASP...42..214T}
}

@ARTICLE{2026ApJS..286...22C,
       author = {{Cukierman}, Ari J. and {Chen}, Shuang-Shuang and {Kang}, Jae Hwan and {Minasyan}, Mary H. and {Murgia}, Giulia and {Bock}, James J. and {Chang}, Tzu-Ching and {Chiang}, Yi-Kuan and {Crill}, Brendan P. and {Dor{\'e}}, Olivier and {Dowell}, C. Darren and {Faisst}, Andreas L. and {Hora}, Joseph L. and {Hui}, Howard and {Kang}, Miju and {Korngut}, Phil M. and {Lee}, Ho-Gyu and {Lee}, Bomee and {Masters}, Daniel C. and {Melnick}, Gary J. and {Mirocha}, Jordan and {Nguyen}, Chi H. and {Rustamkulov}, Zafar and {Tolls}, Volker and {Werner}, Michael W. and {Yang}, Yujin and {Zemcov}, Michael},
        title = "{Spectral Map Making with SPHEREx}",
      journal = {\apjs},
         year = 2026,
        month = sep,
       volume = {286},
       number = {1},
          eid = {22},
        pages = {22},
          doi = {10.3847/1538-4365/ae91df},
archivePrefix = {arXiv},
       eprint = {2603.25790},
 primaryClass = {astro-ph.IM},
       adsurl = {https://ui.adsabs.harvard.edu/abs/2026ApJS..286...22C}
}

@ARTICLE{1995PhRvD..52.4307K,
       author = {{Knox}, Lloyd},
        title = "{Determination of inflationary observables by cosmic microwave background anisotropy experiments}",
      journal = {\prd},
         year = 1995,
        month = oct,
       volume = {52},
       number = {8},
        pages = {4307-4318},
          doi = {10.1103/PhysRevD.52.4307},
archivePrefix = {arXiv},
       eprint = {astro-ph/9504054},
 primaryClass = {astro-ph},
       adsurl = {https://ui.adsabs.harvard.edu/abs/1995PhRvD..52.4307K}
}

@ARTICLE{2026ApJS..285...67A,
       author = {{Akeson}, Rachel and {Dubois-Felsmann}, Gregory P. and {Crill}, Brendan P. and {Faisst}, Andreas L. and {Fatahi}, Tamim and {Fazar}, Candice M. and {Goldina}, Tatiana and {Masters}, Daniel C. and {Nelson}, Christina and {Paladini}, Roberta and {Teplitz}, Harry I. and {Torrini}, Gabriela and {Velicheti}, Phani and {Ashby}, Matthew L.~N. and {Avner}, Dan and {Bach}, Yoonsoo P. and {Bock}, James J. and {Bruton}, Sean and {Bryan}, Sean A. and {Chang}, Tzu-Ching and {Chen}, Shuang-Shuang and {Cooray}, Asantha and {Cukierman}, Ari J. and {Dor{\'e}}, Olivier and {Dowell}, C. Darren and {Everett}, Spencer and {Feder}, Richard M. and {Huai}, Zhaoyu and {Hui}, Howard and {Jeong}, Woong-Seob and {Jo}, Young-Soo and {Korngut}, Phil M. and {Kwon}, Yuna G. and {Lee}, Bomee and {Melnick}, Gary J. and {Murgia}, Giulia and {Nguyen}, Chi H. and {Pourrahmani}, Milad and {Rustamkulov}, Zafar and {Symons}, Teresa and {Tolls}, Volker and {Wang}, Pao-Yu and {Yang}, Yujin and {Zemcov}, Michael},
        title = "{The SPHEREx Image and Spectrophotometry Processing Pipeline}",
      journal = {\apjs},
         year = 2026,
        month = aug,
       volume = {285},
       number = {2},
          eid = {67},
        pages = {67},
          doi = {10.3847/1538-4365/ae8843},
archivePrefix = {arXiv},
       eprint = {2511.15823},
 primaryClass = {astro-ph.IM},
       adsurl = {https://ui.adsabs.harvard.edu/abs/2026ApJS..285...67A}
}

@ARTICLE{2026ApJ..1004..176M,
       author = {{Murgia}, Giulia and {Cukierman}, Ari J. and {Hensley}, Brandon S. and {Ashby}, Matthew L.~N. and {Bock}, James J. and {Chang}, Tzu-Ching and {Chen}, Shuang-Shuang and {Cheng}, Yun-Ting and {Chiang}, Yi-Kuan and {Cooray}, Asantha and {Crill}, Brendan P. and {Dor{\'e}}, Olivier and {Dowell}, C. Darren and {Faisst}, Andreas L. and {Hora}, Joseph L. and {Hui}, Howard and {Kang}, Miju and {Kang}, Jae Hwan and {Korngut}, Phil M. and {Lee}, Dennis and {Lee}, Jeong-Eun and {Lee}, Bomee and {Lisse}, Carey M. and {Masters}, Daniel C. and {Melnick}, Gary J. and {Minasyan}, Mary H. and {Nguyen}, Chi H. and {Paladini}, Roberta and {Tolls}, Volker and {Wen}, Robin Y. and {Werner}, Michael W. and {Zemcov}, Michael},
        title = "{SPHEREx Mapping of Diffuse PAH and H II Emission in the Galactic Plane}",
      journal = {\apj},
         year = 2026,
        month = jun,
       volume = {1004},
       number = {2},
          eid = {176},
        pages = {176},
          doi = {10.3847/1538-4357/ae734b},
archivePrefix = {arXiv},
       eprint = {2603.23620},
 primaryClass = {astro-ph.GA},
       adsurl = {https://ui.adsabs.harvard.edu/abs/2026ApJ..1004..176M}
}

@ARTICLE{2020ApJ...905...55L,
       author = {{Lai}, Thomas S.-Y. and {Smith}, J.~D.~T. and {Baba}, Shunsuke and {Spoon}, Henrik W.~W. and {Imanishi}, Masatoshi},
        title = "{All the PAHs: An AKARI-Spitzer Cross-archival Spectroscopic Survey of Aromatic Emission in Galaxies}",
      journal = {\apj},
         year = 2020,
        month = dec,
       volume = {905},
       number = {1},
          eid = {55},
        pages = {55},
          doi = {10.3847/1538-4357/abc002},
archivePrefix = {arXiv},
       eprint = {2010.05034},
 primaryClass = {astro-ph.GA},
       adsurl = {https://ui.adsabs.harvard.edu/abs/2020ApJ...905...55L}
}

@ARTICLE{2020A&A...639A.138L,
       author = {{Leike}, R.~H. and {Glatzle}, M. and {En{\ss}lin}, T.~A.},
        title = "{Resolving nearby dust clouds}",
      journal = {\aap},
         year = 2020,
        month = jul,
       volume = {639},
          eid = {A138},
        pages = {A138},
          doi = {10.1051/0004-6361/202038169},
archivePrefix = {arXiv},
       eprint = {2004.06732},
 primaryClass = {astro-ph.GA},
       adsurl = {https://ui.adsabs.harvard.edu/abs/2020A&A...639A.138L}
}

@ARTICLE{2025MNRAS.544.3280T,
       author = {{Tokunaga}, Alan T. and {Bernstein}, Lawrence S. and {Onaka}, Takashi},
        title = "{A critical examination of the PAH hypothesis}",
      journal = {\mnras},
         year = 2025,
        month = dec,
       volume = {544},
       number = {4},
        pages = {3280-3289},
          doi = {10.1093/mnras/staf1874},
archivePrefix = {arXiv},
       eprint = {2510.26970},
 primaryClass = {astro-ph.GA},
       adsurl = {https://ui.adsabs.harvard.edu/abs/2025MNRAS.544.3280T}
}

@ARTICLE{2025ApJS..281...10C,
       author = {{Crill}, Brendan P. and {Bach}, Yoonsoo P. and {Bryan}, Sean A. and {Choppin de Janvry}, Jean and {Cukierman}, Ari J. and {Dowell}, C. Darren and {Everett}, Spencer W. and {Fazar}, Candice and {Goldina}, Tatiana and {Huai}, Zhaoyu and {Hui}, Howard and {Jeong}, Woong-Seob and {Kang}, Jae Hwan and {Korngut}, Phillip M. and {Lee}, Jae Joon and {Masters}, Daniel C. and {Nguyen}, Chi H. and {Pyo}, Jeonghyun and {Symons}, Teresa and {Tolls}, Volker and {Yang}, Yujin and {Zemcov}, Michael and {Akeson}, Rachel and {Ashby}, Matthew L.~N. and {Bock}, James J. and {Chang}, Tzu-Ching and {Cheng}, Yun-Ting and {Chiang}, Yi-Kuan and {Cooray}, Asantha and {Dor{\'e}}, Olivier and {Faisst}, Andreas L. and {Feder}, Richard M. and {Werner}, Michael W.},
        title = "{The SPHEREx Sky Simulator: Science Data Modeling for the First All-sky Near-infrared Spectral Survey}",
      journal = {\apjs},
         year = 2025,
        month = nov,
       volume = {281},
       number = {1},
          eid = {10},
        pages = {10},
          doi = {10.3847/1538-4365/ae04cc},
archivePrefix = {arXiv},
       eprint = {2505.24856},
 primaryClass = {astro-ph.IM},
       adsurl = {https://ui.adsabs.harvard.edu/abs/2025ApJS..281...10C}
}

@ARTICLE{2012MNRAS.424..564R,
       author = {{Ross}, Ashley J. and {Percival}, Will J. and {S{\'a}nchez}, Ariel G. and {Samushia}, Lado and {Ho}, Shirley and {Kazin}, Eyal and {Manera}, Marc and {Reid}, Beth and {White}, Martin and {Tojeiro}, Rita and {McBride}, Cameron K. and {Xu}, Xiaoying and {Wake}, David A. and {Strauss}, Michael A. and {Montesano}, Francesco and {Swanson}, Molly E.~C. and {Bailey}, Stephen and {Bolton}, Adam S. and {Dorta}, Antonio Montero and {Eisenstein}, Daniel J. and {Guo}, Hong and {Hamilton}, Jean-Christophe and {Nichol}, Robert C. and {Padmanabhan}, Nikhil and {Prada}, Francisco and {Schlegel}, David J. and {Maga{\~n}a}, Mariana Vargas and {Zehavi}, Idit and {Blanton}, Michael and {Bizyaev}, Dmitry and {Brewington}, Howard and {Cuesta}, Antonio J. and {Malanushenko}, Elena and {Malanushenko}, Viktor and {Oravetz}, Daniel and {Parejko}, John and {Pan}, Kaike and {Schneider}, Donald P. and {Shelden}, Alaina and {Simmons}, Audrey and {Snedden}, Stephanie and {Zhao}, Gong-bo},
        title = "{The clustering of galaxies in the SDSS-III Baryon Oscillation Spectroscopic Survey: analysis of potential systematics}",
      journal = {\mnras},
         year = 2012,
        month = jul,
       volume = {424},
       number = {1},
        pages = {564-590},
          doi = {10.1111/j.1365-2966.2012.21235.x},
archivePrefix = {arXiv},
       eprint = {1203.6499},
 primaryClass = {astro-ph.CO},
       adsurl = {https://ui.adsabs.harvard.edu/abs/2012MNRAS.424..564R}
}

@ARTICLE{2025OJAp....8E..83Z,
       author = {{Zhou}, Rongpu and {Guy}, Julien and {Koposov}, Sergey E. and {Schlafly}, Edward F. and {Schlegel}, David and {Aguilar}, Jessica and {Ahlen}, Steven and {Bailey}, Stephen and {Bianchi}, David and {Brooks}, David and {Chaussidon}, Edmond and {Claybaugh}, Todd and {Dawson}, Kyle and {de la Macorra}, Axel and {Dey}, Arjun and {Dey}, Biprateep and {Eisenstein}, Daniel J. and {Ferraro}, Simone and {Font-Ribera}, Andreu and {Forero-Romero}, Jaime E. and {Gazta{\~n}aga}, Enrique and {Gontcho}, Satya Gontcho A. and {Gutierrez}, Gaston and {Honscheid}, Klaus and {Juneau}, Stephanie and {Kehoe}, Robert and {Kirkby}, David and {Kisner}, Theodore and {Kremin}, Anthony and {Lambert}, Andrew and {Landriau}, Martin and {Guillou}, Laurent Le and {Levi}, Michael E. and {Li}, Ting S. and {Manera}, Marc and {Martini}, Paul and {Meisner}, Aaron and {Miquel}, Ramon and {Moustakas}, John and {Myers}, Adam D. and {Newman}, Jeffrey A. and {Niz}, Gustavo and {Palanque-Delabrouille}, Nathalie and {Percival}, Will J. and {Poppett}, Claire and {Prada}, Francisco and {Raichoor}, Anand and {Ross}, Ashley J. and {Rossi}, Graziano and {Sanchez}, Eusebio and {Saydjari}, Andrew K. and {Schubnell}, Michael and {Sprayberry}, David and {Tarl}, Gregory and {Weaver}, Benjamin A. and {Zarrouk}, Pauline and {Zou}, Hu},
        title = "{Stellar reddening map from DESI imaging and spectroscopy}",
      journal = {The Open Journal of Astrophysics},
         year = 2025,
        month = jul,
       volume = {8},
          eid = {83},
        pages = {83},
          doi = {10.33232/001c.141680},
archivePrefix = {arXiv},
       eprint = {2409.05140},
 primaryClass = {astro-ph.GA},
       adsurl = {https://ui.adsabs.harvard.edu/abs/2025OJAp....8E..83Z}
}

@ARTICLE{2017ApJ...846...38L,
       author = {{Lenz}, Daniel and {Hensley}, Brandon S. and {Dor{\'e}}, Olivier},
        title = "{A New, Large-scale Map of Interstellar Reddening Derived from H I Emission}",
      journal = {\apj},
         year = 2017,
        month = sep,
       volume = {846},
       number = {1},
          eid = {38},
        pages = {38},
          doi = {10.3847/1538-4357/aa84af},
archivePrefix = {arXiv},
       eprint = {1706.00011},
 primaryClass = {astro-ph.GA},
       adsurl = {https://ui.adsabs.harvard.edu/abs/2017ApJ...846...38L}
}

@ARTICLE{2026ApJS..284...10H,
       author = {{Hui}, Howard and {Bock}, James J. and {Condon}, Samuel and {Dowell}, C. Darren and {Jeong}, Woong-Seob and {Jo}, Young-soo and {Korngut}, Phil M. and {Manatt}, Kenneth and {Nguyen}, Chi and {Nguyen}, Hien and {Padin}, Stephen and {Park}, Sung-Joon and {Pyo}, Jeonghyun and {Yang}, Yujin and {Ashby}, Matthew L.~N. and {Bach}, Yoonsoo P. and {Chang}, Tzu-Ching and {Cheng}, Yun-Ting and {Chiang}, Yi-Kuan and {Cooray}, Asantha and {Crill}, Brendan P. and {Cukierman}, Ari J. and {Dor{\'e}}, Olivier and {Faisst}, Andreas L. and {Hora}, Joseph L. and {Kang}, Jae Hwan and {Lee}, BoMee and {Lisse}, Carey M. and {Masters}, Daniel C. and {Paladini}, Roberta and {Rustamkulov}, Zafar and {Tolls}, Volker and {Werner}, Michael W. and {Zemcov}, Michael},
        title = "{Spectral Response of SPHEREx}",
      journal = {\apjs},
         year = 2026,
        month = may,
       volume = {284},
       number = {1},
          eid = {10},
        pages = {10},
          doi = {10.3847/1538-4365/ae522c},
       adsurl = {https://ui.adsabs.harvard.edu/abs/2026ApJS..284...10H}
}

@ARTICLE{2025ApJ...994...61L,
       author = {{Lee}, Dennis and {Hensley}, Brandon S. and {Chang}, Tzu-Ching and {Dor{\'e}}, Olivier},
        title = "{The End of the Road for Far-infrared Reddening Maps? Evidence for Reddening Errors Driven by Changes in Polycyclic Aromatic Hydrocarbon Abundance}",
      journal = {\apj},
         year = 2025,
        month = nov,
       volume = {994},
       number = {1},
          eid = {61},
        pages = {61},
          doi = {10.3847/1538-4357/ae0c12},
archivePrefix = {arXiv},
       eprint = {2508.11746},
 primaryClass = {astro-ph.GA},
       adsurl = {https://ui.adsabs.harvard.edu/abs/2025ApJ...994...61L}
}

@ARTICLE{2025ApJ...993...30C,
       author = {{Cheng}, Yun-Ting and {Hensley}, Brandon S. and {Lai}, Thomas S.-Y.},
        title = "{Feature Intensity Mapping: Polycyclic Aromatic Hydrocarbon Emission from All Galaxies Across Cosmic Time}",
      journal = {\apj},
         year = 2025,
        month = nov,
       volume = {993},
       number = {1},
          eid = {30},
        pages = {30},
          doi = {10.3847/1538-4357/ae044e},
archivePrefix = {arXiv},
       eprint = {2506.13863},
 primaryClass = {astro-ph.CO},
       adsurl = {https://ui.adsabs.harvard.edu/abs/2025ApJ...993...30C}
}

@ARTICLE{2010ApJ...725.1175S,
       author = {{Schlafly}, Edward F. and {Finkbeiner}, Douglas P. and {Schlegel}, David J. and {Juri{\'c}}, Mario and {Ivezi{\'c}}, {\v{Z}}eljko and {Gibson}, Robert R. and {Knapp}, Gillian R. and {Weaver}, Benjamin A.},
        title = "{The Blue Tip of the Stellar Locus: Measuring Reddening with the Sloan Digital Sky Survey}",
      journal = {\apj},
         year = 2010,
        month = dec,
       volume = {725},
       number = {1},
        pages = {1175-1191},
          doi = {10.1088/0004-637X/725/1/1175},
archivePrefix = {arXiv},
       eprint = {1009.4933},
 primaryClass = {astro-ph.GA},
       adsurl = {https://ui.adsabs.harvard.edu/abs/2010ApJ...725.1175S}
}

@ARTICLE{2010ApJ...719..415P,
       author = {{Peek}, J.~E.~G. and {Graves}, Genevieve J.},
        title = "{A Correction to the Standard Galactic Reddening Map: Passive Galaxies as Standard Crayons}",
      journal = {\apj},
         year = 2010,
        month = aug,
       volume = {719},
       number = {1},
        pages = {415-424},
          doi = {10.1088/0004-637X/719/1/415},
archivePrefix = {arXiv},
       eprint = {1006.3310},
 primaryClass = {astro-ph.GA},
       adsurl = {https://ui.adsabs.harvard.edu/abs/2010ApJ...719..415P}
}

@ARTICLE{2014ApJ...783..114G,
       author = {{Green}, Gregory Maurice and {Schlafly}, Edward F. and {Finkbeiner}, Douglas P. and {Juri{\'c}}, Mario and {Rix}, Hans-Walter and {Burgett}, Will and {Chambers}, Kenneth C. and {Draper}, Peter W. and {Flewelling}, Heather and {Kudritzki}, Rolf Peter and {Magnier}, Eugene and {Martin}, Nicolas and {Metcalfe}, Nigel and {Tonry}, John and {Wainscoat}, Richard and {Waters}, Christopher},
        title = "{Measuring Distances and Reddenings for a Billion Stars: Toward a 3D Dust Map from Pan-STARRS 1}",
      journal = {\apj},
         year = 2014,
        month = mar,
       volume = {783},
       number = {2},
          eid = {114},
        pages = {114},
          doi = {10.1088/0004-637X/783/2/114},
archivePrefix = {arXiv},
       eprint = {1401.1508},
 primaryClass = {astro-ph.GA},
       adsurl = {https://ui.adsabs.harvard.edu/abs/2014ApJ...783..114G}
}

@ARTICLE{2014ApJ...781....5M,
       author = {{Meisner}, Aaron M. and {Finkbeiner}, Douglas P.},
        title = "{A Full-sky, High-resolution Atlas of Galactic 12 {\ensuremath{\mu}}m Dust Emission with WISE}",
      journal = {\apj},
         year = 2014,
        month = jan,
       volume = {781},
       number = {1},
          eid = {5},
        pages = {5},
          doi = {10.1088/0004-637X/781/1/5},
archivePrefix = {arXiv},
       eprint = {1312.0947},
 primaryClass = {astro-ph.IM},
       adsurl = {https://ui.adsabs.harvard.edu/abs/2014ApJ...781....5M}
}

@ARTICLE{1991ApJ...380..452T,
       author = {{Tokunaga}, A.~T. and {Sellgren}, K. and {Smith}, R.~G. and {Nagata}, T. and {Sakata}, A. and {Nakada}, Y.},
        title = "{High-Resolution Spectra of the 3.29 Micron Interstellar Emission Feature: A Summary}",
      journal = {\apj},
         year = 1991,
        month = oct,
       volume = {380},
        pages = {452},
          doi = {10.1086/170603},
       adsurl = {https://ui.adsabs.harvard.edu/abs/1991ApJ...380..452T}
}

@ARTICLE{2019A&A...631A..32L,
       author = {{Leike}, R.~H. and {En{\ss}lin}, T.~A.},
        title = "{Charting nearby dust clouds using Gaia data only}",
      journal = {\aap},
         year = 2019,
        month = nov,
       volume = {631},
          eid = {A32},
        pages = {A32},
          doi = {10.1051/0004-6361/201935093},
archivePrefix = {arXiv},
       eprint = {1901.05971},
 primaryClass = {astro-ph.GA},
       adsurl = {https://ui.adsabs.harvard.edu/abs/2019A&A...631A..32L}
}

@ARTICLE{2015ApJ...798...88M,
       author = {{Meisner}, Aaron M. and {Finkbeiner}, Douglas P.},
        title = "{Modeling Thermal Dust Emission with Two Components: Application to the Planck High Frequency Instrument Maps}",
      journal = {\apj},
         year = 2015,
        month = jan,
       volume = {798},
       number = {2},
          eid = {88},
        pages = {88},
          doi = {10.1088/0004-637X/798/2/88},
archivePrefix = {arXiv},
       eprint = {1410.7523},
 primaryClass = {astro-ph.GA},
       adsurl = {https://ui.adsabs.harvard.edu/abs/2015ApJ...798...88M}
}

@ARTICLE{2025JCAP...02..045K,
       author = {{Karim}, Tanveer and {Singh}, Sukhdeep and {Rezaie}, Mehdi and {Eisenstein}, Daniel and {Hadzhiyska}, Boryana and {Speagle}, Joshua S. and {Aguilar}, Jessica Nicole and {Ahlen}, Steven and {Brooks}, David and {Claybaugh}, Todd and {de la Macorra}, Axel and {Ferraro}, Simone and {Forero-Romero}, Jaime E. and {Gazta{\~n}aga}, Enrique and {Gontcho}, Satya Gontcho A. and {Gutierrez}, Gaston and {Guy}, Julien and {Honscheid}, Klaus and {Juneau}, Stephanie and {Kirkby}, David and {Krolewski}, Alex and {Lambert}, Andrew and {Landriau}, Martin and {Levi}, Michael and {Meisner}, Aaron and {Miquel}, Ramon and {Moustakas}, John and {Mu{\~n}oz-Guti{\'e}rrez}, Andrea and {Myers}, Adam and {Niz}, Gustavo and {Palanque-Delabrouille}, Nathalie and {Percival}, Will and {Prada}, Francisco and {Rossi}, Graziano and {Sanchez}, Eusebio and {Schlafly}, Edward and {Schlegel}, David and {Schubnell}, Michael and {Sprayberry}, David and {Tarl{\'e}}, Gregory and {Weaver}, Benjamin Alan and {Zou}, Hu},
        title = "{Measuring {\ensuremath{\sigma}} $_{8}$ using DESI Legacy Imaging Surveys Emission-Line galaxies and Planck CMB lensing, and the impact of dust on parameter inference}",
      journal = {\jcap},
         year = 2025,
        month = feb,
       volume = {2025},
       number = {2},
          eid = {045},
        pages = {045},
          doi = {10.1088/1475-7516/2025/02/045},
archivePrefix = {arXiv},
       eprint = {2408.15909},
 primaryClass = {astro-ph.CO},
       adsurl = {https://ui.adsabs.harvard.edu/abs/2025JCAP...02..045K}
}

@ARTICLE{2022ApJ...933..155S,
       author = {{Saydjari}, Andrew K. and {Finkbeiner}, Douglas P.},
        title = "{Photometry on Structured Backgrounds: Local Pixel-wise Infilling by Regression}",
      journal = {\apj},
         year = 2022,
        month = jul,
       volume = {933},
       number = {2},
          eid = {155},
        pages = {155},
          doi = {10.3847/1538-4357/ac6875},
archivePrefix = {arXiv},
       eprint = {2201.07246},
 primaryClass = {astro-ph.IM},
       adsurl = {https://ui.adsabs.harvard.edu/abs/2022ApJ...933..155S}
}

@ARTICLE{2019ApJ...870..120C,
       author = {{Chiang}, Yi-Kuan and {M{\'e}nard}, Brice},
        title = "{Extragalactic Imprints in Galactic Dust Maps}",
      journal = {\apj},
         year = 2019,
        month = jan,
       volume = {870},
       number = {2},
          eid = {120},
        pages = {120},
          doi = {10.3847/1538-4357/aaf4f6},
archivePrefix = {arXiv},
       eprint = {1808.03294},
 primaryClass = {astro-ph.GA},
       adsurl = {https://ui.adsabs.harvard.edu/abs/2019ApJ...870..120C}
}

@ARTICLE{2014MNRAS.443.1192C,
       author = {{Chen}, B.-Q. and {Liu}, X.-W. and {Yuan}, H.-B. and {Zhang}, H.-H. and {Schultheis}, M. and {Jiang}, B.-W. and {Huang}, Y. and {Xiang}, M.-S. and {Zhao}, H.-B. and {Yao}, J.-S. and {Lu}, H.},
        title = "{A three-dimensional extinction map of the Galactic anticentre from multiband photometry}",
      journal = {\mnras},
         year = 2014,
        month = sep,
       volume = {443},
       number = {2},
        pages = {1192-1210},
          doi = {10.1093/mnras/stu1192},
archivePrefix = {arXiv},
       eprint = {1406.3996},
 primaryClass = {astro-ph.SR},
       adsurl = {https://ui.adsabs.harvard.edu/abs/2014MNRAS.443.1192C}
}

@ARTICLE{2024ApJ...960...96C,
       author = {{C{\'o}rdova Rosado}, Rodrigo and {Hensley}, Brandon S. and {Clark}, Susan E. and {Duivenvoorden}, Adriaan J. and {Atkins}, Zachary and {Battistelli}, Elia Stefano and {Choi}, Steve K. and {Dunkley}, Jo and {Herv{\'\i}as-Caimapo}, Carlos and {Li}, Zack and {Louis}, Thibaut and {Naess}, Sigurd and {Page}, Lyman A. and {Partridge}, Bruce and {Sif{\'o}n}, Crist{\'o}bal and {Staggs}, Suzanne T. and {Vargas}, Cristian and {Wollack}, Edward J.},
        title = "{The Atacama Cosmology Telescope: Galactic Dust Structure and the Cosmic PAH Background in Cross-correlation with WISE}",
      journal = {\apj},
         year = 2024,
        month = jan,
       volume = {960},
       number = {2},
          eid = {96},
        pages = {96},
          doi = {10.3847/1538-4357/ad05cd},
archivePrefix = {arXiv},
       eprint = {2307.06352},
 primaryClass = {astro-ph.GA},
       adsurl = {https://ui.adsabs.harvard.edu/abs/2024ApJ...960...96C}
}

@ARTICLE{2013MNRAS.432.2945H,
       author = {{Huterer}, Dragan and {Cunha}, Carlos E. and {Fang}, Wenjuan},
        title = "{Calibration errors unleashed: effects on cosmological parameters and requirements for large-scale structure surveys}",
      journal = {\mnras},
         year = 2013,
        month = jul,
       volume = {432},
       number = {4},
        pages = {2945-2961},
          doi = {10.1093/mnras/stt653},
archivePrefix = {arXiv},
       eprint = {1211.1015},
 primaryClass = {astro-ph.CO},
       adsurl = {https://ui.adsabs.harvard.edu/abs/2013MNRAS.432.2945H}
}

@ARTICLE{2016ApJ...827...45H,
       author = {{Hensley}, Brandon S. and {Draine}, B.~T. and {Meisner}, Aaron M.},
        title = "{A Case Against Spinning PAHs as the Source of the Anomalous Microwave Emission}",
      journal = {\apj},
         year = 2016,
        month = aug,
       volume = {827},
       number = {1},
          eid = {45},
        pages = {45},
          doi = {10.3847/0004-637X/827/1/45},
archivePrefix = {arXiv},
       eprint = {1505.02157},
 primaryClass = {astro-ph.GA},
       adsurl = {https://ui.adsabs.harvard.edu/abs/2016ApJ...827...45H}
}

@ARTICLE{2007PASJ...59..205Y,
       author = {{Yahata}, Kazuhiro and {Yonehara}, Atsunori and {Suto}, Yasushi and {Turner}, Edwin L. and {Broadhurst}, Tom and {Finkbeiner}, Douglas P.},
        title = "{The Effect of FIR Emission from SDSS Galaxies on the SFD Galactic Extinction Map}",
      journal = {\pasj},
         year = 2007,
        month = feb,
       volume = {59},
        pages = {205-219},
          doi = {10.1093/pasj/59.1.205},
archivePrefix = {arXiv},
       eprint = {astro-ph/0607098},
 primaryClass = {astro-ph},
       adsurl = {https://ui.adsabs.harvard.edu/abs/2007PASJ...59..205Y}
}

@ARTICLE{2014MNRAS.445..256S,
       author = {{Sale}, S.~E. and {Magorrian}, J.},
        title = "{Three-dimensional extinction mapping using Gaussian random fields}",
      journal = {\mnras},
         year = 2014,
        month = nov,
       volume = {445},
       number = {1},
        pages = {256-269},
          doi = {10.1093/mnras/stu1728},
archivePrefix = {arXiv},
       eprint = {1406.1177},
 primaryClass = {astro-ph.IM},
       adsurl = {https://ui.adsabs.harvard.edu/abs/2014MNRAS.445..256S}
}

@ARTICLE{harris2020array,
 title         = {Array programming with {NumPy}},
 author        = {Charles R. Harris and K. Jarrod Millman and St{'{e}}fan J.
                 van der Walt and Ralf Gommers and Pauli Virtanen and David
                 Cournapeau and Eric Wieser and Julian Taylor and Sebastian
                 Berg and Nathaniel J. Smith and Robert Kern and Matti Picus
                 and Stephan Hoyer and Marten H. van Kerkwijk and Matthew
                 Brett and Allan Haldane and Jaime Fern{'{a}}ndez del
                 R{'{\i}}o and Mark Wiebe and Pearu Peterson and Pierre
                 G{'{e}}rard-Marchant and Kevin Sheppard and Tyler Reddy and
                 Warren Weckesser and Hameer Abbasi and Christoph Gohlke and
                 Travis E. Oliphant},
 year          = {2020},
 month         = sep,
 journal       = {Nature},
 volume        = {585},
 number        = {7825},
 pages         = {357--362},
 doi           = {10.1038/s41586-020-2649-2},
 publisher     = {Springer Science and Business Media {LLC}},
 url           = {https://doi.org/10.1038/s41586-020-2649-2}
}
\bibliographystyle{aasjournalv7}
\end{document}